\documentclass[reprint,aps,prb,superscriptaddress]{revtex4-2}
\usepackage[T1]{fontenc}
\usepackage[utf8]{inputenc}
\usepackage{graphicx}
\usepackage{bm}
\usepackage{xcolor}
\usepackage{xr}
\usepackage{amsmath}

\DeclareUnicodeCharacter{2013}{--}  
\DeclareUnicodeCharacter{2014}{---} 
\DeclareUnicodeCharacter{2212}{-}   
\DeclareUnicodeCharacter{00A0}{ }   
\DeclareUnicodeCharacter{2018}{'}   
\DeclareUnicodeCharacter{2019}{'}   
\DeclareUnicodeCharacter{201C}{"}   
\DeclareUnicodeCharacter{201D}{"}   

\begin{document}

\title{Hydrogen in Brownmillerite Perovskites: \\
First-Principles Insights into Energetics and Induced Electronic-Magnetic Changes}

\author{Vladislav Korostelev}
\affiliation{Department of Materials Engineering, Auburn University, Auburn, Alabama, USA}

\author{Pjotrs \v{Z}guns}
\affiliation{Department of Materials Science and Engineering, Massachusetts Institute of Technology, Cambridge, Massachusetts, USA}
\affiliation{Institute of Solid State Physics, University of Latvia, Riga, Latvia}

\author{Konstantin Klyukin}
\email{klyukin@auburn.edu}
\affiliation{Department of Materials Engineering, Auburn University, Auburn, Alabama, USA}


\begin{abstract}

Hydrogen uptake in brownmillerite perovskites A$_2$B$_2$O$_5$ offers an (electro)chemically accessible route to tune functional properties, but mechanistic understanding and design rules for hydrogen-responsive oxides remain limited. Here we employ density functional theory (DFT) to quantify how H absorption affects electronic structure, magnetic exchange and anisotropy in representative Sr$_2$Fe$_2$O$_5$ and Sr$_2$Co$_2$O$_5$ oxides. We find that hydrogenation introduces a localized e$^-$ that stabilizes near the H$^+$, with B-site dependent  preference. The resulting lattice distortions and redistribution of charge density modify exchange coupling and cant the N\'eel vector, giving rise to weak ferromagnetism. We also show that absorption energies are highly sensitive to H$^+$–e$^-$ arrangements and magnetic order, varying by up to 1 eV across different settings. This sensitivity demands consistent treatment of charge localization and spin states, together with careful choice of computational parameters.  Extending to a variety of experimentally reported A$_2$B$_2$O$_5$ compositions, we identify candidates with favorable H uptake and uncover a trend linking more favorable absorption to a higher B-site $d$-electron count. We also demonstrate that preferential H$^+$ absorption site in these materials is governed by local O–O separations and lattice flexibility, which describe ability of the framework to accommodate H$^+$-induced distortions. Finally, benchmarks of universal machine-learning interatomic potentials reveal uncertainties of about 1 eV for site-resolved absorption energies, motivating descriptor-based surrogate models and targeted DFT validation. Together, these results establish practical design rules for hydrogen-responsive oxides relevant to iono-electronic devices, sensors and electrically tunable spin functionality.
\end{abstract}

\maketitle
\section{Introduction}

Dynamically controlling material properties through reversible electrochemical intercalation of small ions, known as iono-electronics, is a rapidly growing research direction~\cite{10.1002/adma.202418484,doi:10.1021/acs.chemrev.4c00512,doi:10.1126/sciadv.adq4696}. When a positive gate voltage is applied, cations migrate from the reservoir through the electrolyte to compensate the flow of electrons and accumulate inside the active-layer material, altering its electric conductance and other properties. Among possible carriers, the small size of H$^+$ makes it uniquely effective, enabling insertion and removal from lattices with minimal structural disruption. Coupled H$^+$ and \textit{e}$^-$ doping has been shown to tune a wide spectrum of functionalities, including mechanical~\cite{OGAWA2020181}, optical~\cite{D1RA01959G,NAGY1989567}, magnetic~\cite{Li2017,doi:10.1021/acs.nanolett.5c01712,Ye2020}, electronic properties ~\cite{10.1063/1.4927322,D1RA01959G,Ding2023}, and thermal transport~\cite{Lu2020,10.1002/adma.201903738}. It can also induce superconductivity~\cite{Drozdov2015,Piatti2023,Ding2023,Kong2021}, ferroelectricity~\cite{SCHULTHEI20242652,Shao2023}, magnetic skyrmions~\cite{Chen2022}, metal--insulator and Mott transitions~\cite{C9CP06522A,PhysRevMaterials.2.114409,yamauchi2022hydrogeninducedmetalinsulatortransitionaccompanied,PhysRevB.109.205124,10.1002/advs.202414991}, and enhanced catalytic activity~\cite{10.1002/asia.201701661,doi:10.1021/jp508313y}.Proton-responsive oxides are also being explored for neuromorphic devices, where ionic dynamics emulate synaptic plasticity and enable electrochemical random-access memory (H-ECRAM)~\cite{doi:10.1021/acs.chemrev.4c00071,C3NR05882D,7845629}. Despite these advances, the microscopic mechanisms of hydrogen incorporation, electron localization, and their influence on correlated electronic states remain insufficiently understood.

Among active layer materials, oxygen-deficient brownmillerite perovskites (A$_2$B$_2$O$_5$) stand out for their compositional flexibility and demonstrated ability to  absorb and conduct hydrogen \cite{FISHER1999355,THABET202091,PhysRevB.92.174111,10.1063/5.0241360}. Combined experimental and calculation efforts on Sr$_2$Co$_2$O$_5$ and Sr$_2$Fe$_2$O$_5$ reveal dramatic hydrogen-induced transformations, including insulator–metal transitions, changes in magnetic order, and tunable thermal transport \cite{Lu2017,Lu2020,10.1002/adfm.202316608}. Yet, systematic atomistic insight into how proton–electron pairs modify electronic and magnetic structures across different brownmillerites remains lacking, and no unified framework currently connects hydrogen insertion to these coupled transformations. Developing such a description is particularly challenging because modeling H–e$^-$ interactions in correlated oxides is highly sensitive to configurational, magnetic, and methodological choices, motivating the discussion of computational complexities that follows.

Theoretical predictions are complicated by the vast configurational space of H$^+$/\textit{e}$^-$ distributions and by strong sensitivity to computational parameters such as the Hubbard $U$ and exchange--correlation functional \cite{PhysRevB.73.195107,Tsang2018TowardsUT,Wang2023AbnormalSO,Teng2022DiffusionSO,Liang2018RoleOI,PhysRevMaterials.2.114409}. Recent computational studies of H in A$_2$B$_2$O$_5$ have primarily focused on investigating ionic transport~\cite{FISHER1999355,chen2024fastlithiumiondiffusion,doi:10.1021/acs.chemmater.0c00544,Lu2022Brownmillerite}, while the theoretical explanation for H-induced phenomena in these oxides remains incomplete, limiting further progress in this field~\cite{10.1002/adfm.202316608}. A broad range of possible concentrations, the large configurational space of H$^+$ and e$^-$ distributions, and their interrelated effects on electronic, magnetic, and crystal structures pose a significant challenge for theoretical studies~\cite{10.1063/5.0241360,Lu2017,Lu2020,10.1002/adfm.202316608}. Recent computational work has further shown that the H$^+$–polaron pair is coupled, and that the proper choice of polaron localization site relative to the H$^+$ bonding site can substantially alter polaron migration barriers by distorting metal–oxygen–metal linkages and reducing orbital overlap, thus strongly affecting electronic transport in oxides such as V$_2$O$_5$, MoO$_3$, and WO$_3$~\cite{hds1-yls4}. The proper choice of computational parameters and their impact on hydrogen absorption energies and the resulting electronic and magnetic properties remain poorly explored. Simulating hydrogen behavior in brownmillerites and other correlated oxides is particularly demanding and computationally intensive, making straightforward high-throughput studies challenging. These limitations highlight the need for efficient yet accurate approaches to capture complex hydrogen--oxide interactions, motivating the search for structure--property relationships or the use of universal machine-learning interatomic potentials (uMLIPs). Although uMLIPs have recently been applied to H–metal systems, accurately predicting absorption and diffusion in alloys and simple metals\cite{Ito2025PredictingHydrogenDiffusionMLIPs,Angeletti2025HydrogenDiffusionMgML,PhysRevMaterials.7.093601,PhysRevMaterials.5.113606}, their performance for predicting hydrogen  energetics in correlated oxides remains limited.

In this work, we employ density functional theory (DFT) and ab initio molecular dynamics (AIMD) with PBEsol+U and SCAN+U to systematically investigate hydrogen absorption and electron localization in representative brownmillerites. We analyze changes in electronic structure, magnetic exchange couplings, and N{\'e}el-vector canting, quantify sensitivity to magnetic order (FM vs AFM) and perform high-throughput screening across 14 compositions to identify promising hosts. We further benchmark selected uMLIPs against DFT calculations for hydrogen absorption in brownmillerites. The resulting analysis offers reliable computational modeling and design guidelines for hydrogen-responsive oxides.

\section{Computational Methods}

First-principles calculations based on density functional theory (DFT) were performed using the VASP \cite{PhysRevB.54.11169,PhysRevB.59.1758,KRESSE199615,PhysRevB.47.558,PhysRevB.49.14251} package to evaluate hydrogen absorption energies and the associated modifications of electronic and magnetic structures in A$_2$B$_2$O$_5$. {Exchange–correlation effects were treated within the meta-GGA SCAN(+U) functional \cite{PhysRevLett.115.036402}, whereas selected benchmark comparisons at the PBEsol(+U) level \cite{PhysRevLett.100.136406} are provided in the Supporting Information.

A plane-wave cutoff of 520 eV and Gamma-centered $k$-meshes of $3 \times 3 \times 2$ (structural relaxations) and $6 \times 6 \times 4$ (density of states, DOS) were employed. Convergence thresholds were set to $10^{-6}$ eV for electronic energies and 0.01 eV/$\mathrm{\AA}$ for ionic forces. Structural optimizations included relaxation of both lattice vectors and atomic positions.  

Hubbard $U$ corrections were applied for PBEsol, consistent with Materials Project recommended values \cite{PhysRevB.70.235121,PhysRevB.73.195107,PhysRevB.84.045115}, while SCAN+U calculations used parameters recommended by Ref.~\cite{PhysRevMaterials.6.035003} In particular,  U values of 
 3.00 eV (Fe) and 3.32 eV (Co) and  1.88 eV (Fe) and 2.08 eV (Co) were used for PBEsol and SCAN calculations, respectively. The benchmarking tests for other U values can be found in the Supporting Information.
Unless otherwise noted, calculations were carried out in a $3 \times 3 \times 1$ supercell (144 atoms).

Hydrogen was introduced by placing a single H atom at each of the seven distinct interstitial sites, followed by full relaxation, corresponding to a dilute concentration of H$_{0.06}$A$_2$B$_2$O$_5$. Hydrogen absorption energies were referenced to half the energy of an isolated H$_2$ molecule, computed in a 20$~\mathrm{\AA}$ cubic box:
\[
E_\text{abs} = E_{\text{A}_2\text{B}_2\text{O}_5+\text{H}} - E_{\text{A}_2\text{B}_2\text{O}_5} - \tfrac{1}{2}E_{\text{H}_2},
\]
where $E_{\text{A}_2\text{B}_2\text{O}_5+\text{H}}$, $E_{\text{A}_2\text{B}_2\text{O}_5}$, and $E_{\text{H}_2}$ are the total energies of the hydrogenated structure, pristine structure, and gas-phase H$_2$, respectively. DOS calculations employed tetrahedron smearing.  

To control electron localization on specific metal sites after hydrogen insertion, we seeded small-polaron formation on a selected transition-metal center either by slightly elongating the local Me–O bonds or by applying a temporary, site-selective Hubbard correction U on its d orbitals. Following ionic relaxation, the same U value was reinstated uniformly across all metal sites (or the local U was reduced to the global value), and the structure was re-optimized to ensure that the localized solution corresponds to a true energy minimum rather than an artificial constraint. Successful localization was confirmed by the emergence and stability of an enhanced local magnetic moment on the targeted metal site and by charge-density-difference isosurfaces revealing excess d-like electron density centered on that atom. A similar approach is described in more detail \cite{PhysRevB.75.195212}.

Magnetic exchange interactions were extracted by mapping DFT energies of four ordered spin states to a Heisenberg Hamiltonian:
\begin{align}
    E_\mathrm{FM} &= E_0 - 4S^2 J_1 - 2S^2 J_2 - 2S^2 J_3, \\
    E_\mathrm{AFM(I)} &= E_0 - 4S^2 J_1 - 2S^2 J_2 + 2S^2 J_3, \\
    E_\mathrm{AFM(II)} &= E_0 - 4S^2 J_1 + 2S^2 J_2, \\
    E_\mathrm{AFM(III)} &= E_0 + 4S^2 J_1 - 2S^2 J_2,
\end{align}
with $S=3/2$. Here, tetrahedrally coordinated Co atoms were assumed to have two nearest neighbors (instead of four, as in octahedral coordination) because of the absence of two O atoms. Solving these equations yields $J_1$, $J_2$, and $J_3$.  

Magnetocrystalline anisotropy (MCA) energies were evaluated only at the PBEsol+$U$ level, as spin–orbit coupling (SOC) is not yet fully supported for meta-GGA functionals in VASP. MCA calculations were performed for Sr$_2$Fe$_2$O$_5$ and Sr$_2$Co$_2$O$_5$ with and without hydrogen at the most stable H$_1$ site, using the 144-atom supercell and a $6 \times 6 \times 4$ $k$-mesh. SOC effects were included non-self-consistently, which is sufficient to capture relative MCA trends at the $\mu$eV scale while maintaining computational efficiency.

For high-throughput screening, we queried the Materials Project for experimentally reported A$_2$B$_2$O$_5$ compounds with closely related symmetries (\texttt{I4/mmm}, \texttt{Ima2}, \texttt{P4/mmm}, or \texttt{Pnma}) and excluded entries containing radioactive elements, yielding 14 compositions listed in the Supporting Information. Because several of these compounds appear in multiple, closely related structural variants, we imposed \texttt{Ima2} space group to all compositions for consistency. We note that the choice of symmetry can affect the computed hydrogen absorption energies, which can be rationalized by the different ability of lattice to accommodate the associated distortions.  Nevertheless, a systematic assessment of this dependence is beyond the scope of the present study.

For universal machine-learning interatomic potentials (uMLIPs), we employed the UMA (OMat24) model implemented in fairchem-core (version 2.3.0), which was trained on PBE+U datasets \cite{BarrosoLuque2024OMat24}. For comparison, we also used CHGNet (version 0.4.1)\cite{Deng2023CHGNet}, a spin-polarized model version obtained via transfer learning on the MP-R2SCAN dataset. Full ionic and cell relaxations were performed using the FIRE optimization algorithm~\cite{PhysRevLett.97.170201}, with a force-convergence threshold of 0.005~eV~\AA$^{-1}$.
All datasets used to generate the figures in this manuscript  and intput files are available via Figshare. \cite{dataset}

\section{Results and Discussion}

We begin by analyzing hydrogen absorption in two representative brownmillerite (BM) perovskites, Sr$_2$Co$_2$O$_5$ and Sr$_2$Fe$_2$O$_5$. In this step, we assess how absorption energies vary with proton sites, electron localization, and computational parameters. Building on these insights, we investigate the influence of hydrogen incorporation on the electronic structure and magnetic ordering across a broader set of compositions. Finally, we extend our analysis to the experimentally reported A$_2$B$_2$O$_5$ compounds, demonstrating that many BM oxides possess favorable energetics for hydrogen uptake and transport. We also evaluate the feasibility of organizing a high-throughput screening campaign based on machine-learning interatomic potentials and discuss possible pitfalls. 

\subsection{Factors Affecting the Calculated Hydrogen Absorption Energies}

\textbf{Preferential sites of H$^+$ and e$^-$ pair in the BM.} Hydrogen is typically introduced into BM oxides in the form of a coupled proton (H$^+$) and electron (e$^-$) pair, either through chemical reactions or electrochemical processes.  To establish the ground-state configurations, we examined the competing arrangements of H$^+$–e$^-$ pairs in Sr$_2$Co$_2$O$_5$ and Sr$_2$Fe$_2$O$_5$. Specifically, we analyze (i) variation of absorption energy across nonequivalent interstitial sites and electron localization sites, (ii) sensitivity to magnetic order, and (iii) dependence on exchange–correlation functional and Hubbard $U$ values.  

As illustrated in Fig.~\ref{fig:structure_BM}(a), the BM lattice consists of alternating octahedral (MeO$_6$) and tetrahedral (MeO$_4$) layers connected by interlayer O$_2$ atoms along $z$. Previous work identified seven non-symmetric hydrogen positions \cite{doi:10.1021/acs.chemmater.0c00544}, shown in Fig.~\ref{fig:structure_BM}(b). H$_1$–H$_3$ are covalently bonded to O$_2$ at the interface between the octahedral and tetrahedral layers; H$_4$–H$_6$ bond to O$_1$ within the octahedral network; and H$_7$ bonds to O$_3$ in the tetrahedral chain. Other trial positions were found to relax into these seven motifs or were significantly less stable. These seven configurations therefore were used for our subsequent absorption-energy analysis.

\begin{figure} 
    \centering
    \includegraphics[width=1\linewidth]{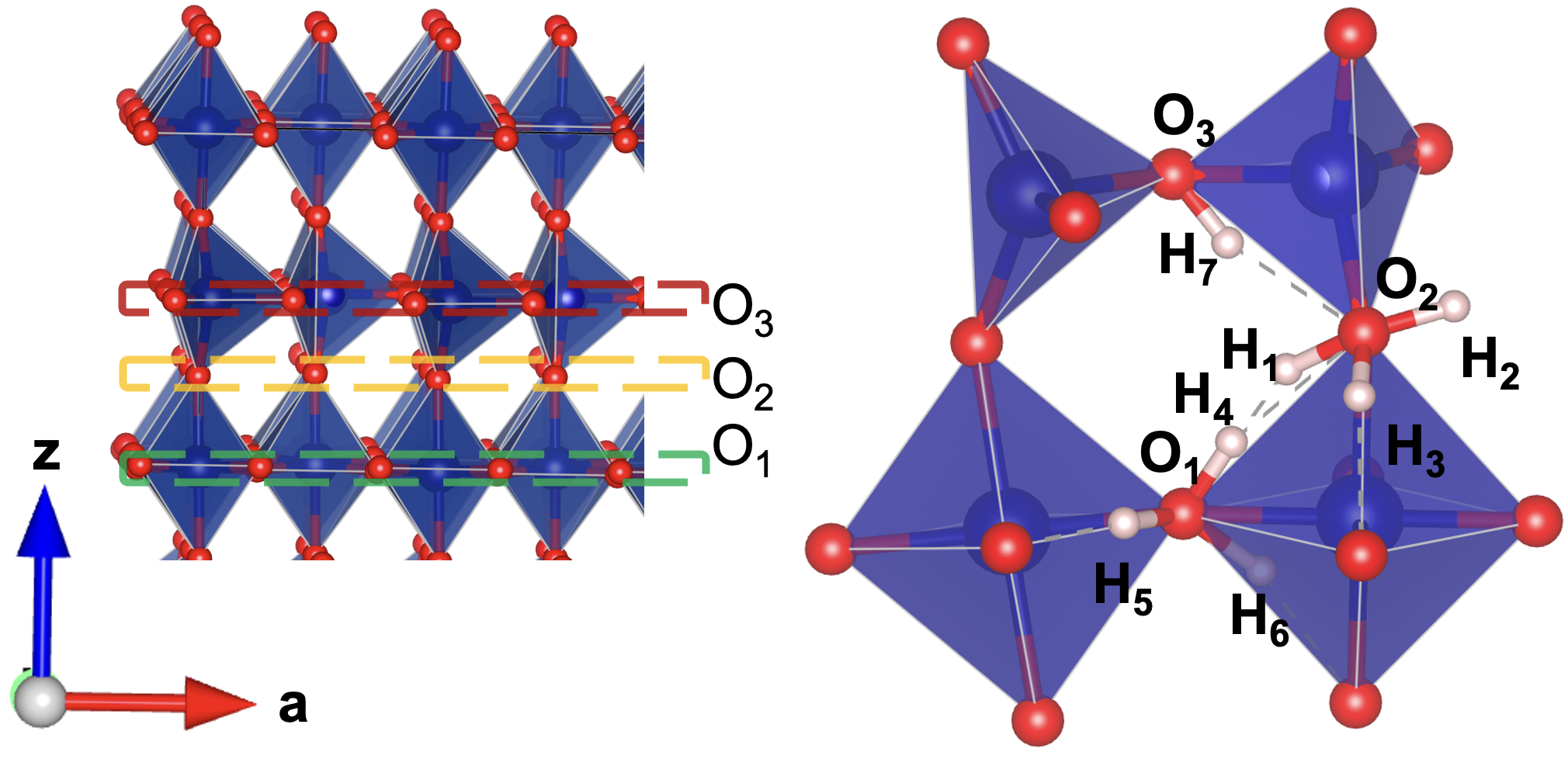}
    \caption{(a) Crystal structure of brownmillerite A$_2$B$_2$O$_5$ (orthorhombic \texttt{Ima2} space group), showing alternating octahedral (MeO$_6$) and tetrahedral (MeO$_4$) layers. (b) Seven nonequivalent interstitial sites (H$_1$–H$_7$) identified for hydrogen incorporation, consistent with Ref.~\cite{doi:10.1021/acs.chemmater.0c00544}.}
    \label{fig:structure_BM}
\end{figure}

To assess the stability of electron localization metal sites, we fixed the most stable proton site (H$_1$) and evaluated possible localization centers for the accompanying electron. Notably, for all configurations considered below, the excess electron forms a small polaron localized on a metal site rather than remaining delocalized in the lattice. As shown in Figs.~\ref{fig:SCO_SFO_e_loc_H1_H1_H7_dif_loc_sites} (a-b), electrons preferentially localize near the proton due to strong Coulomb attraction between the H$^+$ and the polaron. However, the nature of this attraction and the resulting polaron localization differ markedly between the two compounds. In Sr$_2$Co$_2$O$_5$, the polaron–proton attraction is relatively weak, allowing the polaron to localize on multiple nearby Co sites (both  octahedral and tetrahedral) with only small energy differences. In contrast, in Sr$_2$Fe$_2$O$_5$, the polaron is strongly localized on a single tetrahedral Fe site closest to the proton. Localization of  the polaron on any other tetrahedral or octahedral Fe site leads to a substantial energy penalty of approximately 0.2–0.3 eV, depending on the Fe coordination and its distance from the proton, rendering these configurations thermodynamically unfavorable.

We next compare hydrogen absorption energies across the seven nonequivalent interstitial sites (H$_1$–H$_7$) in both compounds (Figs.~\ref{fig:SCO_SFO_e_loc_H1_H1_H7_dif_loc_sites} (c-d)). Our calculations suggest that the H$_1$ site is the most stable, followed closely by H$_6$, which shares the same O$_1$–O$_2$ environment. The least favorable site is H$_7$, located in the tetrahedral layer. The absorption energies depend sensitively on whether the electron localizes at octahedral or tetrahedral metal sites, with red and blue bars in Fig.~\ref{fig:SCO_SFO_e_loc_H1_H1_H7_dif_loc_sites} (c-d) indicating two respective scenarios. The H$_4$ and H$_5$ configurations did not stabilize tetrahedral localization and are therefore not shown.

\begin{figure}[ht]
    \centering
    \includegraphics[width=1\linewidth]{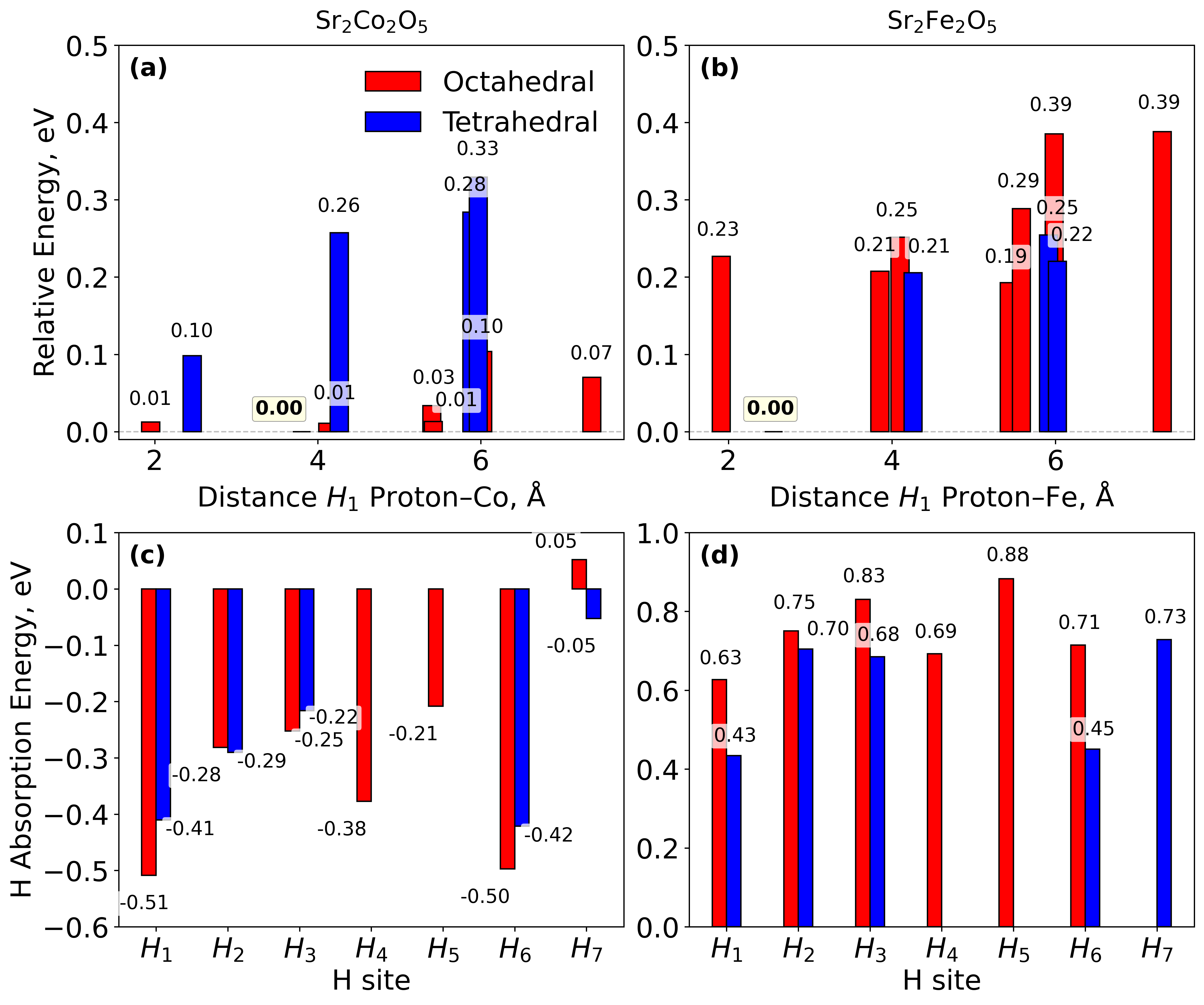}
    \caption{Relative polaron binding energies in Sr$_2$Co$_2$O$_5$ (a) and Sr$_2$Fe$_2$O$_5$ (b) for a fixed most stable H$_1$ proton absorption site. Electrons preferentially localize on tetrahedral Fe sites and on octahedral Co sites, with localization energy increasing as the H–e$^-$ separation grows. Hydrogen absorption energies at sites H$_1$–H$_7$ in Sr$_2$Co$_2$O$_5$ (c) and Sr$_2$Fe$_2$O$_5$ (d). Bars compare electron localization at octahedral (red) and tetrahedral (blue) sites.}
    \label{fig:SCO_SFO_e_loc_H1_H1_H7_dif_loc_sites}
\end{figure}

Across these configurations, the preferred polaron localization sites also differ between the two materials. In Sr$_2$Co$_2$O$_5$, octahedral Co sites are favored by 0.31 eV, while in Sr$_2$Fe$_2$O$_5$, tetrahedral Fe sites are more stable by 0.07 eV. This can be understood from crystal-field theory: Co$^{3+}$ ($d^7$) favors occupation of $t_{2g}$ orbitals on octahedral sites, while Fe$^{3+}$ ($d^6$) favors the less crowded $e_g$ orbitals in a tetrahedral environment \cite{ZHANG202219027,Meng2021} (Fig.~S1). At H$_4$, tetrahedral localization was unstable, with only octahedral solutions obtained. For H$_7$ located in the tetrahedral layer, localization consistently favored tetrahedral sites due to the large separation from octahedral sites. At H$_5$ in the octahedral plane, electrons localized exclusively on octahedral sites. For H$_2$, positioned between layers, both localizations were nearly degenerate, although tetrahedral remained slightly favored. These results demonstrate that electron localization has a substantial impact on hydrogen absorption energetics—accounting to differences of several tenths of an electronvolt (0.1--0.4 eV) and must therefore be explicitly identified and reported. Failure to account for the underlying localization sites can lead to apparent inconsistencies among studies and hinder reproducibility, not only in BM but across transition-metal oxides and related perovskite semiconductors.

We rationalize the variation in absorption energies by explicitly analyzing the roles of dynamical flexibility and key structural factors: (i) the compliance of O–O vibrational modes that accommodate O–H···O bridges, and (ii) the O–O bond length, which reduces steric strain for the inserted proton.
As shown in Figs.~S4–~S5, sites connected by softer phonon modes and larger O–O separations yield more favorable absorption energies. Consistent with this picture, Chung et al.~\cite{Chung2025FlexibilityOxygenSublattice} recently introduced a similar set of descriptors that quantify oxygen-sublattice flexibility and hydrogen-bond geometry to predict proton mobility in ternary oxides. Their ``structural'' (O-O spacing, H–O···O length) and ``dynamic'' (thermal O-O fluctuation) descriptors strongly correlate with proton-transfer barriers, reinforcing our conclusion that both O–O vibrational modes and the local bonding environment critically govern H energetics in BM and related oxygen perovskites.

\textbf{Impact of Magnetic Ordering on Hydrogen Absorption Energies.}  
Because the magnetic structures of brownmillerites can be tuned by strain or temperature \cite{shin2021strain}, we evaluated how hydrogen absorption depends on magnetic ordering. Using Sr$_2$Fe$_2$O$_5$ as a prototype, we compared absorption energies in ferromagnetic (FM) and G-type antiferromagnetic (G-AFM) configurations for the most stable H$^+$–e$^-$ configurations. As shown in Fig.~\ref{fig:FM_VS_AFM_SCAN_U}, hydrogen absorption energy is consistently lower in the FM state, by about 0.2-0.3 eV across most of the sites for Sr$_2$Fe$_2$O$_5$. This trend originates from the fact that AFM has the semiconducting ground state, whereas FM has metallic ground state that favors H absorption. H absorption is also more favorable in FM state compared to AFM-G state for Sr$_2$Co$_2$O$_5$ and Sr$_2$Mn$_2$O$_5$, whereas Sr$_2$Cr$_2$O$_5$ showed nearly identical energetics in both magnetic states. Importantly, the hydrogen absorption energies in different magnetic orderings also depend on the choice of functional: with PBEsol+U, the energy differences between AFM and FM states are slightly larger (0.3–0.6 eV) for the hydrogen sites in Sr$_2$Fe$_2$O$_5$and substantially larger for Sr$_2$Co$_2$O$_5$ (nearly 1 eV), but the FM state consistently exhibits lower absorption energies as shown in Fig.~S21.

\begin{figure}[ht!]
    \centering
    \includegraphics[width=1\linewidth]{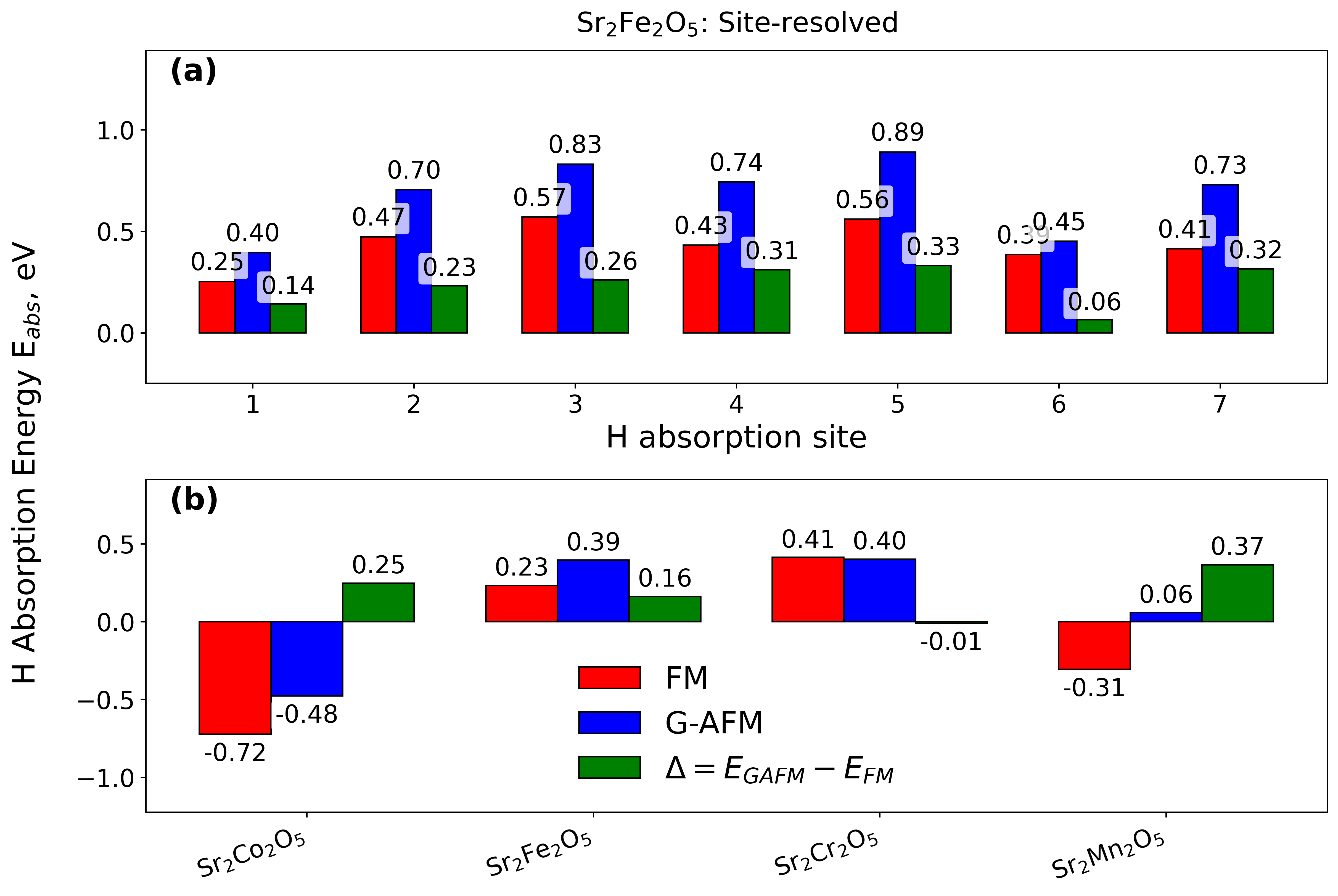}
    \caption{Hydrogen absorption energy variations induced by magnetic ordering (FM and G-AFM). 
Panel (a) shows hydrogen absorption energies for Sr$_2$Fe$_2$O$_5$ at different intercalation sites in FM and G-AFM configurations, together with the energy difference $\Delta$(G-AFM $-$ FM). 
Panel (b) compares hydrogen absorption energies for the most stable H$_1$ site across different brownmillerite oxides (Sr$_2$Co$_2$O$_5$, Sr$_2$Fe$_2$O$_5$, Sr$_2$Cr$_2$O$_5$, and Sr$_2$Mn$_2$O$_5$) in both FM and G-AFM magnetic states.}
    \label{fig:FM_VS_AFM_SCAN_U}
\end{figure}

These results highlight the critical role of magnetic order in high-throughput screening. Assuming a uniform FM state, as is done in the Materials Project \cite{10.1063/1.4812323}, can introduce errors of 0.3–1 eV at the PBEsol+U level of theory and 0.2–0.4 eV at the SCAN+U level of theory in predicted absorption energies. Additional uncertainties stem from incorrect electron localization (0.1–0.4 eV) and proton site identification (up to 0.6 eV). The strong coupling between hydrogen energetics and magnetic order offers an opportunity: by exploiting strain- or temperature-driven spin reordering \cite{PhysRevB.100.174417,10.1063/1.3484147,10.1063/1.5090824,10.1002/adma.201604112,doi:10.1021/cm200454z,PhysRevLett.106.067201}, one can design oxides with tunable hydrogen uptake and catalytic functionality.

\textbf{Role of Computational Parameters.}
To understand sensitivity of hydrogen absorption energies to the choice of computational settings, we examine how the variation of Hubbard $U$ parameter affects the electron localization and proton absorption energies calculated using the most widely used PBE/PBEsol and SCAN functionals. Reported Hubbard $U$ values for PBE+$U$ calculations vary in the range of $3.0$–$6.5$ eV for Sr$_2$Co$_2$O$_5$ \cite{PhysRevB.73.195107,Tsang2018TowardsUT,Wang2023AbnormalSO,Teng2022DiffusionSO,Liang2018RoleOI,PhysRevMaterials.2.114409} and $3.0$–$5.3$ eV for Sr$_2$Fe$_2$O$_5$ \cite{10.1002/adfm.202316608,PhysRevB.92.174111,10.1063/5.0241360,doi:10.1021/acsami.7b17377}, correspondingly. The $U$ values used for SCAN+$U$ are typically smaller, generally not exceeding $\sim$3 eV for either Fe or Co \cite{PhysRevMaterials.6.035003,PhysRevMaterials.2.095401,PhysRevMaterials.4.045401}.
This difference reflects the reduced self-interaction error and improved description of localized $d$-states in SCAN compared to PBE-type GGAs. As demonstrated for transition-metal oxides and multiferroics~\cite{10.1002/wcms.70007}, SCAN can reproduce the correct bandgap and magnetic moments with substantially smaller or even negligible $U$ corrections, highlighting its enhanced ability to capture $d$-orbital physics intrinsically.

Our calculations show that using small or zero $U$ values leads to large deviations in the predicted hydrogen absorption energies because the underlying electronic structure is inaccurately described, most notably at the PBEsol level. For some compounds, the absence of $U$ already results in a metallic electronic structure and an incorrect magnetic ground state, even before hydrogen insertion. In others, the band gap is severely underestimated, and the additional electron introduced upon hydrogen insertion collapses the gap, driving the system metallic (Fig.~S2). In both cases, the polaron is artificially delocalized, leading to erroneous absorption energetics. This behavior does not occur for SCAN, even without a Hubbard correction. Increasing $U$ generally lowers the absorption energies, with changes of up to $\sim$0.5 eV depending on the functional, which can be attributed to enhanced electron localization arising from a larger on-site Coulomb penalty.

To probe how $U$ affects polaron localization site preferences, we examined the energetics of added electron localization in Sr$_2$Co$_2$O$_5$ and Sr$_2$Fe$_2$O$_5$ (Fig.~S3). The right panel of Fig.~S3 shows that the energy cost of electron addition varies nonlinearly with $U$: bell-shaped in Sr$_2$Fe$_2$O$_5$, but monotonically decreasing in Sr$_2$Co$_2$O$_5$. In Sr$_2$Fe$_2$O$_5$, the relative stability of octahedral versus tetrahedral sites depends strongly on $U$: at higher values, the preferred localization site can switch. In contrast, Sr$_2$Co$_2$O$_5$ shows weaker $U$ dependence, with electrons consistently favoring octahedral sites, consistent with the crystal-field splitting arguments discussed above. 
We also considered proton-only insertion (i.e., H$^+$ and neutralizing background), which primarily induces structural distortions (Fig.~S6). The energy penalty for this process increases with $U$, reaching up to 0.45 eV at the highest values tested.  

Thus, both the exchange–correlation functional and the choice of $U$ exert a strong influence on predicted hydrogen absorption energies and electron-localization relative energetics. While PBEsol+$U$ reproduces qualitative trends and SCAN+$U$ is noticeably less sensitive to $U$ variation, neither approach yields fully transferable absolute energies. Moreover, hybrid-functional calculations, often used as a reference, do not provide a consistent baseline: for the most stable H$_1$–electron configurations across the 3 brownmillerites (Sr$_2$Co$_2$O$_5$, Sr$_2$Fe$_2$O$_5$ and Sr$_2$Cr$_2$O$_5$), HSEsol, SCAN, SCAN+$U$, and PBEsol+$U$ disagree by as much as 0.4 eV (Fig.~S7). Rather than recommending a single “correct” $U$, we highlight that significant artificial energetic shifts can arise solely from the choice of functional or $U$. In the absence of experimental benchmarks or higher-level many-body reference data, further validation using hybrid functionals and beyond-DFT approaches such as Dynamical Mean Field Theory (DMFT) is needed to establish more robust and quantitative trends.

\subsection{Best-Practice Protocols for Modeling Hydrogen Intercalation in Transition-Metal Oxides}

Accurate modeling of hydrogen intercalation in correlated oxides requires explicit treatment of the coupled proton--electron--spin degrees of freedom.  As we demonstated above, failure to account for all relevant degrees of freedom, including magnetic order, proton site preference, and polaron localization site preferences, can lead to errors in the calculated hydrogen absorption energies on the order of up to $\sim$1 eV. Establishing a consistent and systematic workflow is therefore essential for obtaining physically meaningful and reproducible absorption energies across different compounds and computational settings. Based on our findings, we outline a practical workflow for identifying  ground-state H$^+$--e$^-$ configuration.

\textit{(i) Establish the ground-state magnetic ordering of the pristine host.}  
This step should be performed prior to hydrogen insertion or in the dilute solubility limit, as a small amount of hydrogen should not determine the magnetic order. Incorrect FM/AFM assumptions can shift absorption energies by 0.2-0.3 eV (Meta-GGA level) and 0.3--1~eV (GGA level).

\textit{(ii) Evaluate preferred polaron-localization sites.}
Examine electron localization on all nonequivalent transition-metal sites (e.g., tetrahedral and octahedral sites in brownmillerites), and assess possible partial or full delocalization solutions. Localization site preference depends on crystal field splitting, which varies for each metal in different local coordination environments (see the example of Sr$_2$Co$_2$O$_5$ vs Sr$_2$Fe$_2$O$_5$ above).
A practical approach is to add one excess electron to the pristine lattice (without a proton) and explicitly converge distinct self-consistent solutions corresponding to localization on different metal sites, as well as possible delocalized states, and then compare and rank their relative energies.

\textit{(iii) Identify all nonequivalent hydrogen intercalation sites.}
With respect to the preferred polaron-localization sites identified in step (ii), relax each trial configuration to determine the set of locally stable H positions. At this stage, probe configurations where the proton H$^+$ is added in proximity to the localized electron (within 2--4~\AA{} of the polaron site). Relative energetics at this point depend on dynamical flexibility and structural factors: the compliance of O--O vibrational modes accommodating O--H$\cdots$O bridges, and the O--O length, which reduces steric strain for the inserted proton.

\textit{(iv) Determine polaron binding energies relative to fixed proton positions.}  
Select the most stable relaxed H sites (within 0--0.25 eV relative to the most stable H site) and, for each identified H position, test all possible polaron localization sites from step (ii) to determine the most stable proton--polaron configuration. In brownmillerites, the polaron typically localizes within 2--4~\AA{} of the proton; metal sites within this range should therefore be examined explicitly. However, in brownmillerites, the preferred polaron formation site (octahedral vs tetrahedral) remains the same across all unique H positions.

This workflow provides consistent treatment of proton--polaron coupling and spin order, minimizing large artificial variations that otherwise arise from incorrect magnetic states, electron localization, or site selection. By explicitly separating the contributions of proton insertion, electron localization, and magnetic order, the approach also improves the reproducibility and comparability of absorption energy calculations across different materials and computational frameworks. An open question we do not address here is how electron-localization preferences, hydrogen-site stability, and H$^+$--e$^-$ coupling vary solely due to computational parameters such as the exchange--correlation functional and Hubbard $U$. These effects for hydrogenated transition-metal oxides, unfortunately, remain insufficiently addressed in the literature at present and will be discussed in the next section.

\subsection{Effect of Hydrogen Absorption on Electronic Properties}

We next investigate the effect of H$^+$ + \textit{e}$^-$ insertion on electronic structure of  Sr$_2$Co$_2$O$_5$ and Sr$_2$Fe$_2$O$_5$, both of which exhibit semiconducting behavior in their pristine AFM-G type ground state with the calculated band gaps of 1.41 and 2.03 eV on SCAN+U matching reported values calculated at PBE+U level \cite{PhysRevMaterials.2.114409,PhysRevMaterials.3.024603,PhysRevB.92.174111}. After introducing H$^+$–e$^-$ pair at the most stable H$_1$ site, projected density of states (PDOS) calculations reveal the formation of localized in-gap states associated with electron occupation of transition-metal $d$ orbitals (Figs.~S9–~S10 a).
In Sr$_2$Co$_2$O$_5$, the electron preferentially localizes at the octahedral Co sites, producing a narrow peak $\sim$0.2 eV above the valence band maximum (Figs.~\ref{fig:El_str_changes_BMs_H} (a)). In Sr$_2$Fe$_2$O$_5$, localization at tetrahedral Fe sites places peaks $\sim$1 eV away from both band edges, resulting in deeper states (Figs.~\ref{fig:El_str_changes_BMs_H} (b)). Localization on tetrahedral sites produces multiple split peaks due to a partial transfer to neighboring sites with antiparallel spins.  

\begin{figure}[ht!]
    \centering
    \includegraphics[width=1\linewidth]{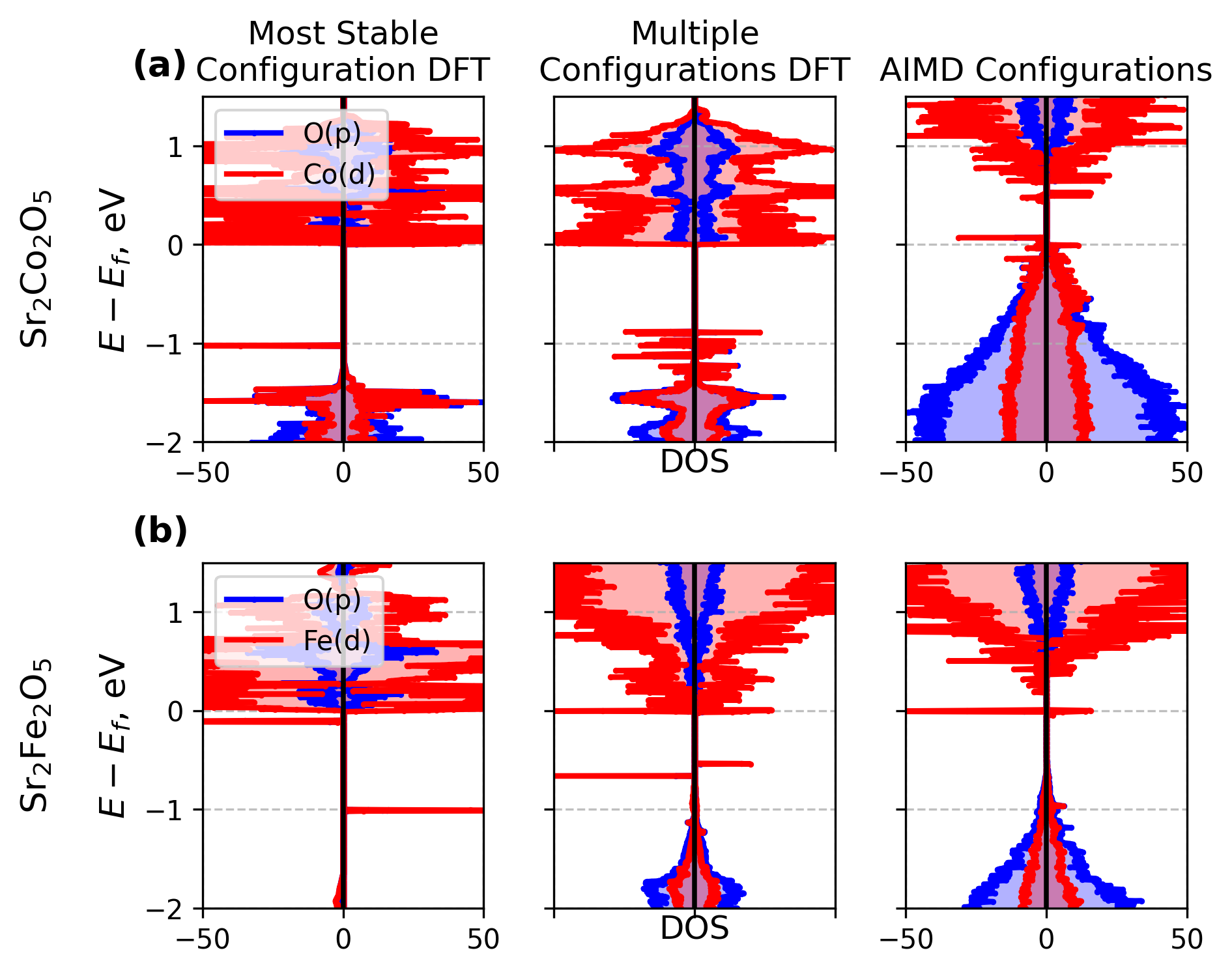}
    \caption{DOS for Sr$_2$Co$_2$O$_5$ (a) Sr$_2$Fe$_2$O$_5$ (b) different electron localization Co sites. DOS from the most stable configuration obtained from static DFT calculations.  Multiple DOS configurations across 9 electron localization sites from static DFT. DOS from AIMD simulation averaged over 40,000 steps with snapshots taken every 2,500 steps after equilibration.}
    \label{fig:El_str_changes_BMs_H}
\end{figure}

Compared to other intrinsic defects, H in oxides and in perovskite oxides can demonstrates rather high mobility even at room temperatures with activation energies ranging from 0.2 to 0.4 eV \cite{doi:10.1021/acs.chemmater.0c00544,doi:10.1021/acs.chemmater.1c02432,10.2138/rmg.2010.72.17,PhysRevB.82.014103,doi:10.1021/acsaelm.2c00711,Lan2020,Zhang2020PerovskiteNeuralTrees}. This means that the distribution of H$^+$ + \textit{e}$^-$ pairs or reorientation of H bonds may fluctuate, requiring special attention to their dynamics.  Considering the combined DOS for a full set of possible H$^+$ + \textit{e}$^-$  configurations in Figs.~\ref{fig:El_str_changes_BMs_H}(a,b) shows that, although peak positions vary and become broader, the states consistently remain well within the band gap. Thus, at dilute hydrogen concentrations, both compounds retain semiconducting character and do not undergo an insulator–metal transition, which occurs at higher concentrations for some complex transitional metal oxides and perovskites \cite{10.1002/advs.202510771,Yoon2016VO2,Wei2012VO2,Cai2024IonDiffusion,10.1002/advs.202414991,10.1002/adma.202101316,10.1002/adfm.202303416}.

Extending these results to other 3$d$ transition-metal brownmillerites (Fig.~S20) reveals consistent trends: compounds containing Sc through Co remain semiconducting upon hydrogenation, while Ni-, Cu-, and Zn-based members exhibit metallic behavior on SCAN level. Hybrid-functional HSEsol calculations recover small residual band gaps, supporting these trends (Fig.~S18). The notable exception is Sr$_2$V$_2$O$_5$, where partial electron delocalization over all octahedral V sites drives an insulator-to-metal transition.

To assess whether the electron-localization configurations identified in static DFT calculations persist under finite-temperature lattice dynamics, and to examine how thermal averaging over multiple configurations modifies the electronic structure, we performed ab initio molecular dynamics (AIMD) simulations at 500 K. Finite-temperature AIMD simulations reveal dynamic electron–lattice effects (Figs.~S11, ~S12). Polaron frequently hop among neighboring metal sites while the proton remains confined within O–H···O bridges (Fig.~S23), producing fluctuating localized states and reducing the effective band gap by approximately 0.5 eV. Tracking the polaron localization during AIMD (Fig.~S13) shows distinct behavior in Sr$_2$Fe$_2$O$_5$ and Sr$_2$Co$_2$O$_5$: in Sr$_2$Co$_2$O$_5$, the excess electron remains almost exclusively on octahedral Co sites, consistent with the larger octahedral–tetrahedral energy splitting (Fig.~S8) and lower octahedral localization energy (Fig.~\ref{fig:SCO_SFO_e_loc_H1_H1_H7_dif_loc_sites}(a)), whereas in Sr$_2$Fe$_2$O$_5$ the electron intermittently hops between both site types.

At the considered timesacles, the thermally activated polaron motion occurs largely without proton diffusion and leads to moderate band-gap narrowing. Both thermal fluctuations and local lattice distortions accompanying polaron hopping dynamically modulate metal–oxygen bond lengths and angles, reducing crystal-field splitting and facilitating thermally activated small-polaron hopping between neighboring metal sites (Fig.~S13). At the considered timescales, the thermally activated polaron motion occurs largely without proton diffusion and leads to moderate band-gap narrowing. Both thermal fluctuations and local lattice distortions accompanying polaron hopping dynamically modulate metal–oxygen bond lengths and angles, reducing crystal-field splitting and facilitating thermally activated small-polaron hopping between neighboring metal sites (Fig.~S13).
In particular, we attribute the band-gap narrowing to angular distortions of the MO$_6$ octahedra and MO$_4$ tetrahedra, which fluctuate from their ground state geometries (Fig.~S24). This sensitivity of the band gap to local bond angles is consistent with prior findings for BM-Sr\(_2\)Co\(_2\)O\(_5\), where changes in Co–O–Co angles result in band gap reductions from 1.37 eV to as low as 0.67 eV~\cite{PhysRevMaterials.3.024603}.
 As a result, hydrogenated brownmillerites remain semiconducting at dilute concentrations, but their electronic gaps become smaller and more fluctuating under finite-temperature conditions. The AIMD trajectories therefore corroborate the static DFT picture: localized H$^+$–\textit{e}$^-$ configurations are stable, yet their dynamic coupling to lattice vibrations and distortions induces fluctuations in band gap position and broadening of the polaronic in-gap states.

At dilute hydrogen contents, the coupled H$^+$–\textit{e}$^-$ pairs thus generate localized in-gap states rather than full metallization, preserving semiconducting behavior while narrowing the band gap through combined thermal and structural effects. As the excess electron localizes on a transition-metal ion (e.g., Co$^{3+}\rightarrow$Co$^{2+}$ or Fe$^{3+}\rightarrow$Fe$^{2+}$), mixed-valence pairs such as Co$^{3+}$–O–Co$^{2+}$ or Fe$^{3+}$–O–Fe$^{2+}$ form, linked through oxygen bridges. Localization of the excess electron converts neighboring transition-metal ions into mixed-valence configurations (e.g., Co$^{3+}$/Co$^{2+}$ or Fe$^{3+}$/Fe$^{2+}$), thereby altering the magnetic exchange interactions, as discussed in the following section.

\subsection{Effect of Hydrogen Absorption on Magnetic Properties}

Hydrogen insertion strongly modifies magnetic interactions in BM by introducing an extra electron and local structural distortions. As shown in Fig.~\ref{fig:exchange-coupling}, the exchange constants $J_1$ (octahedral–octahedral), $J_2$ (tetrahedral–tetrahedral), and $J_3$ (octahedral–tetrahedral) are all negative in pristine Sr$_2$Fe$_2$O$_5$ and Sr$_2$Co$_2$O$_5$, consistent with strong G-type antiferromagnetism, while Sr$_2$Cr$_2$O$_5$ and Sr$_2$Mn$_2$O$_5$ exhibit C-type order with positive $J_3$. Upon hydrogenation, a consistent trend emerges across all compounds: $J_1$ and $J_2$ become less negative (weaker AFM coupling), while $J_3$ decreases. These changes are listed in Tables~S1 and ~S2 for both SCAN+U and PBEsol+U, showing functional-independent behavior. To disentangle the roles of the polaron and the proton, we performed  calculations with electron-only and proton-only configurations. As shown in Figure~S2 and Table~S22, electron doping has the dominant effect on exchange couplings, while the proton introduces secondary, distortion-driven changes. However, their combined effect is not additive, as the polaron remains electrostatically bound to the proton.

\begin{figure}[ht]
    \centering
    \includegraphics[width=1.0\linewidth]{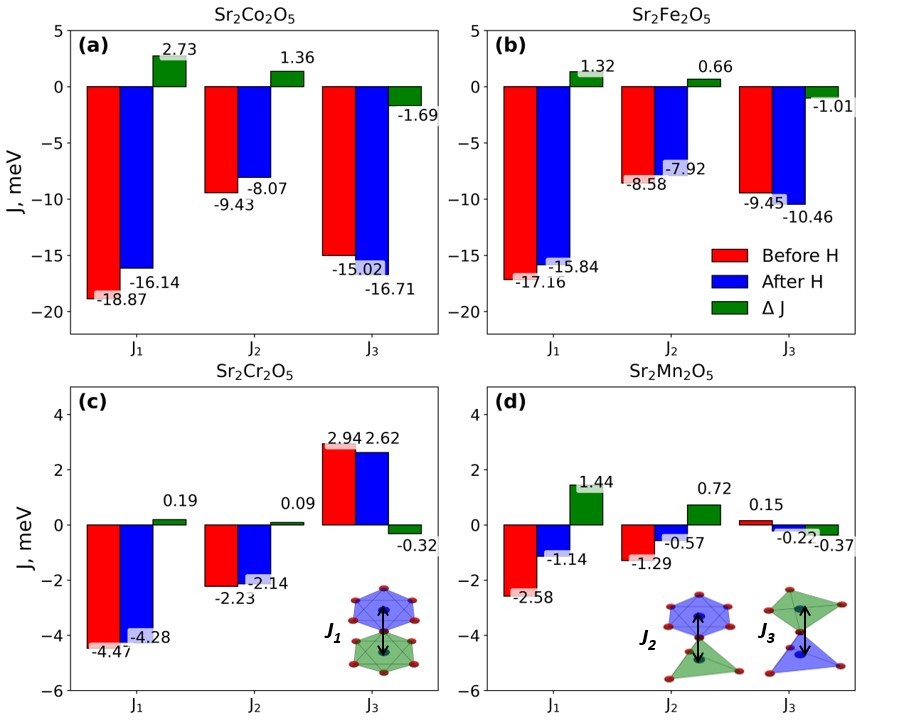}
    \caption{Exchange coupling constants in (a) Sr\(_2\)Co\(_2\)O\(_5\), (b) Sr\(_2\)Fe\(_2\)O\(_5\), (c) Sr\(_2\)Cr\(_2\)O\(_5\) and (d) Sr\(_2\)Mn\(_2\)O\(_5\) before and after hydrogen absorption at the H\(_1\) site. Green bars show the changes in J\(_1\), J\(_2\), and J\(_3\) due to hydrogen absorption.}
    \label{fig:exchange-coupling}
\end{figure}

In the pristine brownmillerite structure, all transition-metal ions possess the same oxidation state (e.g., Co$^{3+}$ or Fe$^{3+}$), and the magnetic coupling is dominated by superexchange interactions through bridging oxygens \cite{PhysRev.79.350}. In this regime, virtual electron hopping between equivalent cations is most favorable when their spins are antiparallel, yielding robust antiferromagnetic (AFM) order. Upon hydrogen insertion, however, the donated electron localizes on one of the B-site cations, generating a mixed-valent configuration such as Co$^{3+}$/Co$^{2+}$ or Fe$^{3+}$/Fe$^{2+}$. Electron hopping between these nonequivalent mixed-valent sites becomes energetically allowed, in contrast to virtual superexchange processes, and is most efficient when neighboring spins are parallel, leading to double-exchange–driven FM coupling \cite{PhysRev.82.403}. This competition between superexchange and double exchange underlies the observed weakening of AFM interactions ($J_1$, $J_2$) and the overall stabilization of the FM state upon hydrogenation.

\begin{figure}[!h]
    \centering
    \includegraphics[width=1.0\linewidth]{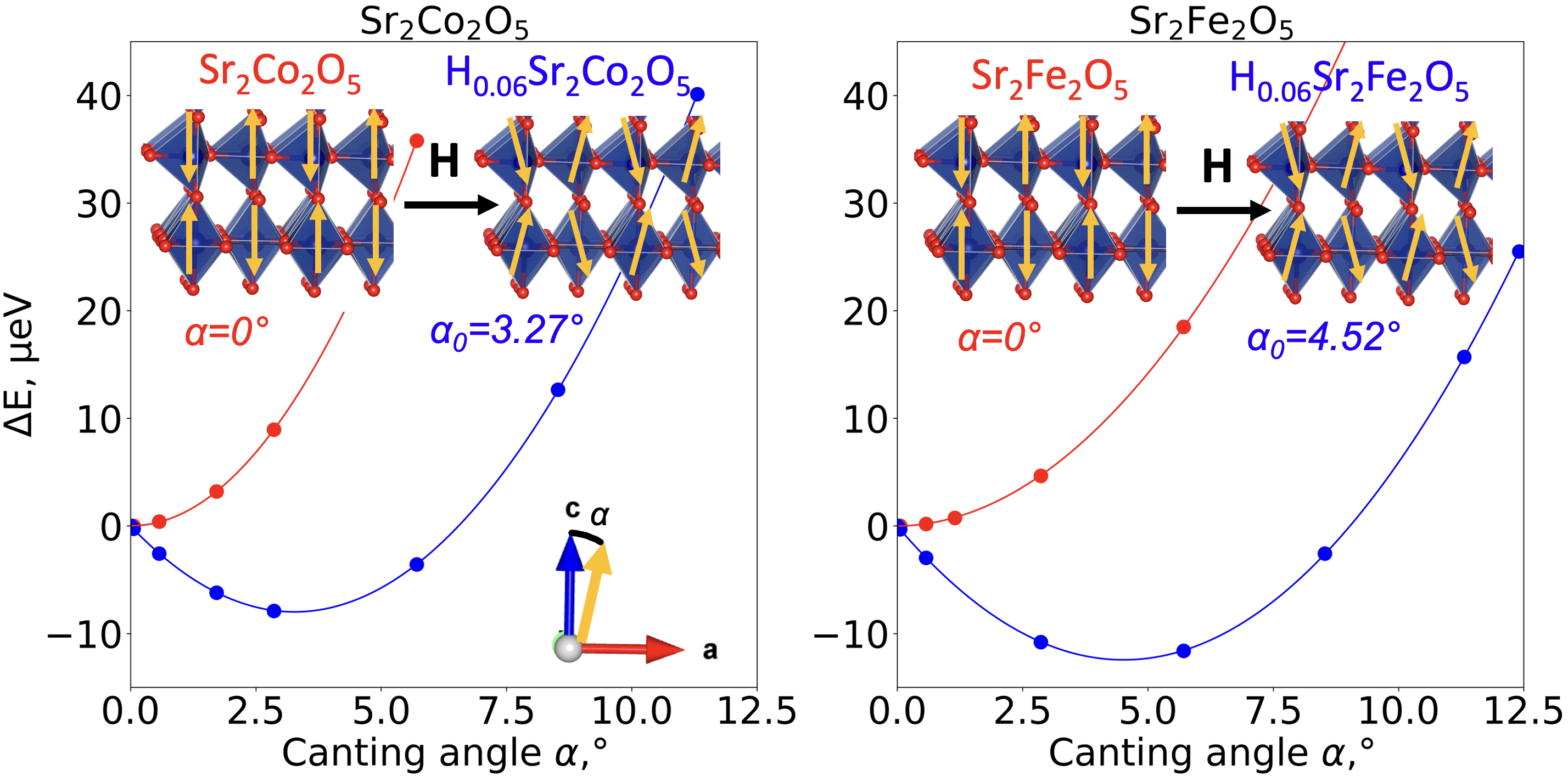}
    \caption{Canting angle energy variations for Sr\(_2\)Fe\(_2\)O\(_5\) and H\(_{0.06}\)Sr\(_2\)Fe\(_2\)O\(_5\). The parabolic fits highlight the effect of hydrogen intercalation on the canting angles. Canting angle energy variations for Sr\(_2\)Co\(_2\)O\(_5\) and H\(_{0.06}\)Sr\(_2\)Co\(_2\)O\(_5\). The parabolic fits highlight the effect of hydrogen intercalation on the canting angles.}
    \label{fig:canting-angles-strain}
\end{figure}

Hydrogen intercalation also induces canting of the N\'eel vector, giving rise to weak ferromagnetism. Figure~\ref{fig:canting-angles-strain} shows canting angle energy profiles for Sr$_2$Fe$_2$O$_5$ and Sr$_2$Co$_2$O$_5$, with hydrogenated systems developing canting angles of 4.5\textdegree and 3.3\textdegree, respectively, compared to zero canting in pristine structures. This is consistent with experimental observations of weak ferromagnetism in hydrogenated Sr$_2$Fe$_2$O$_5$ \cite{10.1002/adfm.202316608}. Microscopically, the canting is consistent with hydrogen-induced changes in magnetic anisotropy driven by orbital hybridization and redistribution of transition-metal $d$-state occupancies, as demonstrated by previous first-principles study. \cite{PhysRevMaterials.4.104416} Thus, hydrogen absorption not only weakens antiferromagnetic exchange but also stabilizes non-collinear spin textures, providing a reversible route to tune magnetic functionality in BM.

\subsection{High-throughput Screening of Hydrogen Absorption Energetics in Brownmillerites}
The possibility of electrochemical H gating or chemical intercalation of H into BM oxides was demonstrated only for a few selected materials, \cite{FISHER1999355,Lu2017,Lu2020,10.1002/adfm.202316608,Jankovic_2011} highlighted in Figure \ref{fig:screening} by red bars,  whereas hydrogen solubility and retention across most other compositions remain unknown.  This motivated us to extend the analysis of hydrogen absorption energies to a broader range of A\(_2\)B\(_2\)O\(_5\) compounds by screening 14 experimentally known BMs reported in the Materials Project database. For selected BMs, we calculated $E_{abs}$ for each of the seven distinct intercalation sites (Fig. \ref{fig:structure_BM}b). The results are summarized in Figure \ref{fig:screening}; the boxes highlight the spread of $E_{abs}$ between different intercalation sites. Blue bars indicate compounds predicted to accommodate hydrogen, with absorption energies comparable to Sr$_2$Fe$_2$O$_5$, which exhibits the highest hydrogen absorption energy among the experimentally confirmed hydrogen-absorbing brownmillerites. Green bars represent compounds with higher absorption energies ($E\mathrm{abs} > 1.5$ eV), suggesting that hydrogen absorption is thermodynamically unfavorable even at low concentrations. The agreement between experimentally demonstrated hydrogen absorption and relatively low calculated absorption energies supports the validity of this classification and highlights additional brownmillerites (Ba$_2$Tl$_2$O$_5$, Sr$_2$Bi$_2$O$_5$, Y$_2$Cu$_2$O$_5$) as promising candidates for further experimental investigation.

\begin{figure}[h]
    \centering
    \includegraphics[width=1\linewidth]{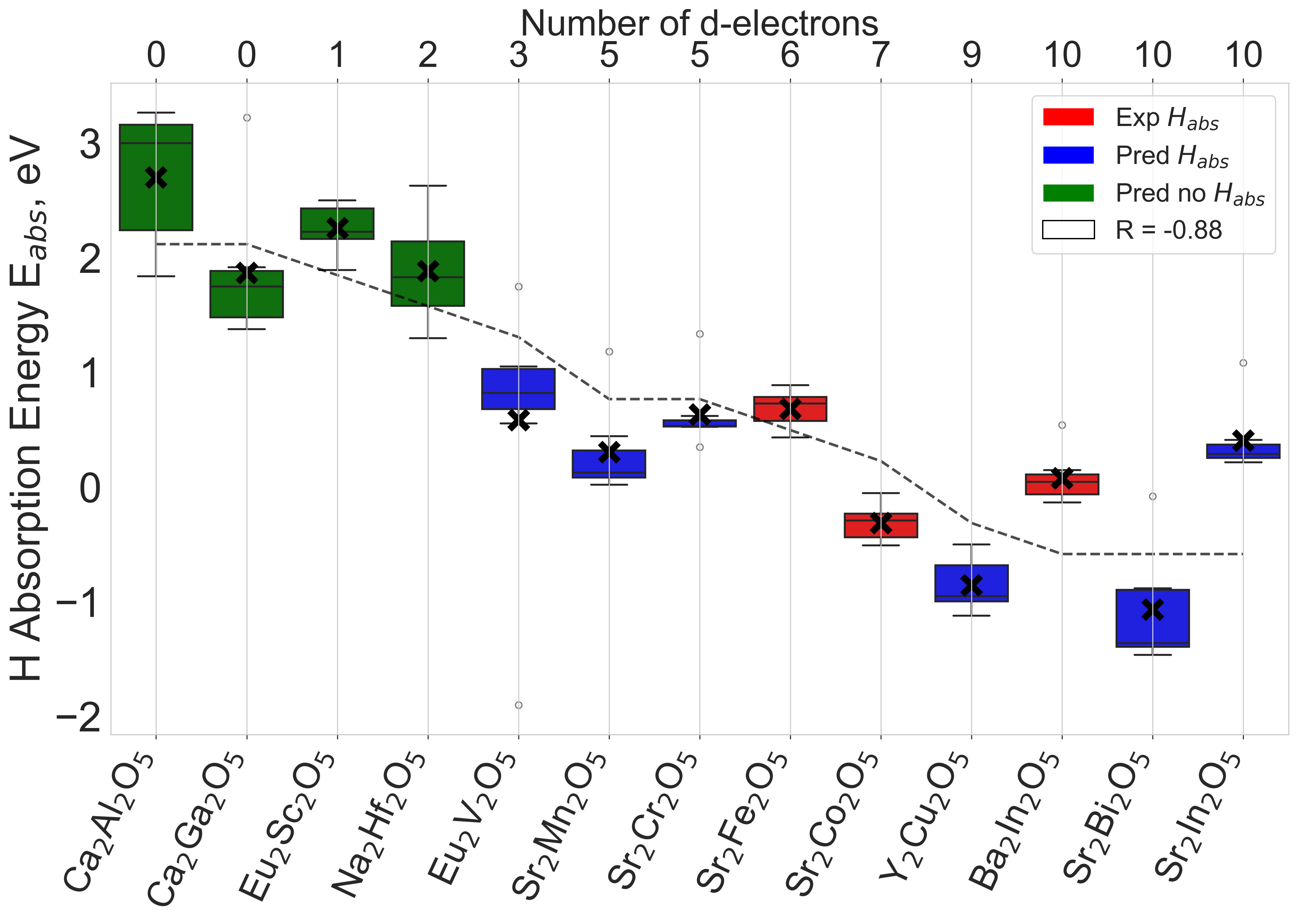}
    \caption{H absorption energy ($E_{\text{abs}}$) of 14 A$_2$B$_2$O$_5$ compounds. The box plot shows the distribution of absorption energies across 7 hydrogen intercalation sites per compound.  Red boxes represent compounds with experimentally observed hydrogen absorption. Blue boxes indicate compounds with favorable absorption energies suggesting potential for hydrogen uptake. Green boxes denote compounds with high absorption energies, making hydrogen absorption unfavorable. Black crosses indicate the mean $E_{\text{abs}}$ for each compound. The dashed black line shows the correlation between $d$-electron count and average absorption energy (R = $-0.88$).}
    \label{fig:screening}
\end{figure}

We find that, as a general trend, hydrogen uptake strongly correlates with the number of B-site $d$ electrons: as the $d$-electron count increases, the average absorption energy across the seven intercalation sites decreases (Fig.~S19, $R = -0.88$). However, this trend does not hold for fully filled $d$ shells of Sr$_2$In$_2$O$_5$ or Ba$_2$Tl$_2$O$_5$ that deviate from this pattern. Regarding the A-site cation, the effect of cation substitution is less pronounced on $E{\mathrm{abs}}$. For example, Ba$_2$In$_2$O$_5$ exhibits just slightly lower average absorption energy than Sr$_2$In$_2$O$_5$, consistent with the larger ionic radius of Ba leading to an expanded lattice and softer vibrational modes that better accommodate hydrogen insertion. However, the limited number of available A-site substitutions prevents drawing general conclusions about A-site chemistry at this stage. 

Compounds ranging from Ba$_2$In$_2$O$_5$ to Sr$_2$Cr$_2$O$_5$ have been successfully synthesized. However, to the best of our knowledge, their potential for hydrogen absorption and diffusion remains unexplored.

\subsection{Assessing uMLIP Performance for Hydrogen Absorption Energetics}
Next, we assessed how selected uMLIPs capture hydrogen $E_{\text{abs}}$ in BM, benchmarking against our first principles calculations at the SCAN(+U) level. To evaluate the performance of uMLIPs, we selected two pretrained neural-network interatomic potentials: CHGNet, trained on a large mixed GGA/GGA+U dataset augmented with r$^2$SCAN calculations (version 0.4.1) \cite{Deng2023CHGNet}, and UMA OMat24, trained on an extensive GGA+U dataset \cite{BarrosoLuque2024OMat24}. As discussed earlier, both GGA+U and SCAN(+U) functionals capture the qualitative absorption trends in BMs, while quantitative differences do not exceed 0.1–0.2 eV (MAE = 0.13 eV; Fig.~S21). 

As shown in Fig.~\ref{fig:CHGnet_vs_DFT_SCAN}, both CHGNet (MP-R2SCAN) and UMA (OMat24) capture the overall energetic trends of hydrogen absorption relative to DFT/SCAN, but exhibit substantial quantitative deviations from parity. Across 95 stable intercalation sites in 14 brownmillerites, CHGNet shows a mean absolute error (MAE) of 1.05 eV, while UMA performs slightly better with an MAE of 0.90 eV, though systematic deviations remain for both models. In both cases the MAE varies significantly from one composition to another (Fig.~S15), underscoring that neither model currently achieves the precision required for quantitative screening.

The deviations from DFT observed for both uMLIPs  are likely a cumulative outcome of multiple factors affecting the energetics of hydrogen absorption. Importantly, neither CHGNet nor UMA explicitly describes electron localization or polaron formation, nor were these models trained on datasets containing polaronic states. However, as we demonstrated  above, variations in H$^+$–e$^-$ separation, polaron localization on transition-metal sites, and magnetic ordering can shift $E_{\text{abs}}$ by 0.5--1 eV even within the same compound and nominal insertion site. The current generation of uMLIPs that do not explicitly encode various structural–electronic–magnetic couplings and therefore inevitably struggles to reproduce robust site-resolved energetics. Both UMA (OMat24) and CHGNet (MP-R2SCAN) perform worst for $|E_{\text{abs}}|\gtrsim1$ eV, suggesting that energetically unfavorable, highly distorted out-of-distribution configurations are underrepresented in their training data. Restricting the dataset to the $|E_\mathrm{abs}|\leq1$ eV range improves quantitative prediction but still yields a MAE of $\approx$ 0.5 eV.

\begin{figure}
    \centering
    \includegraphics[width=0.9\linewidth]{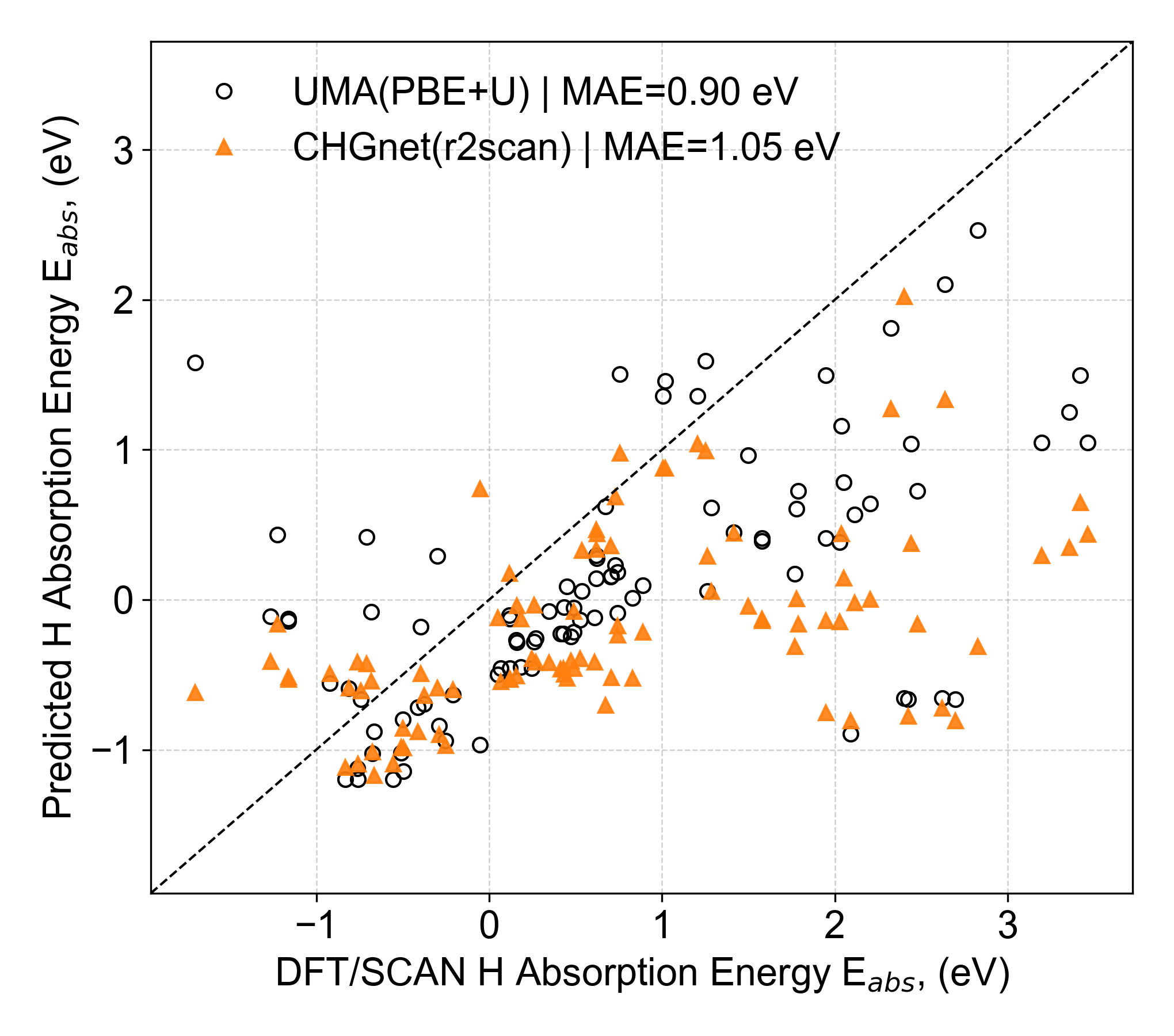}
    \caption{$E_{\text{abs}}$ parity plot: UMA$_{\text{PBE+U}}$ and CHGNet$_{\text{r2SCAN}}$ vs. DFT$_{\text{SCAN}}$ . Data represents 95 intercalation sites across 14 BM compounds. UMA (MAE = 0.90 eV) and CHGNet (MAE = 1.05 eV) show systematic deviations from parity.}
    \label{fig:CHGnet_vs_DFT_SCAN}
\end{figure}

Importantly, the challenge extends beyond predicting the absolute values of absorption energies. uMLIPs struggle to accurately predict and rank the relative site energetics. DFT/SCAN consistently identifies H$_1$ as the most stable interstitial site across Sr$_2$X$_2$O$_5$ (X = Fe, Co, Mn, Cr), whereas both uMLIPs fail to reproduce this ordering reliably (Fig.~S16). In some systems (e.g., Sr$_2$Fe$_2$O$_5$), the models capture the general site trend, but in others, their predictions are inconsistent. For instance, UMA predicts H$_6$ as most stable for Sr$_2$Co$_2$O$_5$ and H$_7$ for Sr$_2$Cr$_2$O$_5$. This irregular behavior indicates that uMLIPs can occasionally yield physically reasonable trends but fail unpredictably across compositions, making them unsuitable for high-throughput screening. These findings are consistent with recent literature. Chung et al. \cite{Chung2025FlexibilityOxygenSublattice} reported qualitatively similar behavior using CHGNet for Sr$_2$Co$_2$O$_5$: while relative site stabilities and migration barriers were partially reproduced, high-energy configurations deviated significantly, and several diffusion pathways failed to converge. Moreover, some H insertion sites relaxed to distinct positions relative to DFT results, illustrating that even advanced uMLIPs face difficulties in capturing the complex energy landscape of hydrogen in oxides. Similarly, the original UMA study \cite{levine2025openmolecules2025omol25} emphasized protonation as a particularly challenging benchmark, with energy errors several times larger than for neutral systems—consistent with our results for transition-metal oxides.

Overall, our results show that both UMA (OMat24) and CHGNet(version 0.4.1, MP-R2SCAN), while capable of reproducing broad energetic trends, remain quantitatively unreliable and occasionally yield qualitatively incorrect predictions. CHGNet exhibits slightly stronger physical awareness of magnetism but larger absolute errors, whereas UMA performs slightly better based on MAE yet fails in predicting correct site orderings. These observations underscore that uMLIPs should not be applied blindly to hydrogen intercalation problems in correlated oxides. Future progress will likely require (i) retraining on datasets that explicitly include proton–polaron pairs and multiple magnetic states, and (ii) hybrid workflows that combine acceleration via uMLIPs for candidate generation with targeted DFT+U/SCAN validation on critical configurations.
In parallel, our findings suggest that surrogate models based on physically motivated descriptors—such as directional phonon modes, O–O distances, and $d$-band characteristics (e.g., $d$-electron count, $d$-band center)—may offer a more transparent and transferable route to predict hydrogen absorption energetics, which warrants exploration in future studies.

\section{Conclusions}

This work identifies an atomic-scale mechanism for how hydrogenation can toggle electronic and magnetic responses in oxygen-deficient perovskite oxides, and  establishes practical design rules for identifying hydrogen-responsive compositions. Our calculations show that hydrogen incorporates as a coupled H$^+$–e$^-$ pair, and the accompanying electron localizes as a small polaron in a chemistry-dependent manner, favoring octahedral sublattices in Co-based compounds and tetrahedral sublattices in Fe-based compounds, consistent with crystal-field splitting. Dilute hydrogen absorption in the considered oxides  produces deep, localized in-gap states that broaden at finite temperature due to polaron mobility near H$^+$ and the associated lattice distortions. In magnetic systems, H$^+$–e$^-$ incorporation further impacts exchange interactions and cants the N\'eel vector, giving rise to weak ferromagnetism. 

We also highlight key methodological challenges for modeling hydrogen absorption in the BM oxides. Our calculations demonstrate that absorption energetics depend strongly on H$^+$–e$^-$ configurations, charge localization, and the underlying magnetic state. Variations in H$^+$–e$^-$ arrangements and magnetic order can shift predicted absorption energies by about 1 eV. Reliable screening therefore requires explicit treatment of competing spin states, accurate control of electron localization, consistent choice of Hubbard $U$ and exchange–correlation functional, and verification that the added electron is captured as a localized polaron rather than spuriously delocalized.

These mechanistic insights translate directly into discovery guidance. Screening across the selected experimentally reported A$_2$B$_2$O$_5$ compositions points to multiple candidates (e.g., Y$_2$Cu$_2$O$_5$, Sr$_2$Bi$_2$O$_5$) with favorable hydrogen uptake and highlights simple, physically interpretable predictors of absorption energy. In particular, more favorable absorption correlates with local structural descriptors that accommodate H$^+$-induced distortions (such as O–O separations and lattice flexibility) and with electronic trends tied to B-site $d$-electron count, providing a compact and transferable set of screening metrics for prioritizing compositions and sites. On the experimental side, the ease of proton insertion in many oxides under applied potential opens opportunities for electrochemically tunable electronic functionality, but it also means that hydrogen can be incorporated unintentionally from ambient moisture. Such ``hidden'' H can complicate interpretation of functional properties and may confound experimental conclusions in catalysis \cite{doi:10.1021/jacs.2c00825,10.1063/5.0082459}, resistive switching \cite{10.1002/aelm.202200353,10.1002/aelm.202200816}, and even superconductivity \cite{PhysRevLett.124.166402,Ding2023}.

Finally, our benchmarking clarifies how further high-throughput discovery campaigns should proceed in this materials class. The current generation of uMLIPs does not reliably reproduce qualitative absorption trends across the considered BM  and site-resolved hydrogen energetics. We attribute these shortcomings to limited representation of relevant configurations in the training data and to the intrinsic complexity of capturing coupled proton–electron incorporation. As a more robust route to identifying hydrogen-responsive materials, we propose descriptor-based screening grounded in identified physically interpretable correlations between absorption energies and materials properties, which offers a transferable framework for identification of hydrogen responsive materials.

\begin{acknowledgments}
This work was supported in part by the NASA Alabama EPSCoR Research Seed Grant program (Grant No.~AL-80NSSC22M0050). 
K.~K. appreciates financial support from the startup package provided by Auburn University. 
V.~K. acknowledges financial support from the Alabama Graduate Research Scholars Program (GRSP), funded through the Alabama Commission on Higher Education and administered by the Alabama EPSCoR. 
Computational resources were provided by the Texas Advanced Computing Center (TACC) at The University of Texas at Austin through allocation DMR22032 on the Frontera supercomputer.
\end{acknowledgments}

\bibliography{apssamp}

@PREAMBLE{
 "\providecommand{\noopsort}[1]{}" 
 # "\providecommand{\singleletter}[1]{#1}%" 
}

@article{hds1-yls4,
  title = {Effect of protons on polaron mobility in transition metal oxides},
  author = {{\v Z}guns, Pjotrs and Yildiz, Bilge},
  journal = {Phys. Rev. Mater.},
  pages = {--},
  year = {2026},
  month = {Jan},
  publisher = {American Physical Society},
  doi = {10.1103/PhysRevMaterials.xx.xxxxxx},
  url = {https://link.aps.org/doi/10.1103/hds1-yls4}
}

@article{PhysRevLett.124.166402,
  title = {Topotactic Hydrogen in Nickelate Superconductors and Akin Infinite-Layer Oxides $AB{\mathrm{O}}_{2}$},
  author = {Si, Liang and Xiao, Wen and Kaufmann, Josef and Tomczak, Jan M. and Lu, Yi and Zhong, Zhicheng and Held, Karsten},
  journal = {Phys. Rev. Lett.},
  volume = {124},
  issue = {16},
  pages = {166402},
  numpages = {8},
  year = {2020},
  month = {Apr},
  publisher = {American Physical Society},
  doi = {10.1103/PhysRevLett.124.166402},
  url = {https://link.aps.org/doi/10.1103/PhysRevLett.124.166402}
}

@article{10.1002/aelm.202200816,
author = {Kunwar, Sundar and Somodi, Chase Bennett and Lalk, Rebecca A. and Rutherford, Bethany X. and Corey, Zachary and Roy, Pinku and Zhang, Di and Hellenbrand, Markus and Xiao, Ming and MacManus-Driscoll, Judith L. and Jia, Quanxi and Wang, Haiyan and Joshua Yang, J. and Nie, Wanyi and Chen, Aiping},
title = {Protons: Critical Species for Resistive Switching in Interface-Type Memristors},
journal = {Advanced Electronic Materials},
volume = {9},
number = {1},
pages = {2200816},
doi = {https://doi.org/10.1002/aelm.202200816},
url = {https://advanced.onlinelibrary.wiley.com/doi/abs/10.1002/aelm.202200816},
year = {2023}
}

@misc{dataset,
  author       = {Korostelev, Vladislav and   \v{Z}guns, Pjotrs and Klyukin, Konstantin},
  title        = {Hydrogen in Brownmillerite Perovskites:First-Principles Insights into Energetics and Induced Electronic-Magnetic Changes, Dataset title on Figshare https://figshare.com/s/d66773e6ca6e96862a19},
  year         = {2025},
  doi          = {10.6084/m9.figshare.30946772},
  publisher    = {Figshare},
  url = {https://figshare.com/s/d66773e6ca6e96862a19}
}

@article{10.1002/aelm.202200353,
author = {Vanka, Srinivas and Shin, Hyungki and Davidson, Bruce A. and Liu, Chong and Zou, Ke},
title = {Hydrogen Atom Doping—A Versatile Method for Modulated Interface Resistive Switching},
journal = {Advanced Electronic Materials},
volume = {8},
number = {10},
pages = {2200353},
doi = {https://doi.org/10.1002/aelm.202200353},
url = {https://advanced.onlinelibrary.wiley.com/doi/abs/10.1002/aelm.202200353},
year = {2022}
}

@article{10.1063/5.0082459,
    author = {Spencer, Michael A. and Fortunato, Jenelle and Augustyn, Veronica},
    title = {Electrochemical proton insertion modulates the hydrogen evolution reaction on tungsten oxides},
    journal = {The Journal of Chemical Physics},
    volume = {156},
    number = {6},
    pages = {064704},
    year = {2022},
    month = {02},
    issn = {0021-9606},
    doi = {10.1063/5.0082459},
    url = {https://doi.org/10.1063/5.0082459},
}

@article{doi:10.1021/jacs.2c00825,
author = {Miu, Evan V. and McKone, James R. and Mpourmpakis, Giannis},
title = {The Sensitivity of Metal Oxide Electrocatalysis to Bulk Hydrogen Intercalation: Hydrogen Evolution on Tungsten Oxide},
journal = {Journal of the American Chemical Society},
volume = {144},
number = {14},
pages = {6420-6433},
year = {2022},
doi = {10.1021/jacs.2c00825},
    note ={PMID: 35289172},

URL = { 
    
        https://doi.org/10.1021/jacs.2c00825
    
    

},
eprint = { 
    
        https://doi.org/10.1021/jacs.2c00825
    
    

}

}

@article{doi:10.1021/acs.chemmater.1c02432,
author = {Islam, Md Shafiqul and Wang, Shuo and Nolan, Adelaide M. and Mo, Yifei},
title = {First-Principles Computational Design and Discovery of Novel Double-Perovskite Proton Conductors},
journal = {Chemistry of Materials},
volume = {33},
number = {21},
pages = {8278-8288},
year = {2021},
doi = {10.1021/acs.chemmater.1c02432},

URL = { 
    
        https://doi.org/10.1021/acs.chemmater.1c02432
    
    

},
eprint = { 
    
        https://doi.org/10.1021/acs.chemmater.1c02432
    
    

}

}

@article{10.2138/rmg.2010.72.17,
    author = {Van Orman, James A. and Crispin, Katherine L.},
    title = {Diffusion in Oxides},
    journal = {Reviews in Mineralogy and Geochemistry},
    volume = {72},
    number = {1},
    pages = {757-825},
    year = {2010},
    month = {01},
    issn = {1529-6466},
    doi = {10.2138/rmg.2010.72.17},
    url = {https://doi.org/10.2138/rmg.2010.72.17},
}

@article{PhysRevB.82.014103,
  title = {Simple descriptors for proton-conducting perovskites from density functional theory},
  author = {Bork, N. and Bonanos, N. and Rossmeisl, J. and Vegge, T.},
  journal = {Phys. Rev. B},
  volume = {82},
  issue = {1},
  pages = {014103},
  numpages = {6},
  year = {2010},
  month = {Jul},
  publisher = {American Physical Society},
  doi = {10.1103/PhysRevB.82.014103},
  url = {https://link.aps.org/doi/10.1103/PhysRevB.82.014103}
}

@article{doi:10.1021/acsaelm.2c00711,
author = {Sidik, Umar and Hattori, Azusa N. and Hattori, Ken and Alaydrus, Musa and Hamada, Ikutaro and Pamasi, Liliany N. and Tanaka, Hidekazu},
title = {Tunable Proton Diffusion in NdNiO3 Thin Films under Regulated Lattice Strains},
journal = {ACS Applied Electronic Materials},
volume = {4},
number = {10},
pages = {4849-4856},
year = {2022},
doi = {10.1021/acsaelm.2c00711},

URL = { 
    
        https://doi.org/10.1021/acsaelm.2c00711
    
    

},
eprint = { 
    
        https://doi.org/10.1021/acsaelm.2c00711
    
    

}

}

@article{Zhang2020PerovskiteNeuralTrees,
  author  = {Zhang, H. T. and Park, T. J. and Zaluzhnyy, I. A. and {et al.}},
  title   = {Perovskite neural trees},
  journal = {Nature Communications},
  volume  = {11},
  pages   = {2245},
  year    = {2020},
  doi     = {10.1038/s41467-020-16105-y},
  url     = {https://doi.org/10.1038/s41467-020-16105-y}
}

@article{Lan2020,
  author  = {Lan, Chunfeng and Li, Huanhuan and Zhao, Shuai},
  title   = {A first-principles study of the proton and oxygen migration behavior in the rare-earth perovskite {SmNiO$_3$}},
  journal = {Journal of Computational Electronics},
  volume  = {19},
  number  = {3},
  pages   = {905--909},
  year    = {2020},
  month   = sep,
  doi     = {10.1007/s10825-020-01501-w},
  url     = {https://doi.org/10.1007/s10825-020-01501-w},
  issn    = {1572-8137}
}

@article{BarrosoLuque2024OMat24,
  title        = {Open Materials 2024 (OMat24) Inorganic Materials Dataset and Models},
  author       = {Barroso-Luque, Luis and Shuaibi, Muhammed and Fu, Xiang and Wood, Brandon M. and Dzamba, Misko and Gao, Meng and Rizvi, Ammar and Zitnick, C. Lawrence and Ulissi, Zachary W.},
  journal      = {arXiv preprint arXiv:2410.12771},
  year         = {2024},
  eprint       = {2410.12771},
  archivePrefix= {arXiv},
  primaryClass = {cond-mat.mtrl-sci},
  doi          = {10.48550/arXiv.2410.12771}
}

@article{PhysRev.82.403,
  title = {Interaction between the $d$-Shells in the Transition Metals. II. Ferromagnetic Compounds of Manganese with Perovskite Structure},
  author = {Zener, Clarence},
  journal = {Phys. Rev.},
  volume = {82},
  issue = {3},
  pages = {403--405},
  numpages = {0},
  year = {1951},
  month = {May},
  publisher = {American Physical Society},
  doi = {10.1103/PhysRev.82.403},
  url = {https://link.aps.org/doi/10.1103/PhysRev.82.403}
}

@article{PhysRev.79.350,
  title = {Antiferromagnetism. Theory of Superexchange Interaction},
  author = {Anderson, P. W.},
  journal = {Phys. Rev.},
  volume = {79},
  issue = {2},
  pages = {350--356},
  numpages = {0},
  year = {1950},
  month = {Jul},
  publisher = {American Physical Society},
  doi = {10.1103/PhysRev.79.350},
  url = {https://link.aps.org/doi/10.1103/PhysRev.79.350}
}

@article{10.1002/adfm.202303416,
author = {Zhou, Xuanchi and Mao, Wei and Cui, Yuchen and Zhang, Hao and Liu, Qi and Nie, Kaiqi and Xu, Xiaoguang and Jiang, Yong and Chen, Nuofu and Chen, Jikun},
title = {Multiple Electronic Phase Transitions of NiO via Manipulating the NiO6 Octahedron and Valence Control},
journal = {Advanced Functional Materials},
volume = {33},
number = {36},
pages = {2303416},
doi = {https://doi.org/10.1002/adfm.202303416},
year = {2023}
}

@article{10.1002/adma.202101316,
author = {Lin, Weinan and Liu, Liang and Liu, Qing and Li, Lei and Shu, Xinyu and Li, Changjian and Xie, Qidong and Jiang, Peiheng and Zheng, Xuan and Guo, Rui and Lim, Zhishiuh and Zeng, Shengwei and Zhou, Guowei and Wang, Han and Zhou, Jing and Yang, Ping and Ariando and Pennycook, Stephen J. and Xu, Xiaohong and Zhong, Zhicheng and Wang, Zhiming and Chen, Jingsheng},
title = {Electric Field Control of the Magnetic Weyl Fermion in an Epitaxial SrRuO3 (111) Thin Film},
journal = {Advanced Materials},
volume = {33},
number = {36},
pages = {2101316},
doi = {https://doi.org/10.1002/adma.202101316},
year = {2021}
}

@article{10.1002/advs.202414991,
author = {Zhou, Xuanchi and Jiao, Yongjie and Lu, Wentian and Guo, Jinjian and Yao, Xiaohui and Ji, Jiahui and Zhou, Guowei and Ji, Huihui and Yuan, Zhe and Xu, Xiaohong},
title = {Hydrogen-Associated Filling-Controlled Mottronics Within Thermodynamically Metastable Vanadium Dioxide},
journal = {Advanced Science},
volume = {12},
number = {14},
pages = {2414991},
doi = {https://doi.org/10.1002/advs.202414991},
year = {2025}
}

@article{Cai2024IonDiffusion,
  author       = {Cai, Y. and Wang, Z. and Wan, J. and et al.},
  title        = {Ion diffusion retarded by diverging chemical susceptibility},
  journal      = {Nature Communications},
  year         = {2024},
  volume       = {15},
  pages        = {5814},
  doi          = {10.1038/s41467-024-50213-3},
  url          = {https://doi.org/10.1038/s41467-024-50213-3},
  issn         = {2041-1723},
  publisher    = {Nature Publishing Group},
  month        = jul
}

@article{Wei2012VO2,
  author       = {Wei, Jing and Ji, Haozhe and Guo, Wei and et al.},
  title        = {Hydrogen stabilization of metallic vanadium dioxide in single-crystal nanobeams},
  journal      = {Nature Nanotechnology},
  year         = {2012},
  volume       = {7},
  pages        = {357--362},
  doi          = {10.1038/nnano.2012.70},
  url          = {https://doi.org/10.1038/nnano.2012.70},
  issn         = {1748-3387},
  publisher    = {Nature Publishing Group},
  month        = jun
}

@article{Yoon2016VO2,
  author       = {Yoon, Heejoon and Choi, Minseok and Lim, Taewoong W. and et al.},
  title        = {Reversible phase modulation and hydrogen storage in multivalent VO$_2$ epitaxial thin films},
  journal      = {Nature Materials},
  year         = {2016},
  volume       = {15},
  pages        = {1113--1119},
  doi          = {10.1038/nmat4692},
  url          = {https://doi.org/10.1038/nmat4692},
  issn         = {1476-1122},
  publisher    = {Nature Publishing Group},
  month        = oct
}

@article{10.1002/advs.202510771,
author = {Zhou, Xuanchi and Yao, Xiaohui and Lu, Wentian and Guo, Jinjian and Ji, Jiahui and Lang, Lili and Zhou, Guowei and Yao, Chunwei and Qiao, Xiaomei and Ji, Huihui and Yuan, Zhe and Xu, Xiaohong},
title = {Manipulating the Hydrogen-Associated Insulator-Metal Transition Through Artificial Microstructure Engineering},
journal = {Advanced Science},
volume = {n/a},
number = {n/a},
pages = {e10771},
doi = {https://doi.org/10.1002/advs.202510771},
year = {2025}
}

@article{10.1002/wcms.70007,
author = {Zhang, Yubo and Ramasamy, Akilan and Pokharel, Kanun and Kothakonda, Manish and Xiao, Bing and Furness, James W. and Ning, Jinliang and Zhang, Ruiqi and Sun, Jianwei},
title = {Advances and Challenges of SCAN and r2SCAN Density Functionals in Transition-Metal Compounds},
journal = {WIREs Computational Molecular Science},
volume = {15},
number = {2},
pages = {e70007},
doi = {https://doi.org/10.1002/wcms.70007},
note = {e70007 CMS-1188.R1},
year = {2025}
}

@article{PhysRevMaterials.4.045401,
  title = {Evaluating optimal $U$ for $3d$ transition-metal oxides within the SCAN+$U$ framework},
  author = {Long, Olivia Y. and Sai Gautam, Gopalakrishnan and Carter, Emily A.},
  journal = {Phys. Rev. Mater.},
  volume = {4},
  issue = {4},
  pages = {045401},
  numpages = {15},
  year = {2020},
  month = {Apr},
  publisher = {American Physical Society},
  doi = {10.1103/PhysRevMaterials.4.045401},
  url = {https://link.aps.org/doi/10.1103/PhysRevMaterials.4.045401}
}

@article{PhysRevMaterials.2.095401,
  title = {Evaluating transition metal oxides within DFT-SCAN and $\text{SCAN}+U$ frameworks for solar thermochemical applications},
  author = {Sai Gautam, Gopalakrishnan and Carter, Emily A.},
  journal = {Phys. Rev. Mater.},
  volume = {2},
  issue = {9},
  pages = {095401},
  numpages = {14},
  year = {2018},
  month = {Sep},
  publisher = {American Physical Society},
  doi = {10.1103/PhysRevMaterials.2.095401},
  url = {https://link.aps.org/doi/10.1103/PhysRevMaterials.2.095401}
}

@article{PhysRevB.75.195212,
  title = {Electron transport via polaron hopping in bulk $\mathrm{Ti}{\mathrm{O}}_{2}$: A density functional theory characterization},
  author = {Deskins, N. Aaron and Dupuis, Michel},
  journal = {Phys. Rev. B},
  volume = {75},
  issue = {19},
  pages = {195212},
  numpages = {10},
  year = {2007},
  month = {May},
  publisher = {American Physical Society},
  doi = {10.1103/PhysRevB.75.195212},
  url = {https://link.aps.org/doi/10.1103/PhysRevB.75.195212}
}

@article{PhysRevMaterials.2.114409,
  title = {Role of interstitial hydrogen in ${\mathrm{SrCoO}}_{2.5}$ antiferromagnetic insulator},
  author = {Liang, Li and Qiao, Shuang and Du, Shiqiao and Zhang, Shunhong and Wu, Jian and Liu, Zheng},
  journal = {Phys. Rev. Mater.},
  volume = {2},
  issue = {11},
  pages = {114409},
  numpages = {5},
  year = {2018},
  month = {Nov},
  publisher = {American Physical Society},
  doi = {10.1103/PhysRevMaterials.2.114409},
  url = {https://link.aps.org/doi/10.1103/PhysRevMaterials.2.114409}
}

@article{PhysRevMaterials.3.024603,
  title = {Towards understanding the special stability of ${\mathrm{SrCoO}}_{2.5}$ and ${\mathrm{HSrCoO}}_{2.5}$},
  author = {Tsang, Sze-Chun and Zhang, Jingzhao and Tse, Kinfai and Zhu, Junyi},
  journal = {Phys. Rev. Mater.},
  volume = {3},
  issue = {2},
  pages = {024603},
  numpages = {9},
  year = {2019},
  month = {Feb},
  publisher = {American Physical Society},
  doi = {10.1103/PhysRevMaterials.3.024603},
  url = {https://link.aps.org/doi/10.1103/PhysRevMaterials.3.024603}
}

@article{PhysRevB.92.174111,
  title = {Crystal structure and electronic properties of bulk and thin film brownmillerite oxides},
  author = {Young, Joshua and Rondinelli, James M.},
  journal = {Phys. Rev. B},
  volume = {92},
  issue = {17},
  pages = {174111},
  numpages = {10},
  year = {2015},
  month = {Nov},
  publisher = {American Physical Society},
  doi = {10.1103/PhysRevB.92.174111},
  url = {https://link.aps.org/doi/10.1103/PhysRevB.92.174111}
}

@article{Chung2025FlexibilityOxygenSublattice,
title = {Flexibility of oxygen sublattice and hydrogen bond length predict proton mobility in ternary metal oxides},
journal = {Matter},
pages = {102568},
year = {2025},
issn = {2590-2385},
doi = {https://doi.org/10.1016/j.matt.2025.102568},
author = {Heejung W. Chung and Pjotrs {\v Z}guns and Ju Li and Bilge Yildiz}
}

@article{PhysRevLett.97.170201,
  title = {Structural Relaxation Made Simple},
  author = {Bitzek, Erik and Koskinen, Pekka and G\"ahler, Franz and Moseler, Michael and Gumbsch, Peter},
  journal = {Phys. Rev. Lett.},
  volume = {97},
  issue = {17},
  pages = {170201},
  numpages = {4},
  year = {2006},
  month = {Oct},
  publisher = {American Physical Society},
  doi = {10.1103/PhysRevLett.97.170201},
  url = {https://link.aps.org/doi/10.1103/PhysRevLett.97.170201}
}

@article{Lu2022Brownmillerite,
  title = {Enhanced low-temperature proton conductivity in hydrogen-intercalated brownmillerite oxide},
  author = {Lu, N. and Zhang, Z. and Wang, Y. and others},
  journal = {Nature Energy},
  volume = {7},
  pages = {1208--1216},
  year = {2022},
  month = {December},
  doi = {10.1038/s41560-022-01166-8},
  url = {https://doi.org/10.1038/s41560-022-01166-8}
}

@article{Deng2023CHGNet,
  title = {CHGNet as a pretrained universal neural network potential for charge-informed atomistic modelling},
  author = {Deng, Bowen and Zhong, Peichen and Jun, KyuJung and others},
  journal = {Nature Machine Intelligence},
  volume = {5},
  pages = {1031--1041},
  year = {2023},
  month = {September},
  doi = {10.1038/s42256-023-00716-3},
  url = {https://doi.org/10.1038/s42256-023-00716-3}
}

@article{PhysRevMaterials.5.113606,
  title = {General-purpose neural network interatomic potential for the $\ensuremath{\alpha}$-iron and hydrogen binary system: Toward atomic-scale understanding of hydrogen embrittlement},
  author = {Meng, Fan-Shun and Du, Jun-Ping and Shinzato, Shuhei and Mori, Hideki and Yu, Peijun and Matsubara, Kazuki and Ishikawa, Nobuyuki and Ogata, Shigenobu},
  journal = {Phys. Rev. Mater.},
  volume = {5},
  issue = {11},
  pages = {113606},
  numpages = {16},
  year = {2021},
  month = {Nov},
  publisher = {American Physical Society},
  doi = {10.1103/PhysRevMaterials.5.113606},
  url = {https://link.aps.org/doi/10.1103/PhysRevMaterials.5.113606}
}

@article{PhysRevMaterials.7.093601,
  title = {Deep neural network potential for simulating hydrogen blistering in tungsten},
  author = {Wang, Xiao-Yang and Wang, Yi-Nan and Xu, Ke and Dai, Fu-Zhi and Liu, Hai-Feng and Lu, Guang-Hong and Wang, Han},
  journal = {Phys. Rev. Mater.},
  volume = {7},
  issue = {9},
  pages = {093601},
  numpages = {20},
  year = {2023},
  month = {Sep},
  publisher = {American Physical Society},
  doi = {10.1103/PhysRevMaterials.7.093601},
  url = {https://link.aps.org/doi/10.1103/PhysRevMaterials.7.093601}
}

@article{Angeletti2025HydrogenDiffusionMgML,
  title        = {Hydrogen diffusion in magnesium using machine learning potentials: a comparative study},
  author       = {Angeletti, Andrea and Leoni, Luca and Massa, Dario and Pasquini, Luca and Papanikolaou, Stefanos and Franchini, Cesare},
  journal      = {npj Computational Materials},
  year         = {2025},
  volume       = {11},
  articleno    = {85},
  pages        = {85},
  month        = mar,
  doi          = {10.1038/s41524-025-01555-z},
  url          = {https://doi.org/10.1038/s41524-025-01555-z}
}

@article{Ito2025PredictingHydrogenDiffusionMLIPs,
  title   = {Predicting hydrogen diffusion in nickel--manganese random alloys using machine learning interatomic potentials},
  author  = {Ito, K. and Matsumura, N. and Iwasaki, Y. and others},
  journal = {Communications Materials},
  year    = {2025},
  volume  = {6},
  pages   = {195},
  month   = aug,
  doi     = {10.1038/s43246-025-00924-x},
  url     = {https://doi.org/10.1038/s43246-025-00924-x}
}

@article{PhysRevB.84.045115,
  title = {Formation enthalpies by mixing GGA and GGA $+$ $U$ calculations},
  author = {Jain, Anubhav and Hautier, Geoffroy and Ong, Shyue Ping and Moore, Charles J. and Fischer, Christopher C. and Persson, Kristin A. and Ceder, Gerbrand},
  journal = {Phys. Rev. B},
  volume = {84},
  issue = {4},
  pages = {045115},
  numpages = {10},
  year = {2011},
  month = {Jul},
  publisher = {American Physical Society},
  doi = {10.1103/PhysRevB.84.045115},
  url = {https://link.aps.org/doi/10.1103/PhysRevB.84.045115}
}

@article{PhysRevB.73.195107,
  title = {Oxidation energies of transition metal oxides within the $\mathrm{GGA}+\mathrm{U}$ framework},
  author = {Wang, Lei and Maxisch, Thomas and Ceder, Gerbrand},
  journal = {Phys. Rev. B},
  volume = {73},
  issue = {19},
  pages = {195107},
  numpages = {6},
  year = {2006},
  month = {May},
  publisher = {American Physical Society},
  doi = {10.1103/PhysRevB.73.195107},
  url = {https://link.aps.org/doi/10.1103/PhysRevB.73.195107}
}

@article{PhysRevB.70.235121,
  title = {First-principles prediction of redox potentials in transition-metal compounds with $\mathrm{LDA}+U$},
  author = {Zhou, F. and Cococcioni, M. and Marianetti, C. A. and Morgan, D. and Ceder, G.},
  journal = {Phys. Rev. B},
  volume = {70},
  issue = {23},
  pages = {235121},
  numpages = {8},
  year = {2004},
  month = {Dec},
  publisher = {American Physical Society},
  doi = {10.1103/PhysRevB.70.235121},
  url = {https://link.aps.org/doi/10.1103/PhysRevB.70.235121}
}

@article{PhysRevLett.115.036402,
  title = {Strongly Constrained and Appropriately Normed Semilocal Density Functional},
  author = {Sun, Jianwei and Ruzsinszky, Adrienn and Perdew, John P.},
  journal = {Phys. Rev. Lett.},
  volume = {115},
  issue = {3},
  pages = {036402},
  numpages = {6},
  year = {2015},
  month = {Jul},
  publisher = {American Physical Society},
  doi = {10.1103/PhysRevLett.115.036402},
  url = {https://link.aps.org/doi/10.1103/PhysRevLett.115.036402}
}

@article{PhysRevLett.100.136406,
  title = {Restoring the Density-Gradient Expansion for Exchange in Solids and Surfaces},
  author = {Perdew, John P. and Ruzsinszky, Adrienn and Csonka, G\'abor I. and Vydrov, Oleg A. and Scuseria, Gustavo E. and Constantin, Lucian A. and Zhou, Xiaolan and Burke, Kieron},
  journal = {Phys. Rev. Lett.},
  volume = {100},
  issue = {13},
  pages = {136406},
  numpages = {4},
  year = {2008},
  month = {Apr},
  publisher = {American Physical Society},
  doi = {10.1103/PhysRevLett.100.136406},
  url = {https://link.aps.org/doi/10.1103/PhysRevLett.100.136406}
}

@article{PhysRevB.49.14251,
  title = {Ab initio molecular-dynamics simulation of the liquid-metal--amorphous-semiconductor transition in germanium},
  author = {Kresse, G. and Hafner, J.},
  journal = {Phys. Rev. B},
  volume = {49},
  issue = {20},
  pages = {14251--14269},
  numpages = {0},
  year = {1994},
  month = {May},
  publisher = {American Physical Society},
  doi = {10.1103/PhysRevB.49.14251},
  url = {https://link.aps.org/doi/10.1103/PhysRevB.49.14251}
}

@article{PhysRevB.47.558,
  title = {Ab initio molecular dynamics for liquid metals},
  author = {Kresse, G. and Hafner, J.},
  journal = {Phys. Rev. B},
  volume = {47},
  issue = {1},
  pages = {558--561},
  numpages = {0},
  year = {1993},
  month = {Jan},
  publisher = {American Physical Society},
  doi = {10.1103/PhysRevB.47.558},
  url = {https://link.aps.org/doi/10.1103/PhysRevB.47.558}
}

@article{KRESSE199615,
title = {Efficiency of ab-initio total energy calculations for metals and semiconductors using a plane-wave basis set},
journal = {Computational Materials Science},
volume = {6},
number = {1},
pages = {15-50},
year = {1996},
issn = {0927-0256},
doi = {https://doi.org/10.1016/0927-0256(96)00008-0},
url = {https://www.sciencedirect.com/science/article/pii/0927025696000080},
author = {G. Kresse and J. Furthmüller}
}

@article{PhysRevB.59.1758,
  title = {From ultrasoft pseudopotentials to the projector augmented-wave method},
  author = {Kresse, G. and Joubert, D.},
  journal = {Phys. Rev. B},
  volume = {59},
  issue = {3},
  pages = {1758--1775},
  numpages = {0},
  year = {1999},
  month = {Jan},
  publisher = {American Physical Society},
  doi = {10.1103/PhysRevB.59.1758},
  url = {https://link.aps.org/doi/10.1103/PhysRevB.59.1758}
}

@article{PhysRevB.54.11169,
  title = {Efficient iterative schemes for ab initio total-energy calculations using a plane-wave basis set},
  author = {Kresse, G. and Furthm\"uller, J.},
  journal = {Phys. Rev. B},
  volume = {54},
  issue = {16},
  pages = {11169--11186},
  numpages = {0},
  year = {1996},
  month = {Oct},
  publisher = {American Physical Society},
  doi = {10.1103/PhysRevB.54.11169},
  url = {https://link.aps.org/doi/10.1103/PhysRevB.54.11169}
}

@misc{levine2025openmolecules2025omol25,
      title={The Open Molecules 2025 (OMol25) Dataset, Evaluations, and Models}, 
      author={Daniel S. Levine and Muhammed Shuaibi and Evan Walter Clark Spotte-Smith and Michael G. Taylor and Muhammad R. Hasyim and Kyle Michel and Ilyes Batatia and Gábor Csányi and Misko Dzamba and Peter Eastman and Nathan C. Frey and Xiang Fu and Vahe Gharakhanyan and Aditi S. Krishnapriyan and Joshua A. Rackers and Sanjeev Raja and Ammar Rizvi and Andrew S. Rosen and Zachary Ulissi and Santiago Vargas and C. Lawrence Zitnick and Samuel M. Blau and Brandon M. Wood},
      year={2025},
      eprint={2505.08762},
      archivePrefix={arXiv},
      primaryClass={physics.chem-ph},
      url={https://arxiv.org/abs/2505.08762}, 
}

@article{
doi:10.1126/sciadv.adq4696,
author = {Jia-Wei Zhao  and Yunxiang Li  and Deyan Luan  and Xiong Wen (David) Lou },
title = {Structural evolution and catalytic mechanisms of perovskite oxides in electrocatalysis},
journal = {Science Advances},
volume = {10},
number = {39},
pages = {eadq4696},
year = {2024},
doi = {10.1126/sciadv.adq4696},
URL = {https://www.science.org/doi/abs/10.1126/sciadv.adq4696}}

@article{Kong2021,
  author    = {P. Kong and V. S. Minkov and M. A. Kuzovnikov and others},
  title     = {Superconductivity up to 243 K in the yttrium-hydrogen system under high pressure},
  journal   = {Nature Communications},
  volume    = {12},
  pages     = {5075},
  year      = {2021},
  doi       = {10.1038/s41467-021-25372-2},
  url       = {https://doi.org/10.1038/s41467-021-25372-2},
  publisher = {Springer Nature},
  note      = {Published: 20 August 2021}
}

@article{Drozdov2015,
  author    = {A. Drozdov and M. Eremets and I. Troyan and others},
  title     = {Conventional superconductivity at 203 kelvin at high pressures in the sulfur hydride system},
  journal   = {Nature},
  volume    = {525},
  pages     = {73--76},
  year      = {2015},
  doi       = {10.1038/nature14964},
  url       = {https://doi.org/10.1038/nature14964},
  publisher = {Springer Nature},
  note      = {Published: 17 August 2015}
}

@article{Ye2020,
  author    = {X. Ye and H. K. Singh and H. Zhang and others},
  title     = {Giant voltage-induced modification of magnetism in micron-scale ferromagnetic metals by hydrogen charging},
  journal   = {Nature Communications},
  volume    = {11},
  pages     = {4849},
  year      = {2020},
  doi       = {10.1038/s41467-020-18552-z},
  url       = {https://doi.org/10.1038/s41467-020-18552-z},
  publisher = {Springer Nature},
  note      = {Published: 24 September 2020}
}

@article{doi:10.1021/acs.nanolett.5c01712,
author = {Wei, Long and Jiang, Peiheng and Chen, Pan and Hong, Yuhao and Deng, Zhixiong and Wang, Tianyang and Xiao, Wen and Wang, Lei and Gan, Yulin and Tian, Xuezeng and Zhong, Zhicheng and Chen, Kai and Liao, Zhaoliang},
title = {Hydrogen-Driven Ferromagnetic Insulator in Cobalt Perovskite above Room Temperature},
journal = {Nano Letters},
volume = {25},
number = {26},
pages = {10478-10486},
year = {2025},
doi = {10.1021/acs.nanolett.5c01712},
    note ={PMID: 40545651},

URL = { 
    
        https://doi.org/10.1021/acs.nanolett.5c01712
    
    

},
eprint = { 
    
        https://doi.org/10.1021/acs.nanolett.5c01712
    
    

}

}

@article{Li2017,
  author    = {H. B. Li and N. Lu and Q. Zhang and others},
  title     = {Electric-field control of ferromagnetism through oxygen ion gating},
  journal   = {Nature Communications},
  volume    = {8},
  pages     = {2156},
  year      = {2017},
  doi       = {10.1038/s41467-017-02359-6},
  url       = {https://doi.org/10.1038/s41467-017-02359-6},
  publisher = {Springer Nature},
  note      = {Published: 18 December 2017}
}

@article{Ding2023,
  author    = {X. Ding and C. C. Tam and X. Sui and others},
  title     = {Critical role of hydrogen for superconductivity in nickelates},
  journal   = {Nature},
  volume    = {615},
  pages     = {50--55},
  year      = {2023},
  doi       = {10.1038/s41586-022-05657-2},
  url       = {https://doi.org/10.1038/s41586-022-05657-2},
  publisher = {Springer Nature},
  note      = {Published: 01 March 2023}
}

@article{PhysRevLett.106.067201,
  title = {High Temperature Magnetic Ordering in the $4d$ Perovskite ${\mathrm{SrTcO}}_{3}$},
  author = {Rodriguez, Efrain E. and Poineau, Fr\'ed\'eric and Llobet, Anna and Kennedy, Brendan J. and Avdeev, Maxim and Thorogood, Gordon J. and Carter, Melody L. and Seshadri, Ram and Singh, David J. and Cheetham, Anthony K.},
  journal = {Phys. Rev. Lett.},
  volume = {106},
  issue = {6},
  pages = {067201},
  numpages = {4},
  year = {2011},
  month = {Feb},
  publisher = {American Physical Society},
  doi = {10.1103/PhysRevLett.106.067201},
  url = {https://link.aps.org/doi/10.1103/PhysRevLett.106.067201}
}

@article{doi:10.1021/cm200454z,
author = {Chang, Jaewan and Lee, Kyujoon and Jung, Myung Hwa and Kwon, Ji-Hwan and Kim, Miyoung and Kim, Sang-Koog},
title = {Emergence of Room-Temperature Magnetic Ordering in Artificially Fabricated Ordered-Double-Perovskite Sr2FeRuO6},
journal = {Chemistry of Materials},
volume = {23},
number = {11},
pages = {2693-2696},
year = {2011},
doi = {10.1021/cm200454z},

URL = { 
    
        https://doi.org/10.1021/cm200454z
    
    

},
eprint = { 
    
        https://doi.org/10.1021/cm200454z
    
    

}

}

@article{10.1002/adma.201604112,
author = {Copie, Olivier and Varignon, Julien and Rotella, Hélène and Steciuk, Gwladys and Boullay, Philippe and Pautrat, Alain and David, Adrian and Mercey, Bernard and Ghosez, Philippe and Prellier, Wilfrid},
title = {Chemical Strain Engineering of Magnetism in Oxide Thin Films},
journal = {Advanced Materials},
volume = {29},
number = {22},
pages = {1604112},
doi = {https://doi.org/10.1002/adma.201604112},
year = {2017}
}

@article{10.1063/1.5090824,
    author = {Maurel, L. and Marcano, N. and Langenberg, E. and Guzmán, R. and Prokscha, T. and Magén, C. and Pardo, J. A. and Algarabel, P. A.},
    title = {Engineering the magnetic order in epitaxially strained Sr1−xBaxMnO3 perovskite thin films},
    journal = {APL Materials},
    volume = {7},
    number = {4},
    pages = {041117},
    year = {2019},
    month = {04},
    issn = {2166-532X},
    doi = {10.1063/1.5090824},
    url = {https://doi.org/10.1063/1.5090824},
}

@article{10.1063/1.3484147,
    author = {Yang, F. and Kemik, N. and Biegalski, M. D. and Christen, H. M. and Arenholz, E. and Takamura, Y.},
    title = {Strain engineering to control the magnetic and magnetotransport properties of La0.67Sr0.33MnO3 thin films},
    journal = {Applied Physics Letters},
    volume = {97},
    number = {9},
    pages = {092503},
    year = {2010},
    month = {08},
    issn = {0003-6951},
    doi = {10.1063/1.3484147},
    url = {https://doi.org/10.1063/1.3484147},
}

@article{PhysRevB.100.174417,
  title = {Strain engineering of magnetic and orbital order in perovskite ${\mathrm{LuMnO}}_{3}$ epitaxial films},
  author = {Ji, Chao and Wang, Yancheng and Guo, Bixiang and Shen, Xiaofan and Luo, Qunyong and Wang, Jianli and Meng, Xianwen and Zhang, Junting and Lu, Xiaomei and Zhu, Jinsong},
  journal = {Phys. Rev. B},
  volume = {100},
  issue = {17},
  pages = {174417},
  numpages = {9},
  year = {2019},
  month = {Nov},
  publisher = {American Physical Society},
  doi = {10.1103/PhysRevB.100.174417},
  url = {https://link.aps.org/doi/10.1103/PhysRevB.100.174417}
}

@article{10.1063/1.4812323,
    author = {Jain, Anubhav and Ong, Shyue Ping and Hautier, Geoffroy and Chen, Wei and Richards, William Davidson and Dacek, Stephen and Cholia, Shreyas and Gunter, Dan and Skinner, David and Ceder, Gerbrand and Persson, Kristin A.},
    title = {Commentary: The Materials Project: A materials genome approach to accelerating materials innovation},
    journal = {APL Materials},
    volume = {1},
    number = {1},
    pages = {011002},
    year = {2013},
    month = {07},
    issn = {2166-532X},
    doi = {10.1063/1.4812323},
    url = {https://doi.org/10.1063/1.4812323},
}

@article{doi:10.1021/acsami.7b17377,
author = {Khare, Amit and Lee, Jaekwang and Park, Jaeseoung and Kim, Gi-Yeop and Choi, Si-Young and Katase, Takayoshi and Roh, Seulki and Yoo, Tae Sup and Hwang, Jungseek and Ohta, Hiromichi and Son, Junwoo and Choi, Woo Seok},
title = {Directing Oxygen Vacancy Channels in SrFeO2.5 Epitaxial Thin Films},
journal = {ACS Applied Materials \& Interfaces},
volume = {10},
number = {5},
pages = {4831-4837},
year = {2018},
doi = {10.1021/acsami.7b17377},
    note ={PMID: 29327588},

URL = { 
    
        https://doi.org/10.1021/acsami.7b17377
    
    

},
eprint = { 
    
        https://doi.org/10.1021/acsami.7b17377
    
    

}

}

@article{Lu2020,
  author    = {Lu, Q. and Huberman, S. and Zhang, H. and others},
  title     = {Bi-directional tuning of thermal transport in SrCoO\(_x\) with electrochemically induced phase transitions},
  journal   = {Nature Materials},
  volume    = {19},
  pages     = {655--662},
  year      = {2020},
  month     = jun,
  doi       = {10.1038/s41563-020-0612-0},
  url       = {https://doi.org/10.1038/s41563-020-0612-0},
  note      = {Published online 24 February 2020}
}

@article{Lu2017,
  author    = {Lu, N. and Zhang, P. and Zhang, Q. and others},
  title     = {Electric-field control of tri-state phase transformation with a selective dual-ion switch},
  journal   = {Nature},
  volume    = {546},
  pages     = {124--128},
  year      = {2017},
  doi       = {10.1038/nature22389},
  url       = {https://doi.org/10.1038/nature22389},
  received  = {2016-10-11},
  accepted  = {2017-04-12},
  published = {2017-05-31},
  issue_date= {2017-06-01}
}

@article{PhysRevMaterials.4.104416,
  title = {Hydrogen tunes magnetic anisotropy by affecting local hybridization at the interface of a ferromagnet with nonmagnetic metals},
  author = {Klyukin, Konstantin and Beach, Geoffrey and Yildiz, Bilge},
  journal = {Phys. Rev. Mater.},
  volume = {4},
  issue = {10},
  pages = {104416},
  numpages = {7},
  year = {2020},
  month = {Oct},
  publisher = {American Physical Society},
  doi = {10.1103/PhysRevMaterials.4.104416},
  url = {https://link.aps.org/doi/10.1103/PhysRevMaterials.4.104416}
}

@article{Jankovic_2011,
doi = {10.1149/1.3511787},
url = {https://dx.doi.org/10.1149/1.3511787},
year = {2010},
month = {nov},
publisher = {The Electrochemical Society, Inc.},
volume = {158},
number = {1},
pages = {B61},
author = {Jankovic, Jasna and Wilkinson, David P. and Hui, Rob},
title = {Proton Conductivity and Stability of  Ba2In2O5 in Hydrogen Containing Atmospheres},
journal = {Journal of The Electrochemical Society}
}

@misc{chen2024fastlithiumiondiffusion,
      title={Fast Lithium Ion Diffusion in Brownmillerite $\mathrm{Li}_{x}\mathrm{{Sr}_{2}{Co}_{2}{O}_{5}}$}, 
      author={Xin Chen and Xixiang Zhang and Jie-Xiang Yu and Jiadong Zang},
      year={2024},
      eprint={2402.17557},
      archivePrefix={arXiv},
      primaryClass={cond-mat.mtrl-sci},
      url={https://arxiv.org/abs/2402.17557}, 
}

@incollection{THABET202091,
title = {Chapter 4 - Protonic-based ceramics for fuel cells and electrolyzers},
editor = {Massimiliano {Lo Faro}},
booktitle = {Solid Oxide-Based Electrochemical Devices},
publisher = {Academic Press},
pages = {91-122},
year = {2020},
isbn = {978-0-12-818285-7},
doi = {https://doi.org/10.1016/B978-0-12-818285-7.00004-6},
url = {https://www.sciencedirect.com/science/article/pii/B9780128182857000046},
author = {Kawther Thabet and Annie {Le Gal La Salle} and Eric Quarez and Olivier Joubert}
}

@article{doi:10.1021/acs.chemrev.4c00071,
  author  = {Yuan, Yifan and Patel, Ranjan Kumar and Banik, Suvo and Reta, Tadesse Billo and Bisht, Ravindra Singh and Fong, Dillon D. and Sankaranarayanan, Subramanian K. R. S. and Ramanathan, Shriram},
  title   = {Proton Conducting Neuromorphic Materials and Devices},
  journal = {Chemical Reviews},
  year    = {2024},
  volume  = {124},
  number  = {16},
  pages   = {9733--9784},
  doi     = {10.1021/acs.chemrev.4c00071},
  url     = {https://doi.org/10.1021/acs.chemrev.4c00071}
}

@article{doi:10.1021/acs.chemrev.4c00512,
author = {Talin, A. Alec and Meyer, Jordan and Li, Jingxian and Huang, Mantao and Schwacke, Miranda and Chung, Heejung W. and Xu, Longlong and Fuller, Elliot J. and Li, Yiyang and Yildiz, Bilge},
title = {Electrochemical Random-Access Memory: Progress, Perspectives, and Opportunities},
journal = {Chemical Reviews},
volume = {125},
number = {4},
pages = {1962-2008},
year = {2025},
doi = {10.1021/acs.chemrev.4c00512},
    note ={PMID: 39960411},

URL = { 
    
        https://doi.org/10.1021/acs.chemrev.4c00512
    
    

},
eprint = { 
    
        https://doi.org/10.1021/acs.chemrev.4c00512
    
    

}

}

@article{Chen2022,
  author = {Chen, G. and Ophus, C. and Quintana, A. and others},
  title = {Reversible writing/deleting of magnetic skyrmions through hydrogen adsorption/desorption},
  journal = {Nature Communications},
  volume = {13},
  pages = {1350},
  year = {2022},
  doi = {10.1038/s41467-022-28968-4},
  url = {https://doi.org/10.1038/s41467-022-28968-4},
  received = {2021-06-11},
  accepted = {2022-02-17},
  published = {2022-03-15}
}

@article{Piatti2023,
  author = {Piatti, E. and Prando, G. and Meinero, M. and others},
  title = {Superconductivity induced by gate-driven hydrogen intercalation in the charge-density-wave compound 1T-TiSe$_2$},
  journal = {Communications Physics},
  volume = {6},
  pages = {202},
  year = {2023},
  doi = {10.1038/s42005-023-01330-w},
  url = {https://doi.org/10.1038/s42005-023-01330-w},
  received = {2023-03-21},
  accepted = {2023-07-28},
  published = {2023-08-05}
}

@article{NAGY1989567,
title = {Hydrogen in tungsten bronzes: mechanism of hydrogen intercalation},
journal = {International Journal of Hydrogen Energy},
volume = {14},
number = {8},
pages = {567-572},
year = {1989},
note = {WHEC-VII papers not included in the proceedings},
issn = {0360-3199},
doi = {https://doi.org/10.1016/0360-3199(89)90115-8},
url = {https://www.sciencedirect.com/science/article/pii/0360319989901158},
author = {G. Nagy and R. Schiller}
}

@Article{D1RA01959G,
author ="Yu, Ting and Zhang, He and Li, Dan and Lu, Yanwu",
title  ="Electronic and optical properties of silicene on GaAs(111) with hydrogen intercalation: a first-principles study",
journal  ="RSC Adv.",
year  ="2021",
volume  ="11",
issue  ="26",
pages  ="16040-16050",
publisher  ="The Royal Society of Chemistry",
doi  ="10.1039/D1RA01959G",
url  ="http://dx.doi.org/10.1039/D1RA01959G"}

@article{OGAWA2020181,
title = {Hydrogen, as an alloying element, enables a greater strength-ductility balance in an Fe-Cr-Ni-based, stable austenitic stainless steel},
journal = {Acta Materialia},
volume = {199},
pages = {181-192},
year = {2020},
issn = {1359-6454},
doi = {https://doi.org/10.1016/j.actamat.2020.08.024},
url = {https://www.sciencedirect.com/science/article/pii/S1359645420306212},
author = {Yuhei Ogawa and Hyuga Hosoi and Kaneaki Tsuzaki and Timothée Redarce and Osamu Takakuwa and Hisao Matsunaga}
}

@article{10.1002/adma.201903738,
author = {Ning, Shuai and Huberman, Samuel C. and Ding, Zhiwei and Nahm, Ho-Hyun and Kim, Yong-Hyun and Kim, Hyun-Suk and Chen, Gang and Ross, Caroline A.},
title = {Anomalous Defect Dependence of Thermal Conductivity in Epitaxial WO3 Thin Films},
journal = {Advanced Materials},
volume = {31},
number = {43},
pages = {1903738},
doi = {https://doi.org/10.1002/adma.201903738},
year = {2019}
}

@article{shin2021strain,
  title={Strain-Induced Magnetic Transitions in SrMO2. 5 (M= Mn, Fe) Thin Films with Ordered Oxygen Vacancies},
  author={Shin, Yongjin and Rondinelli, James M},
  journal={Inorganic chemistry},
  volume={60},
  number={17},
  pages={13161--13167},
  year={2021},
  publisher={ACS Publications}
}

@article{10.1002/adma.202418484,
author = {Huang, Mantao and Xu, Longlong and del Alamo, Jesús A. and Li, Ju and Yildiz, Bilge},
title = {Nonlinear Ion Dynamics Enable Spike Timing Dependent Plasticity of Electrochemical Ionic Synapses},
journal = {Advanced Materials},
volume = {37},
number = {10},
pages = {2418484},
doi = {https://doi.org/10.1002/adma.202418484},
year = {2025}
}

@article{10.1063/5.0241360,
    author = {Ou, Jiahui and Zhou, Haiping and Huang, Haoliang and Rao, Feng and Zeng, Xierong and Chen, Lang and Shao, Ruiwen and Duan, Manyi and Huang, Chuanwei},
    title = {Above 400 K robust ferromagnetic insulating phase in hydrogenated brownmillerite iron oxide films with distinct coordinate},
    journal = {Applied Physics Reviews},
    volume = {12},
    number = {1},
    pages = {011412},
    year = {2025},
    month = {02},
    issn = {1931-9401},
    doi = {10.1063/5.0241360},
    url = {https://doi.org/10.1063/5.0241360}
}

@article{FISHER1999355,
title = {Defect, protons and conductivity in brownmillerite-structured Ba2In2O5},
journal = {Solid State Ionics},
volume = {118},
number = {3},
pages = {355-363},
year = {1999},
issn = {0167-2738},
doi = {https://doi.org/10.1016/S0167-2738(98)00391-9},
url = {https://www.sciencedirect.com/science/article/pii/S0167273898003919},
author = {C.A.J. Fisher and M.S. Islam}
}

@Article{C3NR05882D,
author ="Wan, Chang Jin and Zhu, Li Qiang and Zhou, Ju Mei and Shi, Yi and Wan, Qing",
title  ="Inorganic proton conducting electrolyte coupled oxide-based dendritic transistors for synaptic electronics",
journal  ="Nanoscale",
year  ="2014",
volume  ="6",
issue  ="9",
pages  ="4491-4497",
publisher  ="The Royal Society of Chemistry",
doi  ="10.1039/C3NR05882D",
url  ="http://dx.doi.org/10.1039/C3NR05882D"}

@ARTICLE{7845629,
  author={Wan, Xiang and Yang, Yi and He, Yongli and Feng, Ping and Li, Wenjun and Wan, Qing},
  journal={IEEE Electron Device Letters}, 
  title={Neuromorphic Simulation of Proton Conductors Laterally Coupled Oxide-Based Transistors With Multiple in-Plane Gates}, 
  year={2017},
  volume={38},
  number={4},
  pages={525-528},
  doi={10.1109/LED.2017.2665578}}

@article{Liang2018RoleOI,
  title={Role of interstitial hydrogen in 
SrCoO2.5
 antiferromagnetic insulator},
  author={Li Liang and Shuang Qiao and Shunhong Zhang and Jian Wu and Zheng Liu},
  journal={Physical Review Materials},
  year={2018},
  url={https://api.semanticscholar.org/CorpusID:67838527}
}

@article{Tsang2018TowardsUT,
  title={Towards understanding the special stability of 
SrCoO2.5
 and 
HSrCoO2.5},
  author={Sze-Chun Tsang and J. Zhang and Kinfai Tse and Junyi Zhu},
  journal={Physical Review Materials},
  year={2018},
  url={https://api.semanticscholar.org/CorpusID:119068535}
}

@article{Wang2023AbnormalSO,
  title={Abnormal stability of hydrogenic defects and magnetism near the HSrCoO2.5(0 0 1) surface},
  author={Yupu Wang and Gaofeng Teng and Chun To Yiu and Junyi Zhu},
  journal={Applied Surface Science},
  year={2023},
  url={https://api.semanticscholar.org/CorpusID:264890329}
}

@article{Teng2022DiffusionSO,
  title={Diffusion Studies of H of Different Charge States and Their Interplay with Co Spin States in SrCoO2.5},
  author={Gaofeng Teng and Yupu Wang and J. Zhang and Sze-Chun Tsang and Junyi Zhu},
  journal={The Journal of Physical Chemistry C},
  year={2022},
  url={https://api.semanticscholar.org/CorpusID:252316181}
}

@article{SCHULTHEI20242652,
title = {Two unexpected companions: Ferroelectricity and proton conductivity},
journal = {Matter},
volume = {7},
number = {8},
pages = {2652-2654},
year = {2024},
issn = {2590-2385},
doi = {https://doi.org/10.1016/j.matt.2024.05.005},
url = {https://www.sciencedirect.com/science/article/pii/S2590238524002303},
author = {J. Schultheiß}
}

@article{C9CP06522A,
author ="Yoo, Pilsun and Liao, Peilin",
title  ="First principles study on hydrogen doping induced metal-to-insulator transition in rare earth nickelates RNiO3 (R = Pr{,} Nd{,} Sm{,} Eu{,} Gd{,} Tb{,} Dy{,} Yb)",
journal  ="Phys. Chem. Chem. Phys.",
year  ="2020",
volume  ="22",
issue  ="13",
pages  ="6888-6895",
publisher  ="The Royal Society of Chemistry",
doi  ="10.1039/C9CP06522A",
url  ="http://dx.doi.org/10.1039/C9CP06522A"}

@article{10.1063/1.4927322,
    author = {Chen, Jikun and Zhou, You and Middey, Srimanta and Jiang, Jun and Chen, Nuofu and Chen, Lidong and Shi, Xun and D{\"o}beli, Max and Shi, Jian and Chakhalian, Jak and Ramanathan, Shriram},
    title = {Self-limited kinetics of electron doping in correlated oxides},
    journal = {Applied Physics Letters},
    volume = {107},
    number = {3},
    pages = {031905},
    year = {2015},
    month = {07},
    issn = {0003-6951},
    doi = {10.1063/1.4927322},
    url = {https://doi.org/10.1063/1.4927322},
}

@misc{yamauchi2022hydrogeninducedmetalinsulatortransitionaccompanied,
      title={Hydrogen-Induced Metal-Insulator Transition Accompanied by Inter-Layer Charge Ordering in SmNiO$_3$}, 
      author={Kunihiko Yamauchi and Ikutaro Hamada},
      year={2022},
      eprint={2210.07656},
      archivePrefix={arXiv},
      primaryClass={cond-mat.str-el},
      url={https://arxiv.org/abs/2210.07656}, 
}

@article{Shao2023,
  author = {Shao, Yuzhou and Liu, Jiale and Li, Peng and Zhao, Xinyu and Lin, Zhaoliang and Chen, Yuan and Li, Jing and Ding, Ning},
  title = {Tuning hydrogen chemisorption by structural engineering in perovskite oxides},
  journal = {Nature Communications},
  year = {2023},
  volume = {14},
  pages = {6799},
  doi = {10.1038/s41467-024-49213-0},
  url = {https://www.nature.com/articles/s41467-024-49213-0}
}

@article{PhysRevB.109.205124,
  title = {Dynamical correlations leading to site and orbital selective Mott insulator transition in hydrogen doped ${\mathrm{SmNiO}}_{3}$},
  author = {Bhat, Soumya S. and Singh, Vijay and Herath, Uthpala and Varughese, Bilvin and Sankaranarayanan, Subramanian K. R. S. and Park, Hyowon and Romero, Aldo H.},
  journal = {Phys. Rev. B},
  volume = {109},
  issue = {20},
  pages = {205124},
  numpages = {7},
  year = {2024},
  month = {May},
  publisher = {American Physical Society},
  doi = {10.1103/PhysRevB.109.205124},
  url = {https://link.aps.org/doi/10.1103/PhysRevB.109.205124}
}

@article{doi:10.1021/acs.chemmater.0c00544,
author = {Islam, Md Shafiqul and Nolan, Adelaide M. and Wang, Shuo and Bai, Qiang and Mo, Yifei},
title = {A Computational Study of Fast Proton Diffusion in Brownmillerite Sr2Co2O5},
journal = {Chemistry of Materials},
volume = {32},
number = {12},
pages = {5028-5035},
year = {2020},
doi = {10.1021/acs.chemmater.0c00544},

URL = { 
    
        https://doi.org/10.1021/acs.chemmater.0c00544
    
    

},
eprint = { 
    
        https://doi.org/10.1021/acs.chemmater.0c00544
}
}

@article{10.1002/asia.201701661,
author = {Xie, Lifang and Chen, Ting and Chan, Hang Cheong and Shu, Yijin and Gao, Qingsheng},
title = {Hydrogen Doping into MoO3 Supports toward Modulated Metal–Support Interactions and Efficient Furfural Hydrogenation on Iridium Nanocatalysts},
journal = {Chemistry – An Asian Journal},
volume = {13},
number = {6},
pages = {641-647},
doi = {https://doi.org/10.1002/asia.201701661},
year = {2018}
}

@article{doi:10.1021/jp508313y,
author = {Guan, Jing and Peng, Gongming and Cao, Quan and Mu, Xindong},
title = {Role of MoO3 on a Rhodium Catalyst in the Selective Hydrogenolysis of Biomass-Derived Tetrahydrofurfuryl Alcohol into 1,5-Pentanediol},
journal = {The Journal of Physical Chemistry C},
volume = {118},
number = {44},
pages = {25555-25566},
year = {2014},
doi = {10.1021/jp508313y},

URL = { 
    
        https://doi.org/10.1021/jp508313y
},
eprint = { 
    
        https://doi.org/10.1021/jp508313y
}
}

@article{10.1002/adfm.202316608,
author = {Yan, Fengbo and Korostelev, Vladislav and Cho, Eunsoo and Yang, Kaichuang and Wu, Jingrui and Lu, Qiyang and Luo, Feng and Ross, Caroline A. and Klyukin, Konstantin and Ning, Shuai},
title = {Ionic-Liquid-Gating-Induced Hydrogenation in Epitaxial Strontium Ferrite},
journal = {Advanced Functional Materials},
volume = {34},
number = {27},
pages = {2316608},
doi = {https://doi.org/10.1002/adfm.202316608},
year = {2024}
}

@article{ZHANG202219027,
title = {Optimizing electronic state in Sr2Co2O5-x with ferromagnetic state by improving oxygen vacancies for oxygen evolution reaction},
journal = {International Journal of Hydrogen Energy},
volume = {47},
number = {44},
pages = {19027-19037},
year = {2022},
issn = {0360-3199},
doi = {https://doi.org/10.1016/j.ijhydene.2022.04.095},
url = {https://www.sciencedirect.com/science/article/pii/S0360319922016081},
author = {Qian Zhang and Maosong Sun and Jie Zhu and Sudong Yang and Lin Chen and Xulin Yang and Pan Wang and Kui Li and Peng Zhao}
}

@article{Meng2021,
  author       = {Meng, M. and Sun, Y. and Li, Y. and others},
  title        = {Three dimensional band-filling control of complex oxides triggered by interfacial electron transfer},
  journal      = {Nature Communications},
  volume       = {12},
  pages        = {2447},
  year         = {2021},
  doi          = {10.1038/s41467-021-22790-0},
  url          = {https://doi.org/10.1038/s41467-021-22790-0},
  publisher    = {Springer Nature},
  received     = {2020-08-21},
  accepted     = {2021-03-29},
  published    = {2021-04-27}
}

@article{PhysRevMaterials.6.035003,
  title = {Data-driven approach to parameterize $\mathrm{SCAN}+U$ for an accurate description of $3d$ transition metal oxide thermochemistry},
  author = {Artrith, Nongnuch and Garrido Torres, Jos\'e Antonio and Urban, Alexander and Hybertsen, Mark S.},
  journal = {Phys. Rev. Mater.},
  volume = {6},
  issue = {3},
  pages = {035003},
  numpages = {13},
  year = {2022},
  month = {Mar},
  publisher = {American Physical Society},
  doi = {10.1103/PhysRevMaterials.6.035003},
  url = {https://link.aps.org/doi/10.1103/PhysRevMaterials.6.035003}
}

\end{document}


\title{Supporing Information for "Hydrogen in Brownmillerite Perovskites: \\
First-Principles Insights into Energetics and Induced Electronic-Magnetic Changes"}

\author[1]{Vladislav Korostelev}
\author[2,3]{Pjotrs \v{Z}guns}
\author[1]{Konstantin Klyukin}

\affil[1]{Department of Materials Engineering, Auburn University, Auburn, Alabama, USA}
\affil[2]{Department of Materials Science and Engineering, Massachusetts Institute of Technology,Cambridge, Massachusetts, USA}
\affil[3]{Institute of Solid State Physics, University of Latvia, Riga, Latvia}

\date{}

\maketitle

\noindent Corresponding author: \texttt{klyukin@auburn.edu}

\maketitle

\renewcommand{\thefigure}{S\arabic{figure}}
\setcounter{figure}{0}
\renewcommand{\thetable}{S\arabic{table}}
\setcounter{table}{0}

\begin{figure*}
    \centering
    \includegraphics[width=0.9\linewidth]{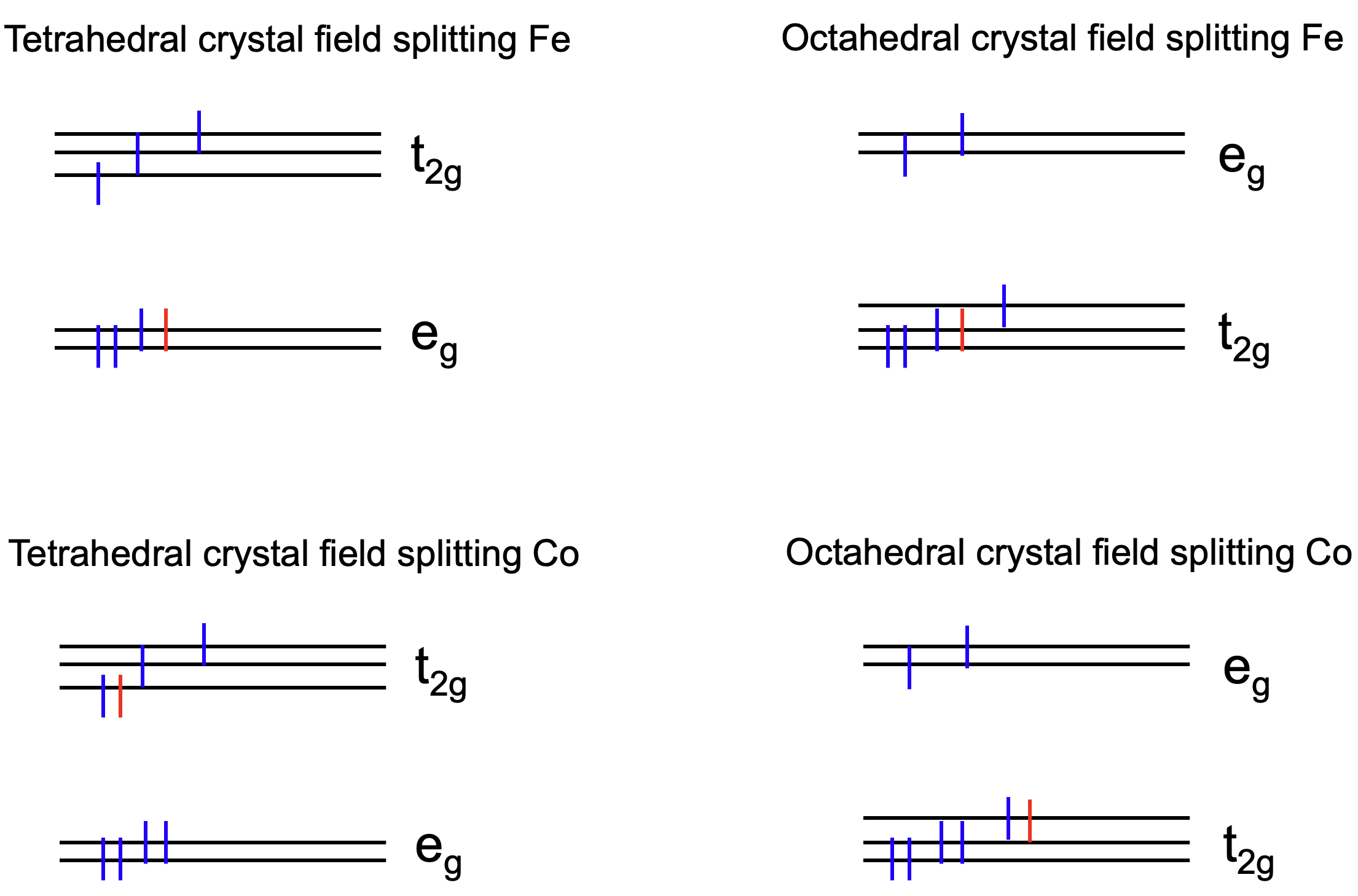}
    \caption{Schematic representation of tetrahedral and octahedral crystal field splitting for Fe and Co. Blue lines represent the d\(^6\) electrons of Fe and d\(^7\) electrons of Co. The red line indicates the additional electron introduced by hydrogen, which localizes on either octahedral or tetrahedral metal atoms.}
    \label{fig:tetrahedral_octahedral_splitting}
\end{figure*}

\begin{figure*}
    \centering
    \includegraphics[width=1.0\linewidth]{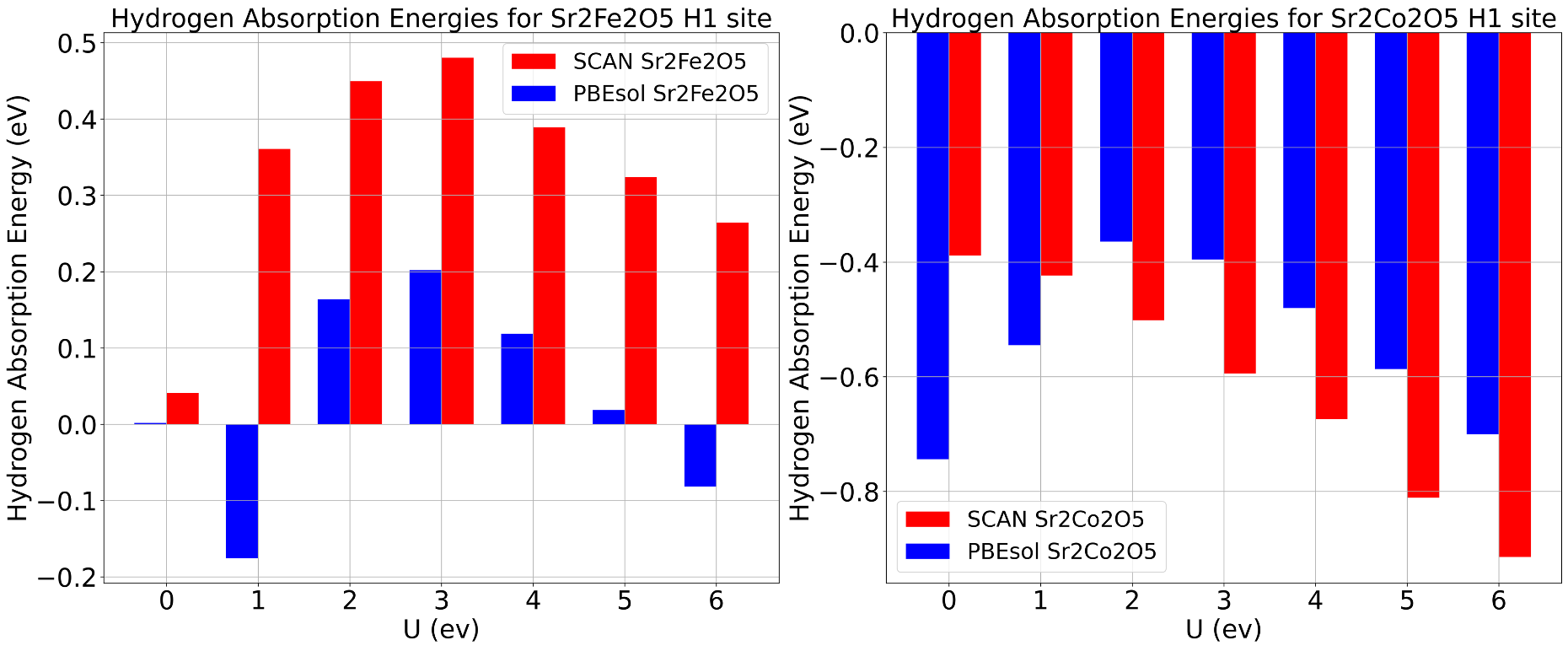}
    \caption{Hydrogen absorption energy variation as a function of chosen U for Fe in Sr\(_2\)Fe\(_2\)O\(_5\) and Sr\(_2\)Cr\(_2\)O\(_5\) perovskite}
    \label{fig:GGA_META_GGA_U}
\end{figure*}

\begin{figure*}
    \centering
    \includegraphics[width=1\linewidth]{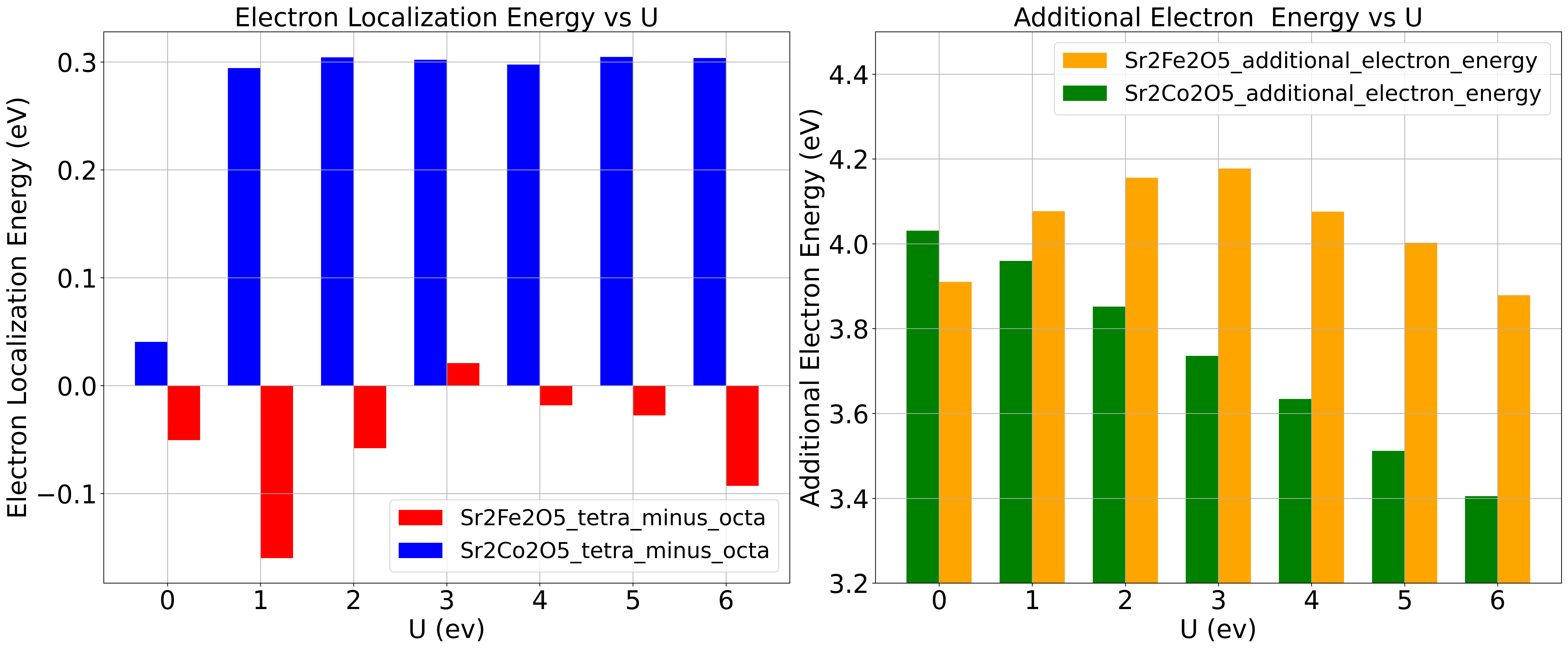}
    \caption{Impact of the Hubbard U parameter on electron localization and additional electron energy for SrCoO$_{2.5}$ and SrFeO$_{2.5}$. The left graph (Electron Localization Energy vs U) shows the difference in electron localization energy between tetrahedral and octahedral sites in SrFeO$_{2.5}$ and SrCoO$_{2.5}$, represented in red and blue respectively. The right graph (Additional Electron Energy vs U) displays the energy required to add an electron to the system at various U values, illustrating how this parameter influences the electronic structure of these brownmillerite compounds. The patterns reveal a stronger dependency of electron addition energy on U for SrFeO$_{2.5}$ compared to SrCoO$_{2.5}$.}
    \label{fig:U_influence_on_electron_localization}
\end{figure*}

\begin{table}
\centering
\caption{Exchange Coupling constants in pure and hydrogenated Sr\(_2\)Co\(_2\)O\(_5\), Sr\(_2\)Fe\(_2\)O\(_5\),  Sr\(_2\)Cr\(_2\)O\(_5\) and Sr\(_2\)Mn\(_2\)O\(_5\). SCAN+U}
\begin{tabular}{|l|c|c|c|}
\hline
\textbf{System} & \textbf{J\(_1\), meV} & \textbf{J\(_2\), meV} & \textbf{J\(_3\), meV} \\ \hline
Sr\(_2\)Co\(_2\)O\(_5\) & -18.87 & -9.43 & -15.02 \\ \hline
H + Sr\(_2\)Co\(_2\)O\(_5\) & -16.14 & -8.07 & -16.71 \\ \hline
Sr\(_2\)Fe\(_2\)O\(_5\) & -17.16 & -8.58 & -9.45 \\ \hline
H + Sr\(_2\)Fe\(_2\)O\(_5\) & -15.84 & -7.92 & -10.46 \\ \hline
Sr\(_2\)Cr\(_2\)O\(_5\) & -4.47  & -2.23 & 2.94 \\ \hline
H + Sr\(_2\)Cr\(_2\)O\(_5\) & -4.28  & -2.14 & 2.62 \\ \hline

Sr\(_2\)Mn\(_2\)O\(_5\) & -2.58  & -1.29 & 0.15 \\ \hline
H + Sr\(_2\)Mn\(_2\)O\(_5\) & -1.14  & -0.57 & -0.22 \\ \hline
\end{tabular}
\label{table:SCAN_exchange-coupling_table}
\end{table}

\begin{table}[h!]
\centering
\caption{Magnetic exchange couplings in Sr\(_2\)Co\(_2\)O\(_5\) and Sr\(_2\)Fe\(_2\)O\(_5\), including pristine, hydrogenated, electron-only, and proton-only cases. All values in meV. PBE\(_{sol}\)+U.}
\begin{tabular}{|l|c|c|c|}
\hline
\textbf{System} & \textbf{J\(_1\), meV} & \textbf{J\(_2\), meV} & \textbf{J\(_3\), meV} \\ \hline
Sr\(_2\)Co\(_2\)O\(_5\) U6.5 & -19.27 & -9.64 & -12.25 \\ \hline
H + Sr\(_2\)Co\(_2\)O\(_5\) U6.5 & -17.72 & -8.86 & -14.08 \\ \hline
Sr\(_2\)Co\(_2\)O\(_5\) U3.32 & -18.81 & -9.41 & -20.28 \\ \hline
H + Sr\(_2\)Co\(_2\)O\(_5\) U3.32 & -14.00 & -7.00 & -19.45 \\ \hline
e\(^-\) + Sr\(_2\)Co\(_2\)O\(_5\) U3.32 (e alone) & -14.68 & -7.34 & -18.49 \\ \hline
H\(^+\) + Sr\(_2\)Co\(_2\)O\(_5\) U3.32 (proton alone) & -17.17 & -8.59 & -19.22 \\ \hline
Sr\(_2\)Fe\(_2\)O\(_5\) & -20.07 & -10.03 & -11.92 \\ \hline
H + Sr\(_2\)Fe\(_2\)O\(_5\) & -18.57 & -9.28 & -12.66 \\ \hline
e\(^-\) + Sr\(_2\)Fe\(_2\)O\(_5\) U3.32 (e alone) & -16.66 & -8.33 & -9.82 \\ \hline
H\(^+\) + Sr\(_2\)Fe\(_2\)O\(_5\) U3.32 (proton alone) & -17.96 & -8.98 & -10.66 \\ \hline
\end{tabular}
\label{table:exchange_Js}
\end{table}

\begin{figure*}[htbp]
  \centering
  \includegraphics[width=\linewidth]{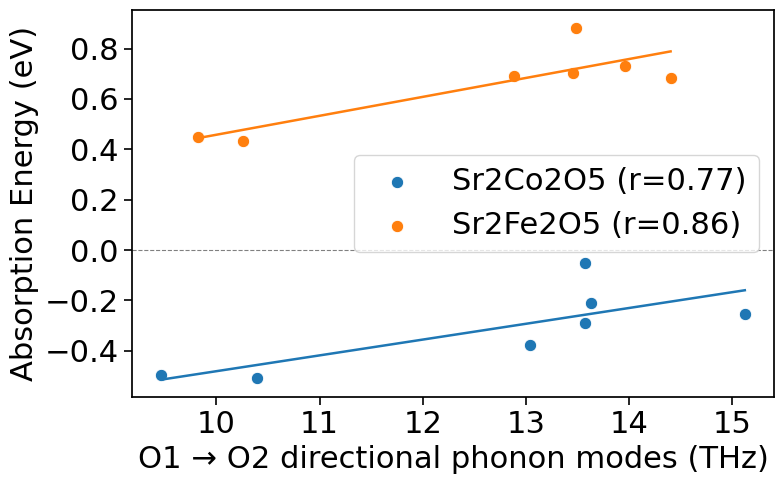}
  \caption{Correlation between the O\(_1\) → O\(_2\) directional phonon modes
           and the hydrogen-absorption energy for brownmillerite
           Sr\(_2\)Co\(_2\)O\(_5\) (blue) and Sr\(_2\)Fe\(_2\)O\(_5\) (orange).
           Here, O\(_1\) denotes the oxygen atom forming a covalent O–H bond
           with the absorbed H, whereas O\(_2\) is the hydrogen-bonded partner
           in the O–H···O linkage. The abscissa gives the averaged vibrational
           frequency (THz) of the O1-to-O2 mode at each crystallographic site;
           the ordinate shows the corresponding DFT-calculated
           hydrogen-absorption energy (eV). Solid lines are least-squares fits,
           with Pearson correlation coefficients \(r = 0.77\) (Co) and
           \(r = 0.86\) (Fe), illustrating that softer O\(_1\) → O\(_2\) modes
           systematically stabilize H absorption in both compounds.
           \label{fig:Directional_phonons}}
\end{figure*}

\begin{figure*}[htbp]
  \centering
  \includegraphics[width=\linewidth]{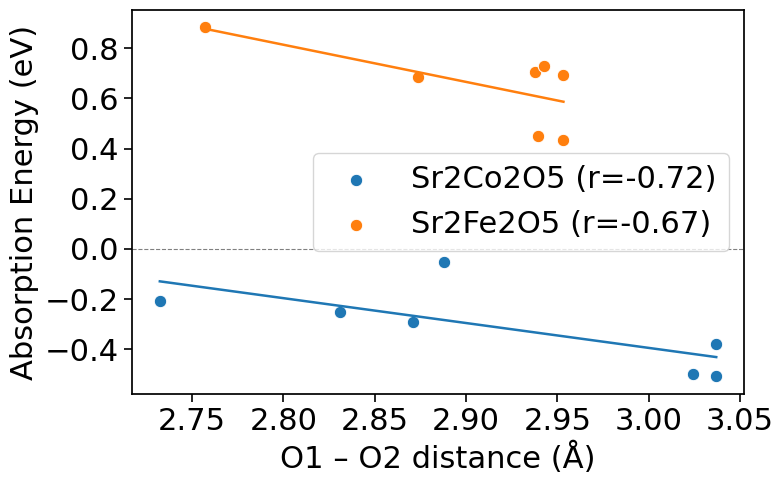}
  \caption{Correlation between the O$_1$–O$_2$ inter-site distance and the hydrogen-absorption energy for Sr$_2$Co$_2$O$_5$ (blue) and Sr$_2$Fe$_2$O$_5$ (orange).  More space in between O$_1$ and O$_2$, the lower the absorption energy.  Solid lines are least-squares fits to guide the eye.
  \label{fig:O1toO2distance}}
\end{figure*}

\begin{figure*}
    \centering
    \includegraphics[width=1\linewidth]{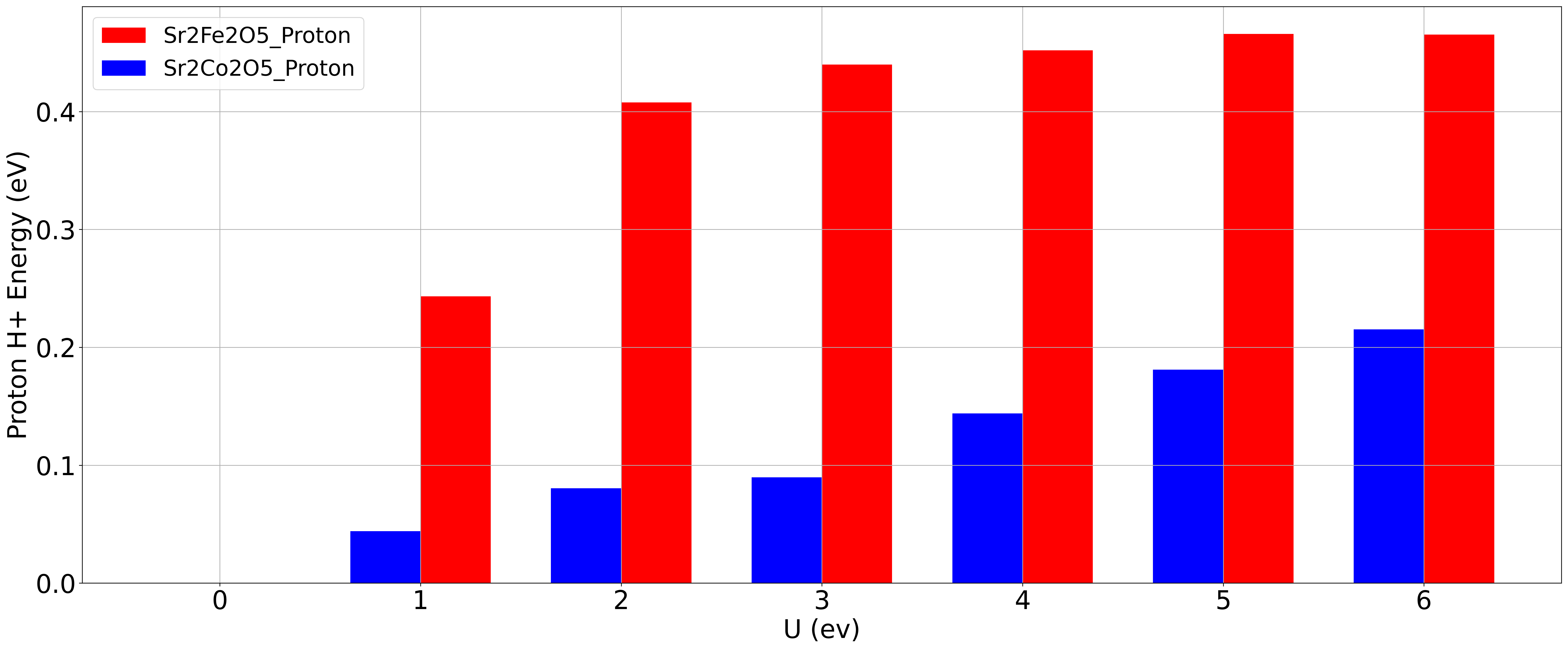}
    \caption{Energy cost (relative to the pristine structure) associated with introducing a proton (H\(^+\)) alone, plotted as a function of the Hubbard U value for Sr\(_2\)FeO\(_{2.5}\) (red) and Sr\(_2\)CoO\(_{2.5}\) (blue). As U increases, the energy difference arising from structural distortions grows, reaching up to 0.45 eV at the highest U values tested. }
    \label{fig:SCO_SFO_Hplus_no_e}
\end{figure*}

\begin{figure*}
    \centering
    \includegraphics[width=1.0\linewidth]{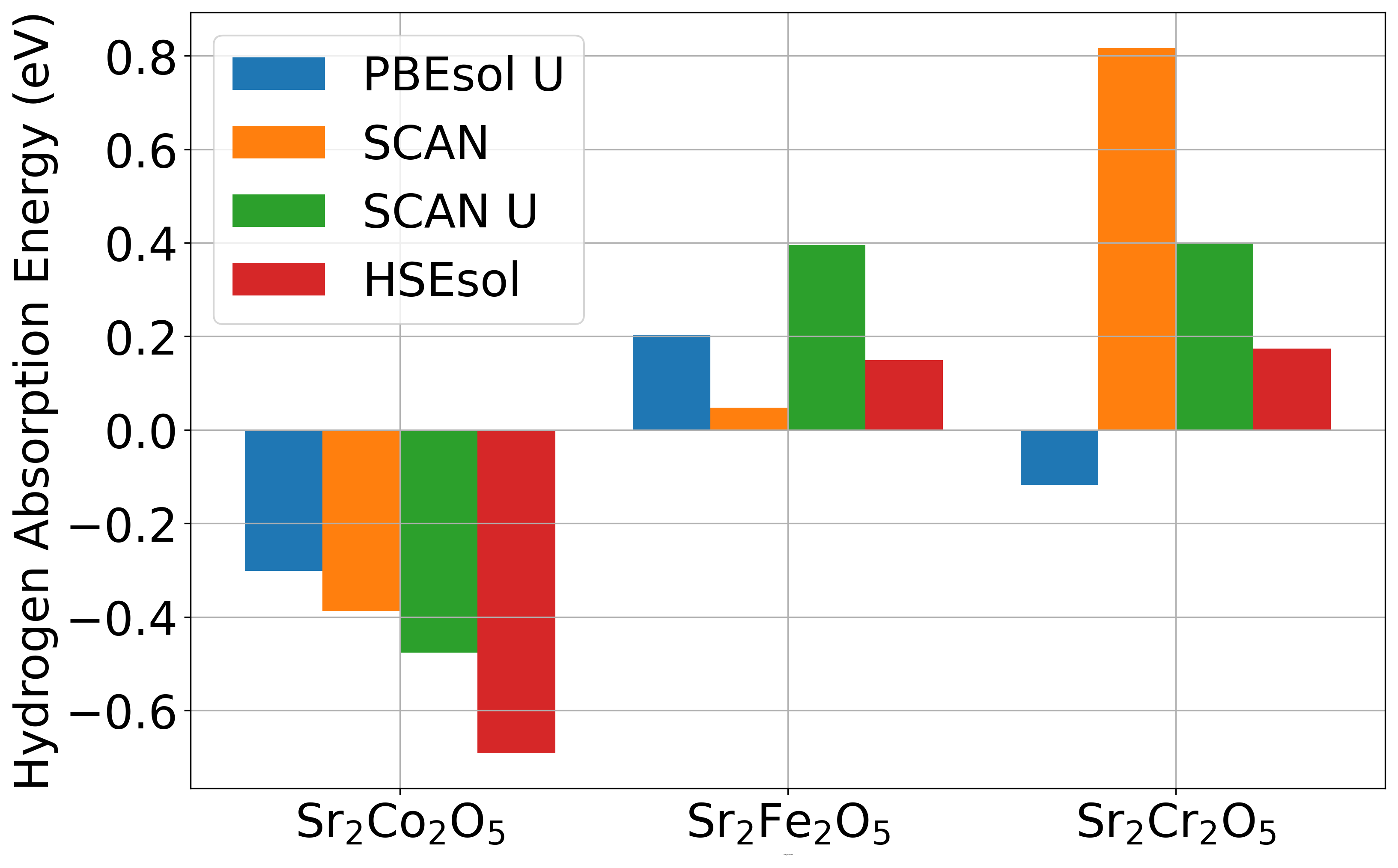}
    \caption{Influence of different functionals on hydrogen absorption energy for Sr\(_2\)Co\(_2\)O\(_5\), Sr\(_2\)Fe\(_2\)O\(_5\), and Sr\(_2\)Cr\(_2\)O\(_5\). Most stable H\(_1\) site and most stable electron localization metal site were selected. The functionals compared are PBEsol+U, SCAN, SCAN+U, and HSEsol.}
    \label{fig:functionals_H1_influence}
\end{figure*}

\begin{figure}
    \centering
    \includegraphics[width=1.0\linewidth]{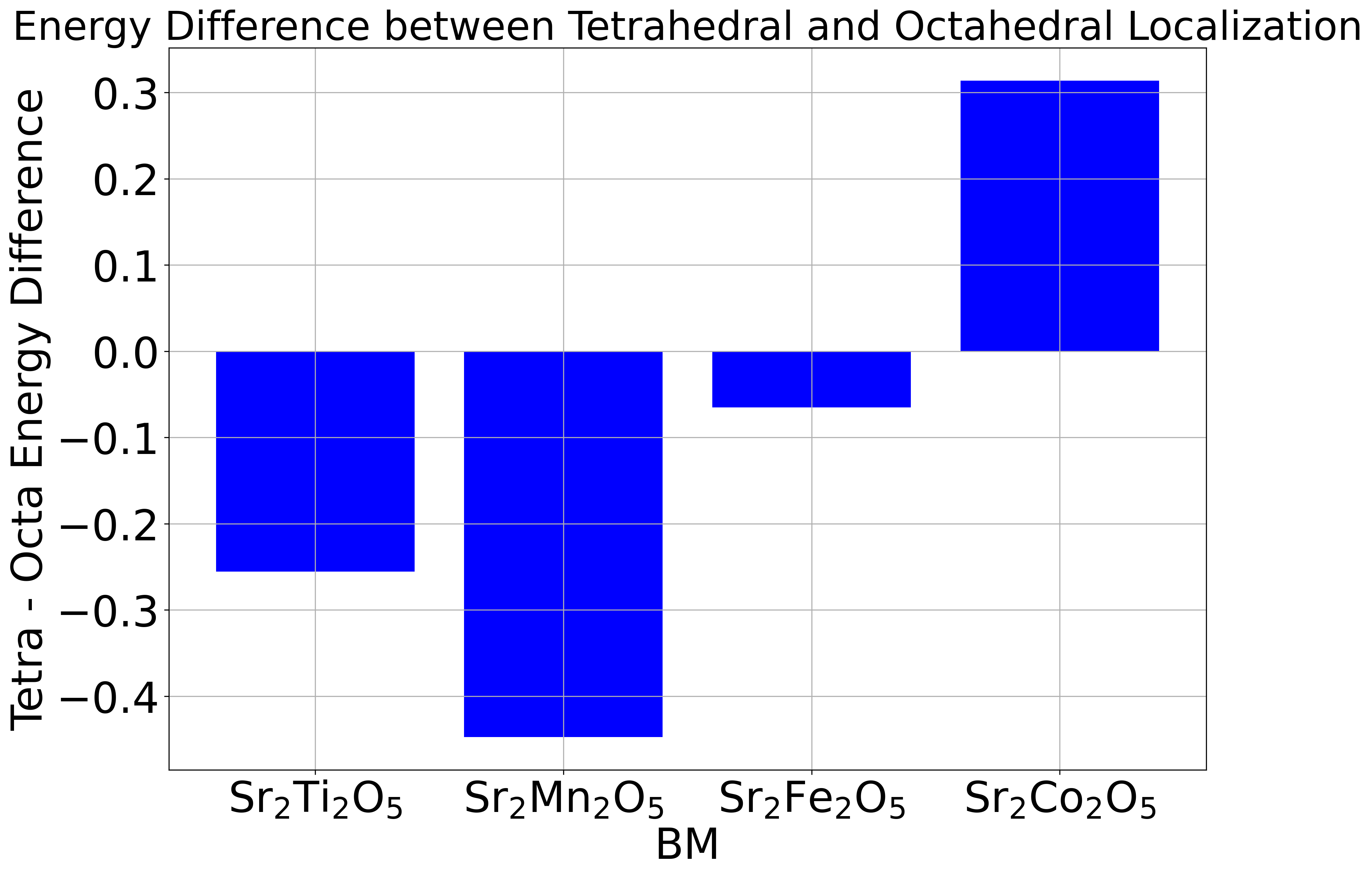}
    \caption{Energy difference for the fixed H\(_1\) position between electron localization on the tetrahedral site and the nearest octahedral site for various brownmillerite compounds. Negative values indicate a preference for tetrahedral localization, which is observed for elements up to d\(^6\) (e.g., Ti, Mn, Fe), whereas the d\(^7\) Co-based compound favors octahedral localization.}
    \label{fig:SrMeO_octa_tetra}
\end{figure}

\begin{figure*}[ht!]
    \centering
    \includegraphics[width=1.0\linewidth]{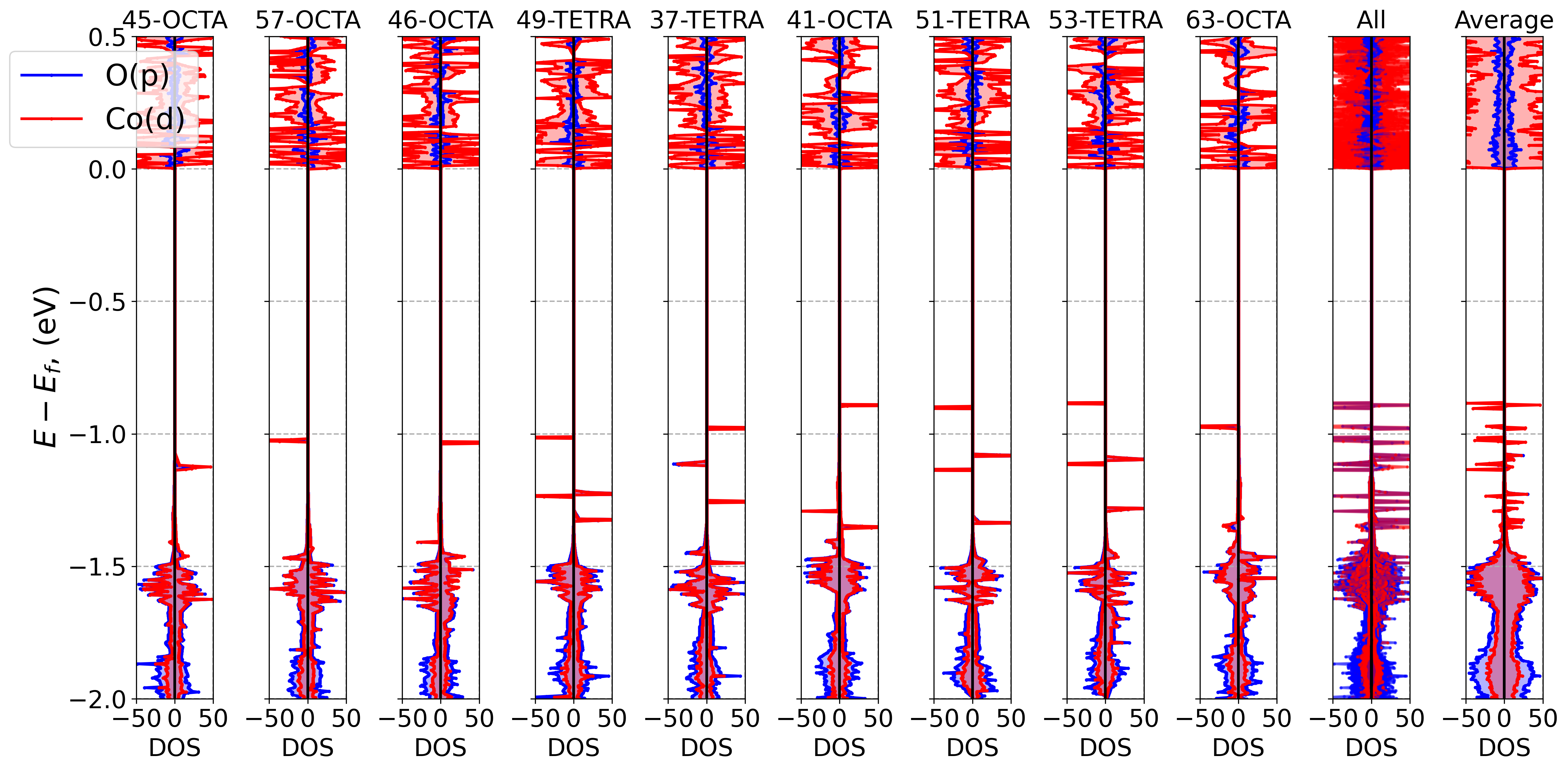}
    \caption{DOS for different electron localization Co sites}
    \label{fig:SCO_H1_eloc_DFT_STATIC}
\end{figure*}

\begin{figure*}[ht!]
    \centering
    \includegraphics[width=1.0\linewidth]{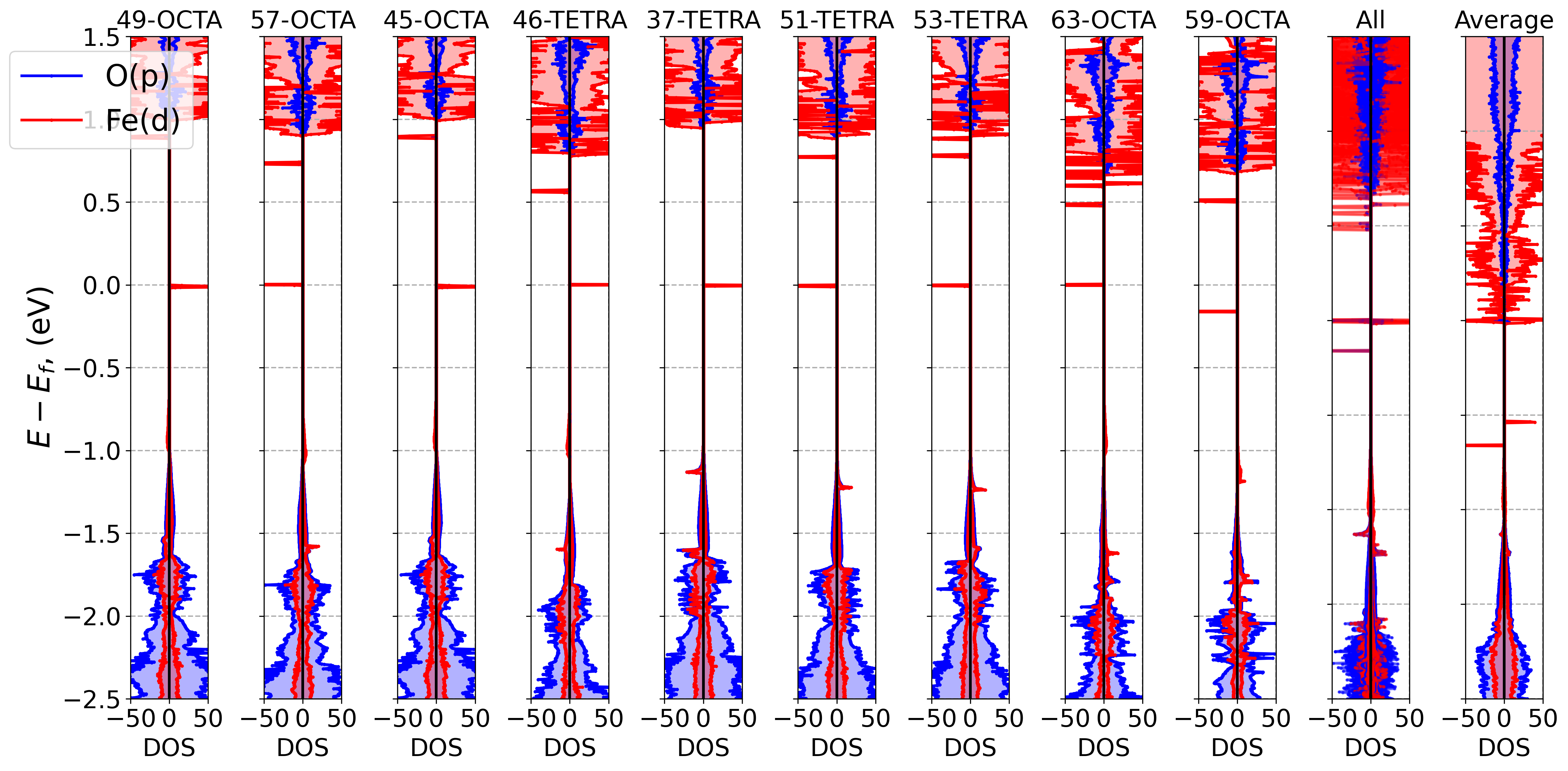}
    \caption{DOS for different electron localization Fe sites}
    \label{fig:SFO_H1_eloc_DFT_STATIC}
\end{figure*}

\begin{figure*}[ht!]
    \centering
    \includegraphics[width=1.0\linewidth]{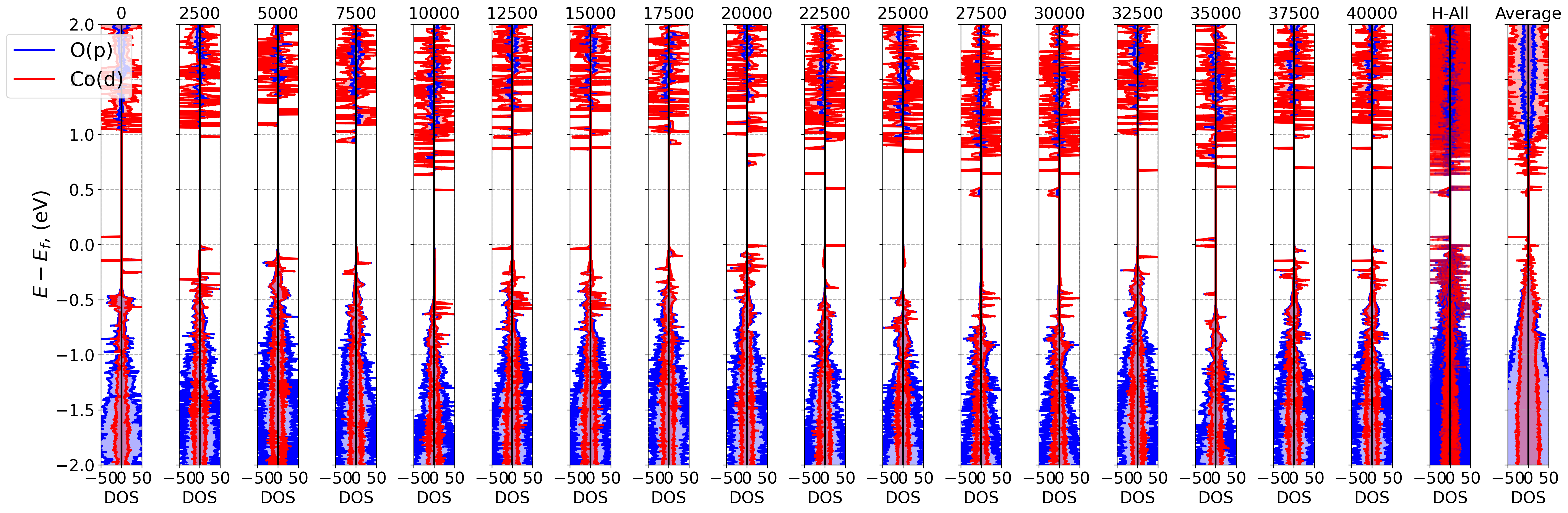}
    \caption{DOS for SrCoO$_{2.5}$ calculated for configurations for the AIMD simulations. The electron hops between sites, indicating polaron dynamics.}
    \label{fig:SCO_MD}
\end{figure*}

\begin{figure*}[ht!]
    \centering
    \includegraphics[width=1.0\linewidth]{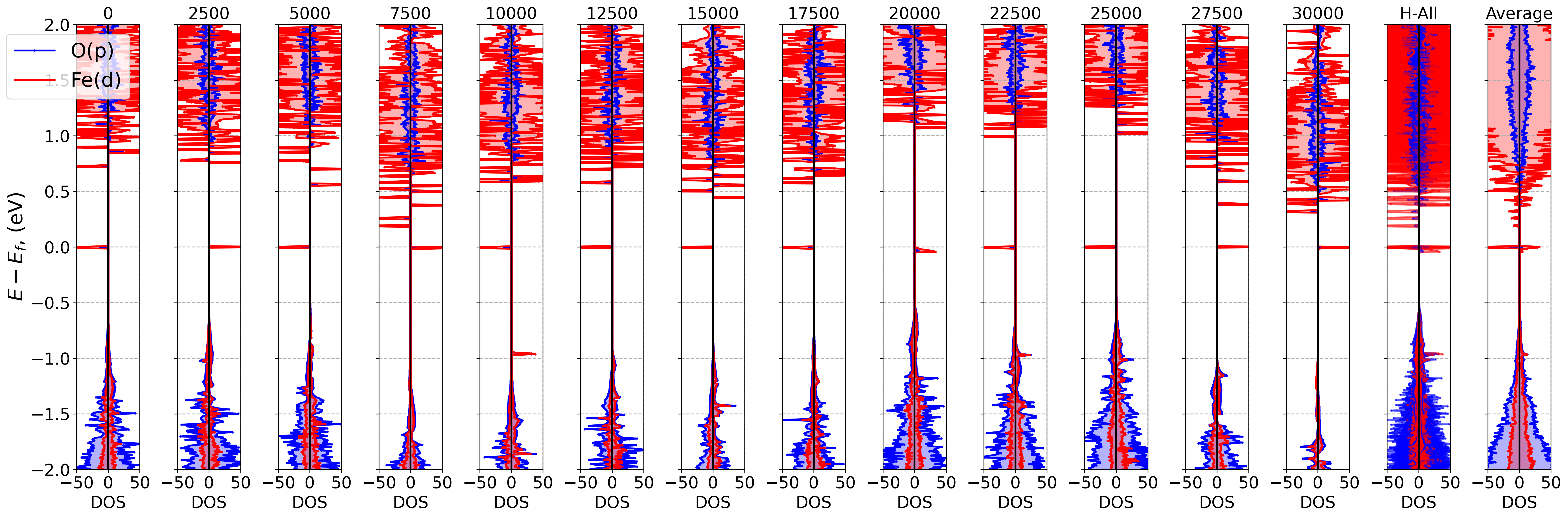}
    \caption{DOS for SrFeO$_{2.5}$ calculated for configurations for the AIMD simulations. The electron hops between sites, indicating polaron dynamics.}
    \label{fig:SFO_MD}
\end{figure*}

\begin{figure*}[ht!]
    \centering
    \includegraphics[width=1.0\linewidth]{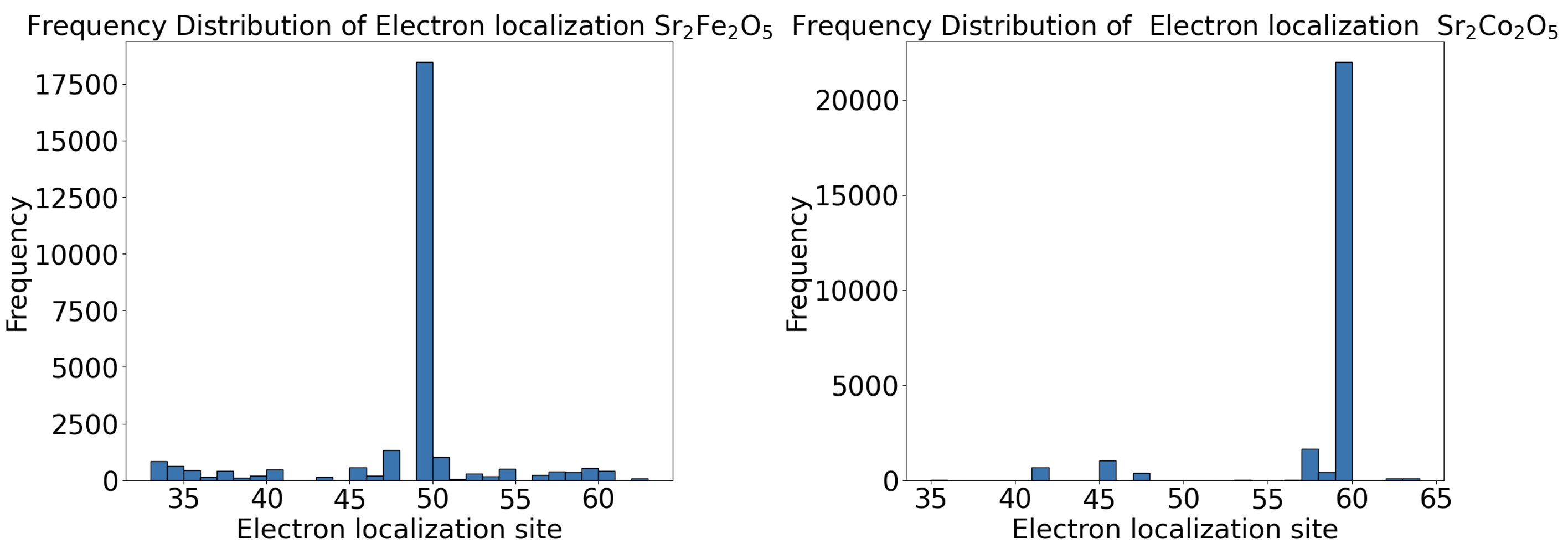}
    \caption{Frequency distribution of electron localization sites observed during AIMD runs for Sr\(_2\)Fe\(_2\)O\(_5\) (left) and Sr\(_2\)Co\(_2\)O\(_5\) (right). Although the electron occasionally hops among different metal sites, the most frequently occupied site corresponds to the one identified as most stable by static DFT calculations.}
    \label{fig:El_loc_site_MD}
\end{figure*}


\begin{figure*}[ht!]
    \centering
    \includegraphics[width=1.0\linewidth]{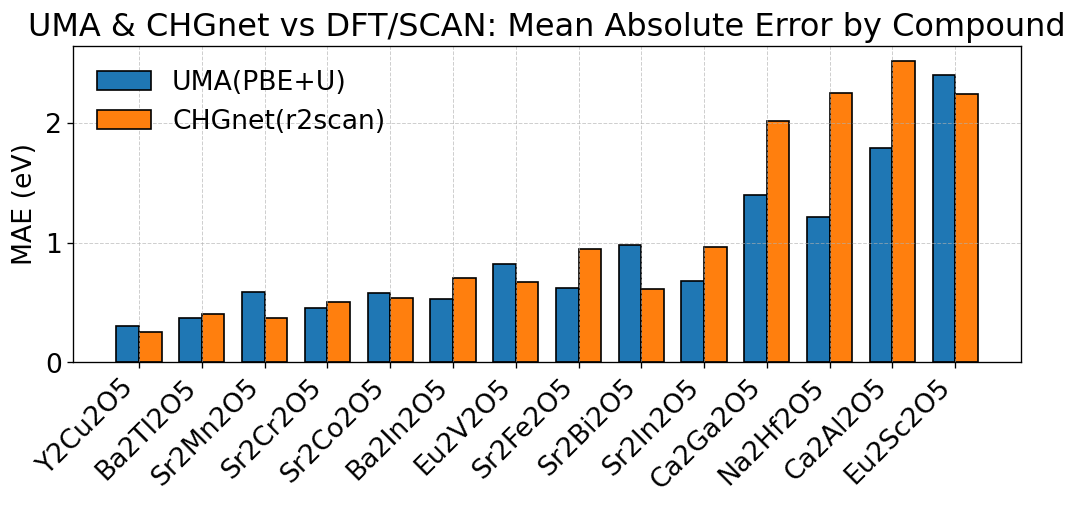}
    \caption{\textbf{Per–compound mean absolute error (MAE) against DFT/SCAN.}
    Side-by-side bars compare UMA and CHGnet for each brownmillerite; compounds are ordered by increasing MAE (averaged across models).
    The panel highlights composition-dependent transferability: UMA generally achieves lower MAE across most materials, while both models show larger errors for a few challenging chemistries.}
    \label{fig:mae_by_material}
\end{figure*}

\begin{figure}[!h]
    \centering
    \includegraphics[width=1.0\linewidth]{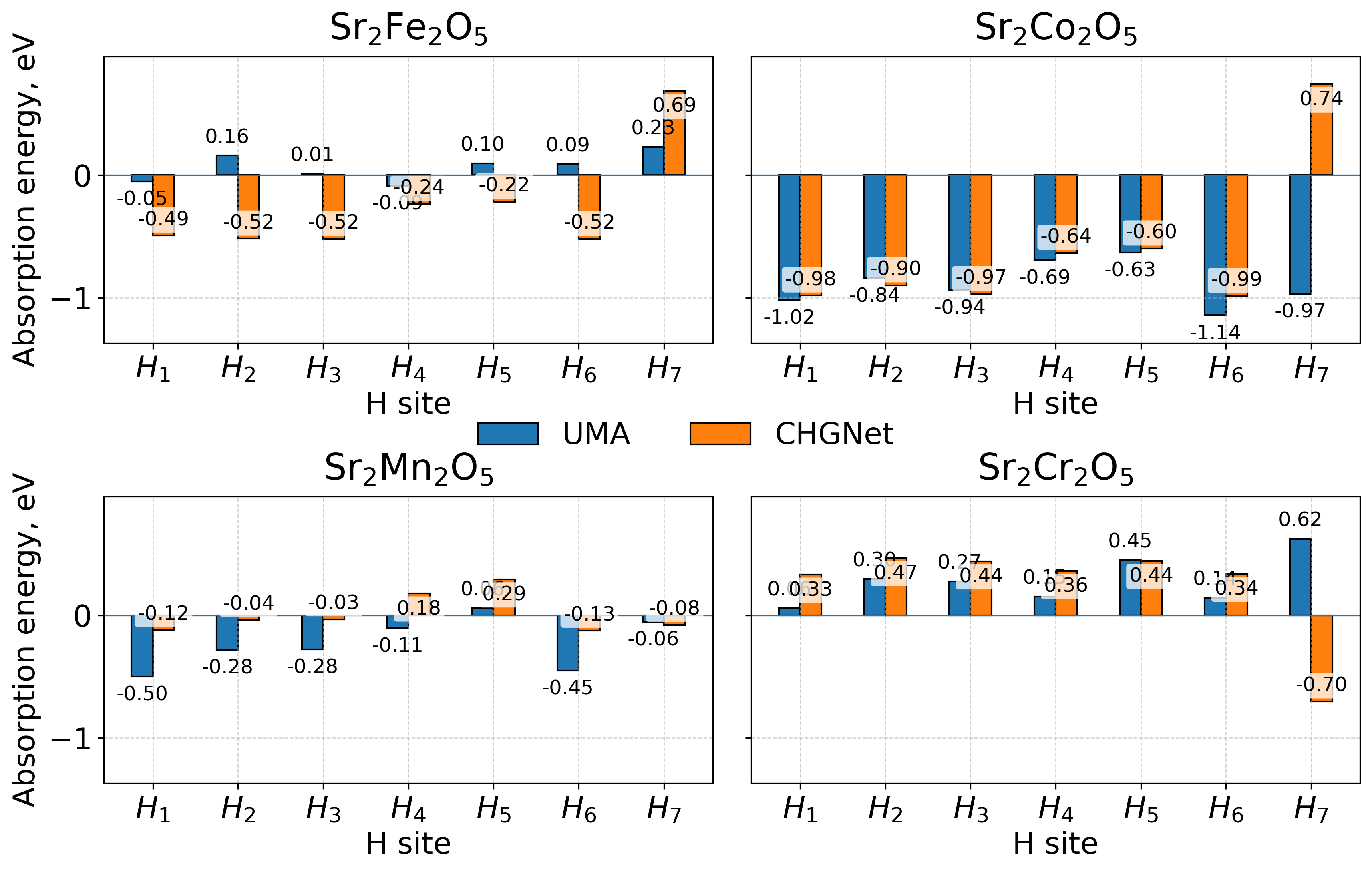}
    \caption{Site-resolved comparison of hydrogen absorption energies predicted by UMA and CHGnet for Sr$_2$X$_2$O$_5$ (X = Fe, Co, Mn, Cr). Each bar corresponds to one of the seven inequivalent interstitial hydrogen sites (H$_1$–H$_7$). DFT/SCAN consistently identifies H$_1$ as the most stable configuration, whereas UMA and CHGnet fails to reproduce this trend and yields qualitatively inconsistent site energetics. These discrepancies highlight the current limitations of MLIPs in describing the interplay of electron localization and magnetic interactions in complex oxides.}
    \label{fig:Site_to_site_CHGnet_and_UMA}
\end{figure}

\begin{figure*}[ht!]
    \centering
    \includegraphics[width=1.0\linewidth]{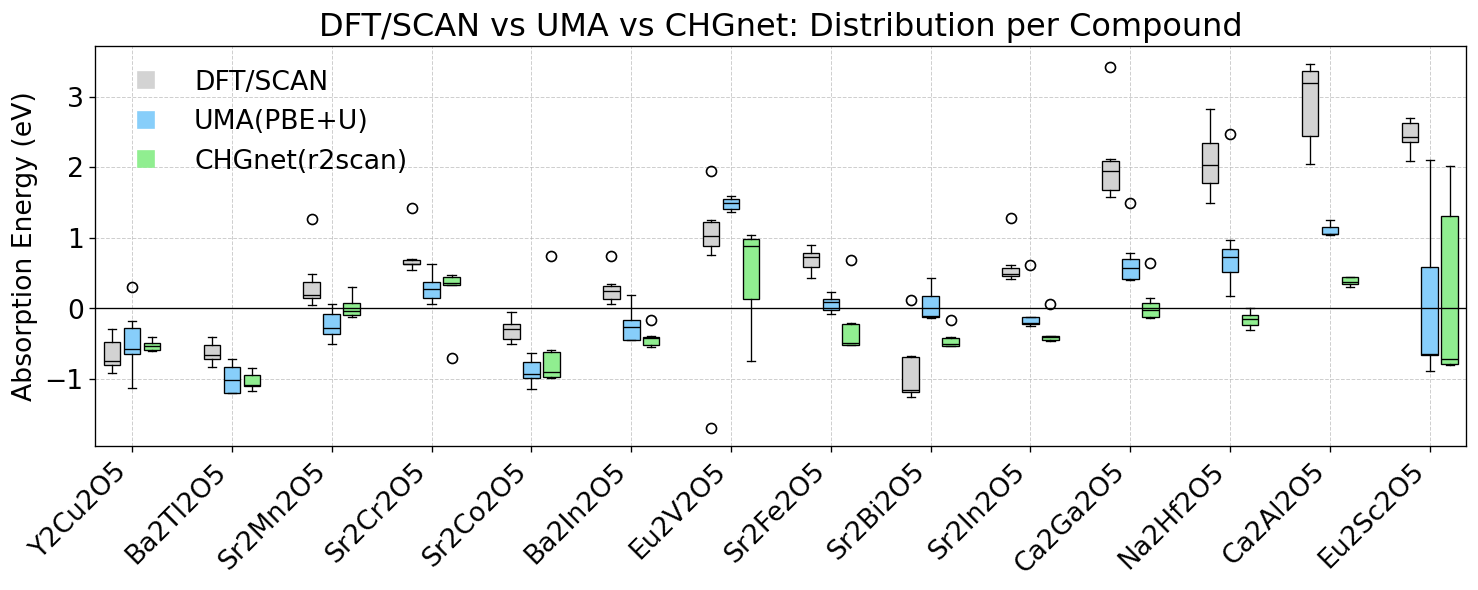}
    \caption{\textbf{Distribution of site-resolved absorption energies per compound.}
    For each material, boxplots summarize DFT/SCAN (gray), UMA (blue), and CHGnet (green) over all unique sites included in the filtered merges. Boxes show the interquartile range with medians; whiskers denote the central; points indicate outliers.
    The comparison reveals how each model reproduces the median and spread of DFT energies: UMA typically tracks DFT central tendency and variability more closely than CHGnet, while discrepancies are most apparent for compounds exhibiting broad DFT distributions. Despite reproducing the qualitative spread of site energetics, both models display compound-specific and site-specific failures, especially in cases with broad DFT distributions or extreme binding energies.}

    \label{fig:triple_boxplots}
\end{figure*}

\begin{figure*}[ht!]
    \centering
    \includegraphics[width=1.0\linewidth]{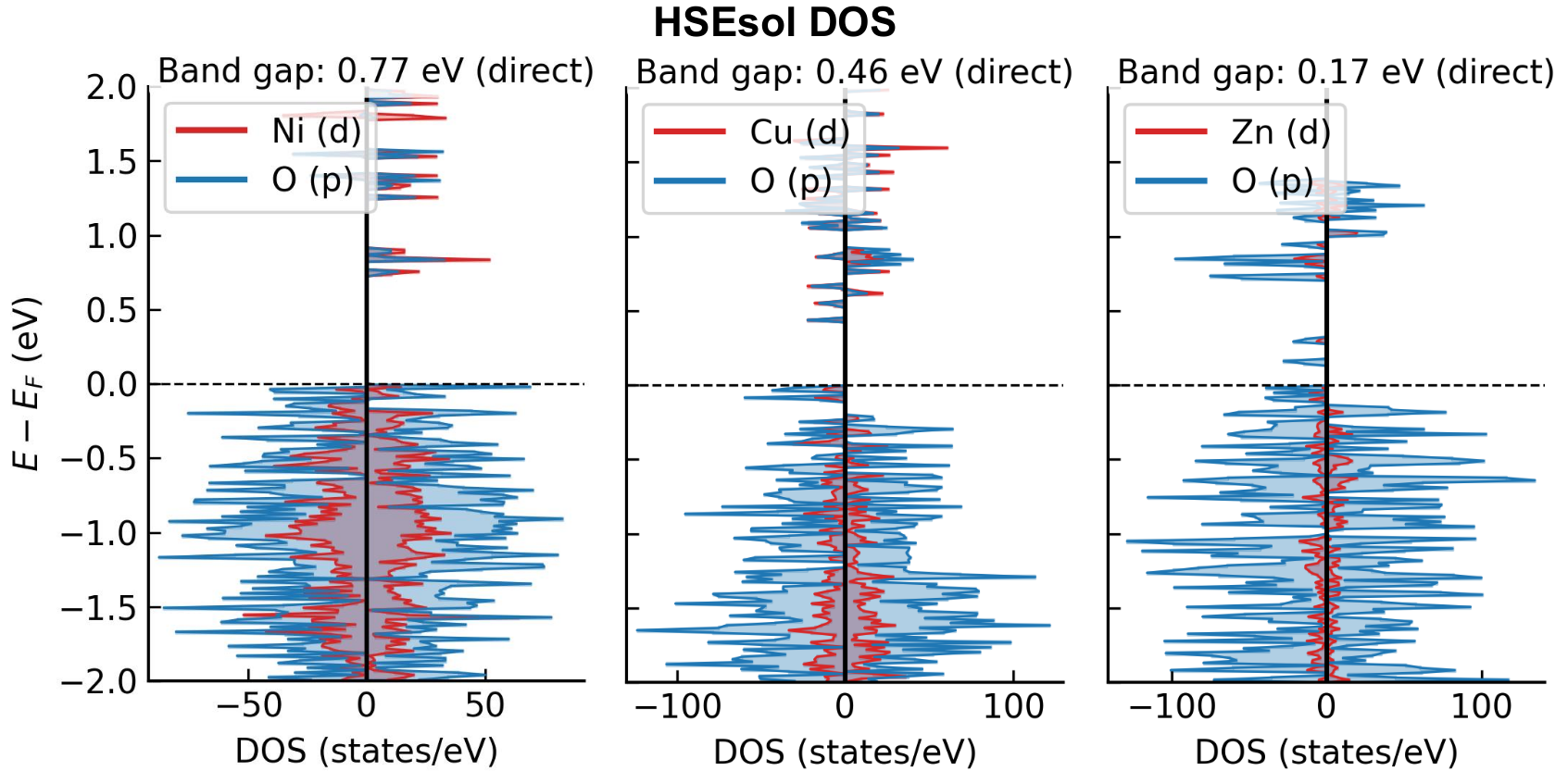}
    \caption{Electronic density of states calculated with the HSEsol functional for hydrogenated brownmillerites Sr$_2$Me$_2$O$_5$ (Me = Ni, Cu, Zn) containing H$_1$–e$^-$ pair. Small band gaps are preserved upon hydrogen insertion, in contrast to the SCAN and PBEsol results where no band gap is observed.}

    \label{fig:HSEsol_DOS}
\end{figure*}

\begin{figure*}[ht!]
    \centering
    \includegraphics[width=1.0\linewidth]{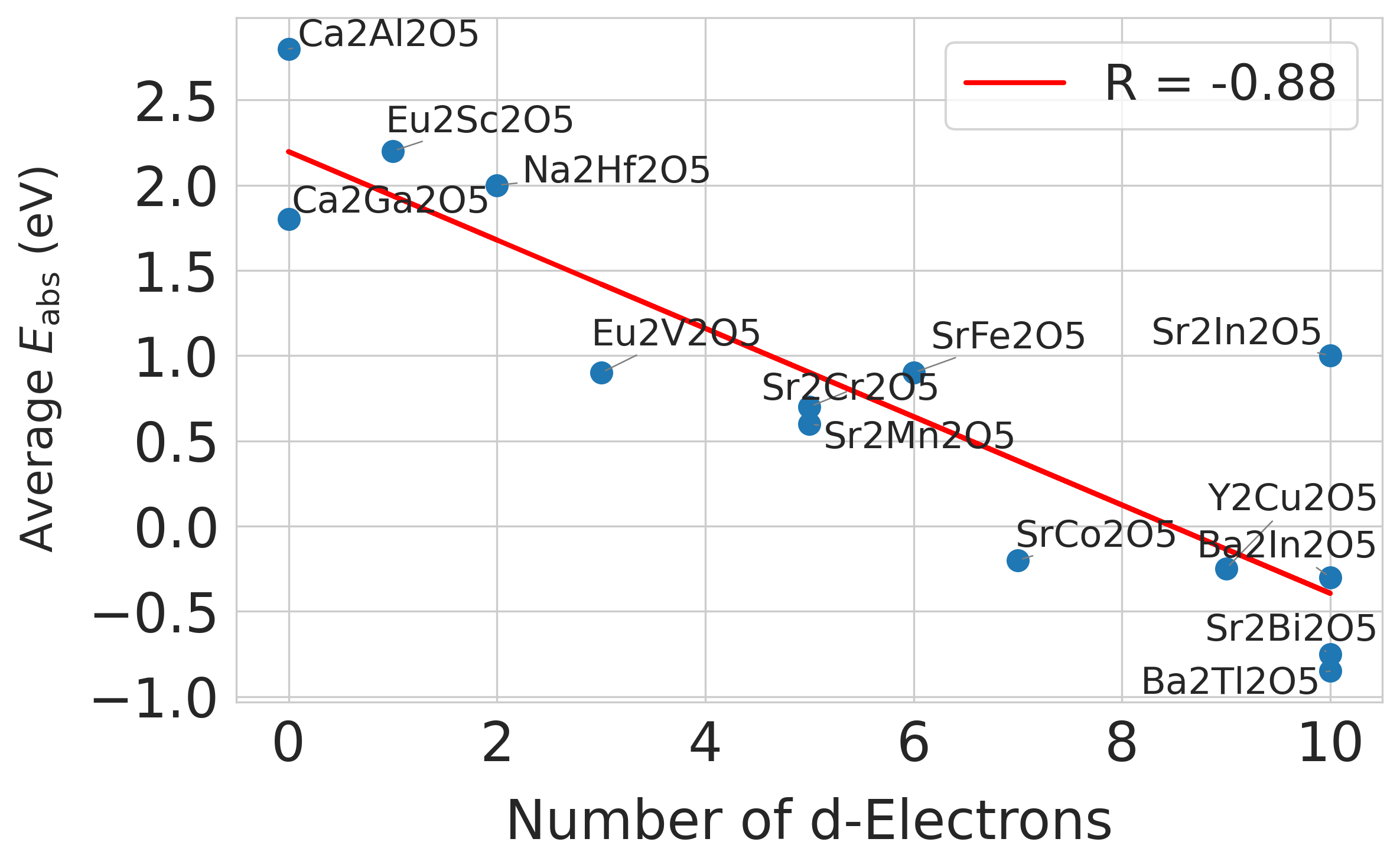}
    \caption{Correlation between B-site $d$-electron count and the average hydrogen absorption energy across 14 brownmillerite oxides. Each point represents the average absorption energy over 7 intercalation sites for a given compound.}
    \label{fig:avarage_H_abs_across_sites}
\end{figure*}

\begin{figure*}[ht!]
    \centering
    \includegraphics[width=1.0\linewidth]{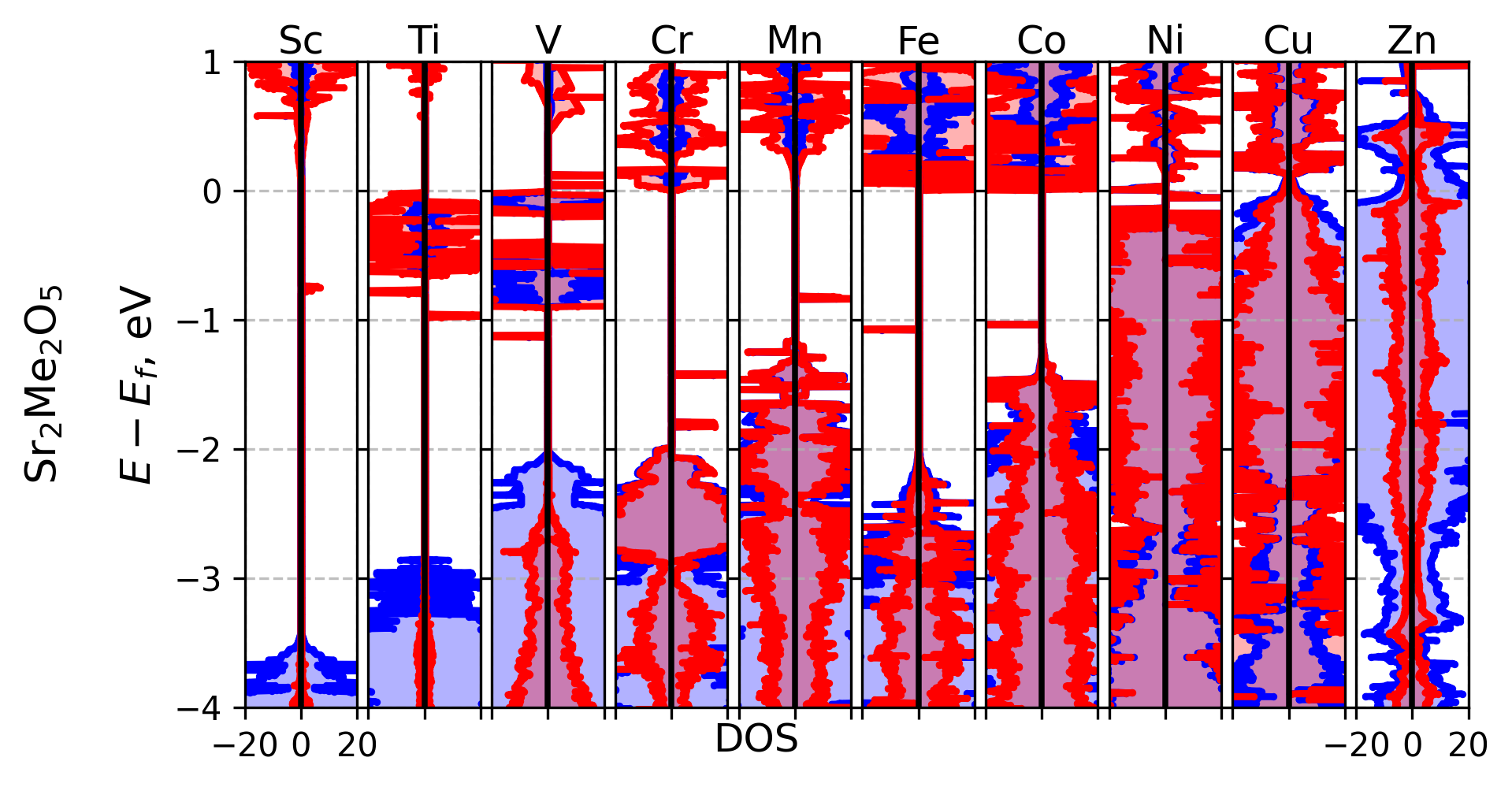}
    \caption{DOS for 3d transition metal brownmillerite oxides with hydrogen absorbed at the H$_{1}$ position. The projected DOS plots display the contributions from oxygen p-orbitals (blue) and metal d-orbitals (red) for each transition metal from Sc to Zn}
    \label{fig:3d_elem_DOS}
\end{figure*}

\begin{figure}[ht!]
    \centering
    \includegraphics[width=1\linewidth]{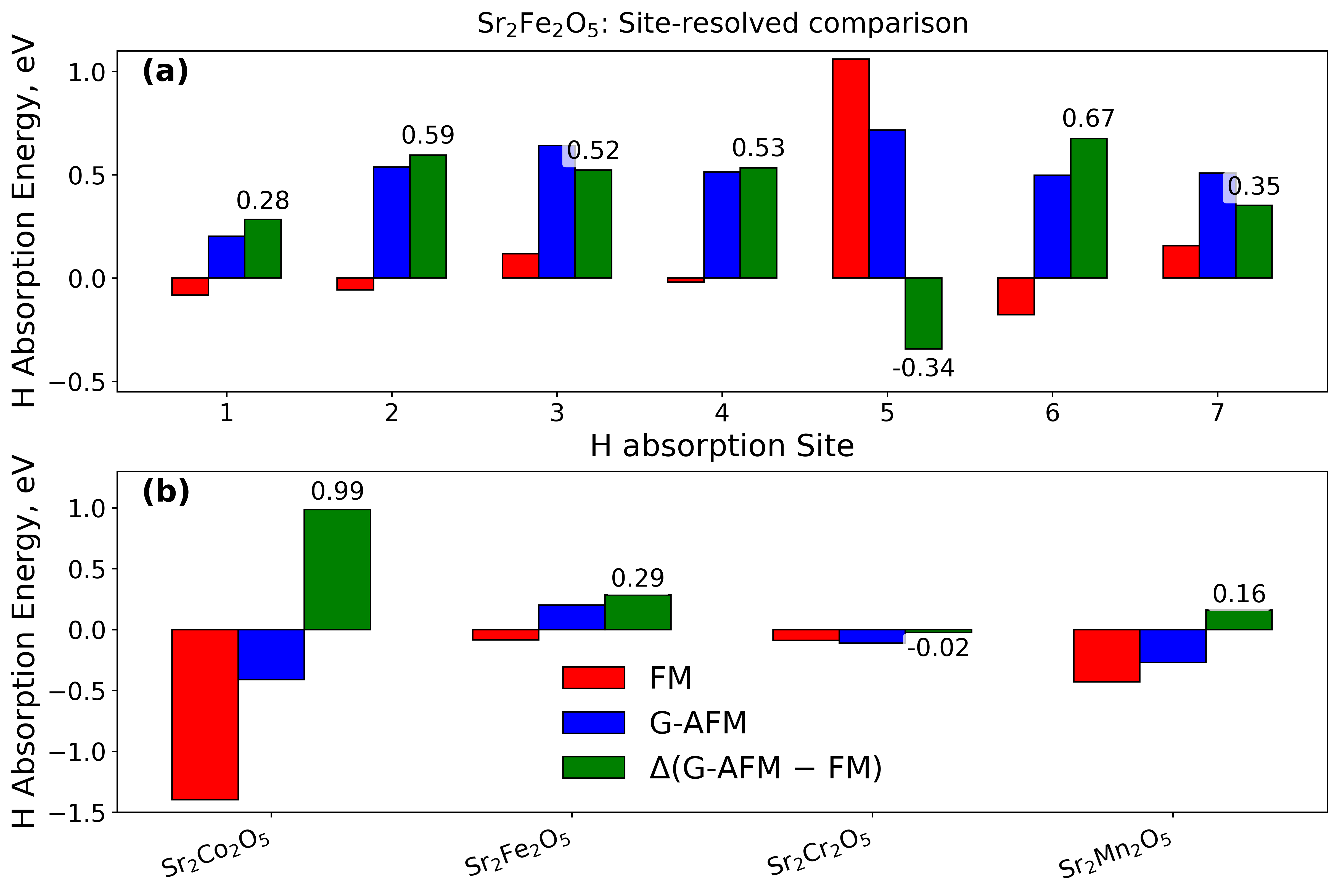}
    \caption{Hydrogen absorption energy variations induced by magnetic ordering (FM and G-AFM) (PBEsol(U) data). 
Panel (a) shows hydrogen absorption energies for Sr$_2$Fe$_2$O$_5$ at different intercalation sites in FM and G-AFM configurations, together with the energy difference $\Delta$(G-AFM $-$ FM). 
Panel (b) compares hydrogen absorption energies for the most stable H$_1$ site across different brownmillerite oxides (Sr$_2$Co$_2$O$_5$, Sr$_2$Fe$_2$O$_5$, Sr$_2$Cr$_2$O$_5$, and Sr$_2$Mn$_2$O$_5$) in both FM and G-AFM magnetic states.}
    \label{fig:FM_VS_AFM_PBEsolU}
\end{figure}

\begin{figure*}[ht!]
    \centering
    \includegraphics[width=1.0\linewidth]{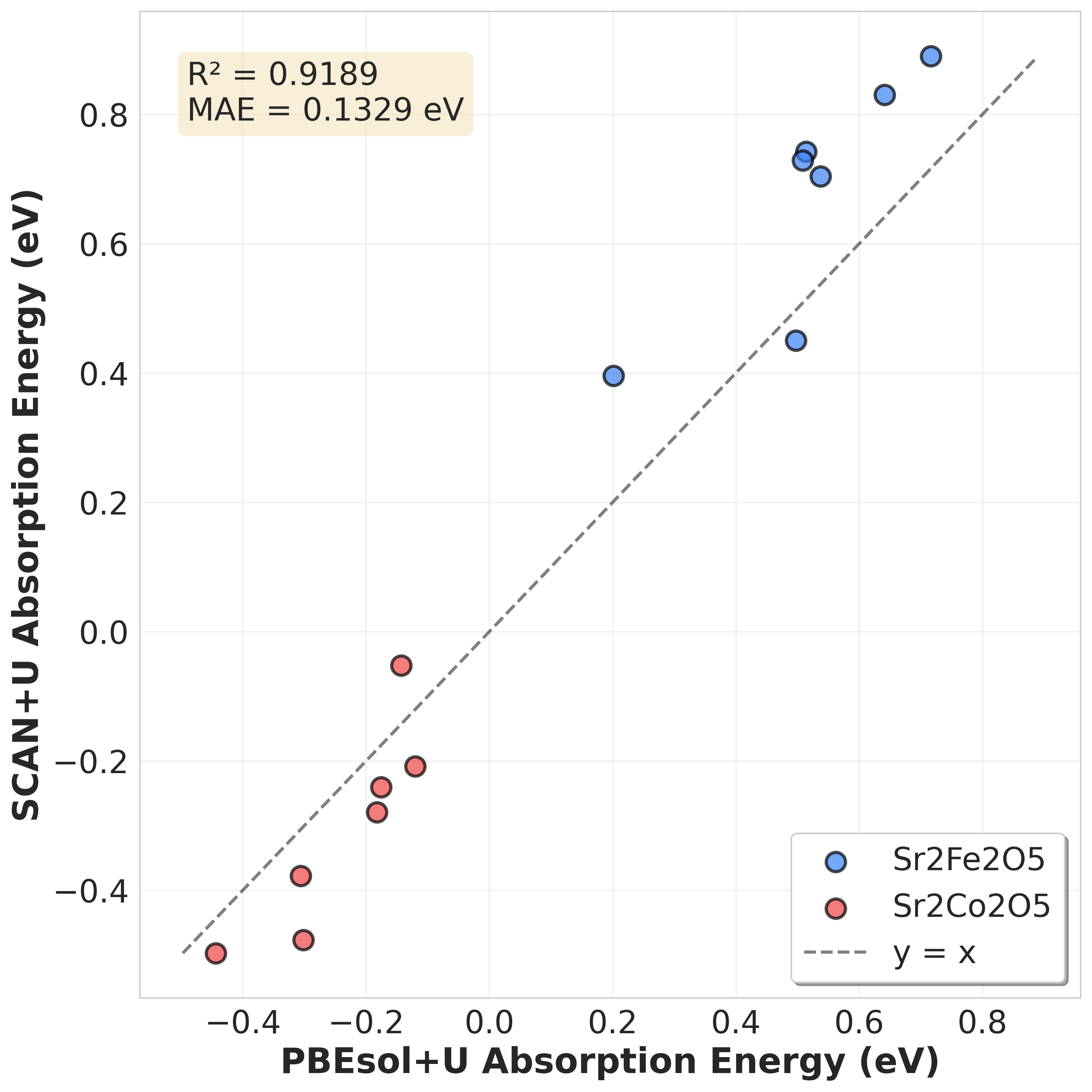}
    \caption{Comparison of hydrogen absorption energies calculated using PBEsol+$U$ and SCAN+$U$ for Sr$_2$Fe$_2$O$_5$ and Sr$_2$Co$_2$O$_5$ across different hydrogen intercalation sites. The dashed line indicates perfect agreement ($y=x$). Both functionals capture consistent qualitative trends, while quantitative differences remain modest, with a mean absolute error of 0.13 eV and typical deviations on the order of 0.1–0.2 eV.}
    \label{fig:SCANUvsPBEsolU}
\end{figure*}

\begin{table}[h]
\centering
\caption{Materials Project IDs for the 14 A$_2$B$_2$O$_5$ compounds investigated in this study.}
\label{tab:mp_ids}
\begin{tabular}{lll}
\hline
\textbf{Compound} & \textbf{Materials Project ID} \\
\hline
Ba$_2$In$_2$O$_5$ & mp-20546\\
Ca$_2$Ga$_2$O$_5$ & mp-16433\\
Sr$_2$Fe$_2$O$_5$ & mp-542018\\
Ca$_2$Al$_2$O$_5$ & mp-10444\\
Sr$_2$Co$_2$O$_5$ & mp-542899\\
Sr$_2$In$_2$O$_5$ & mp-22512\\
Y$_2$Cu$_2$O$_5$ & mp-510475\\
Ca$_2$Fe$_2$O$_5$ & mp-22113\\
Eu$_2$V$_2$O$_5$ & mp-1099750\\
Sr$_2$Mn$_2$O$_5$ & mp-1076718\\
Sr$_2$Bi$_2$O$_5$ & mp-23357\\
Ba$_2$Tl$_2$O$_5$ & mp-18327\\
Na$_2$Hf$_2$O$_5$ & mp-865500 \\
Eu$_2$Sc$_2$O$_5$ & mp-1076123 \\
\hline
\end{tabular}
\end{table}

\begin{figure*}[ht!]
    \centering
    \includegraphics[width=1.0\linewidth]{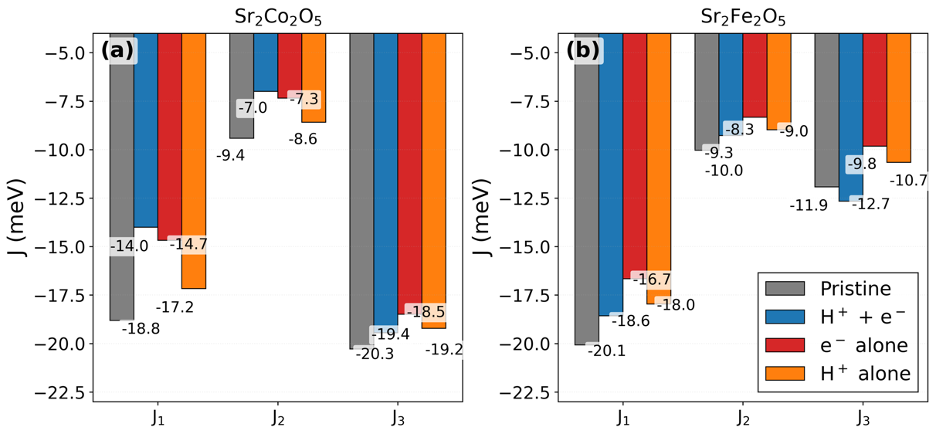}
    \caption{Decomposition of hydrogen defect effects on magnetic exchange couplings in Sr\(_2\)Co\(_2\)O\(_5\) and Sr\(_2\)Fe\(_2\)O\(_5\). Comparison of exchange coupling constants (\(J_1\), \(J_2\), \(J_3\)) for (a) Sr\(_2\)Co\(_2\)O\(_5\) and (b) Sr\(_2\)Fe\(_2\)O\(_5\) in four configurations: pristine (gray), full hydrogen defect (blue), electron-only doping (red), and proton-only doping (orange). All exchange constants are in meV.}
    \label{fig:Proton_and_e_effects}
\end{figure*}

\begin{figure}[h!]
    \centering
    \includegraphics[width=\linewidth]{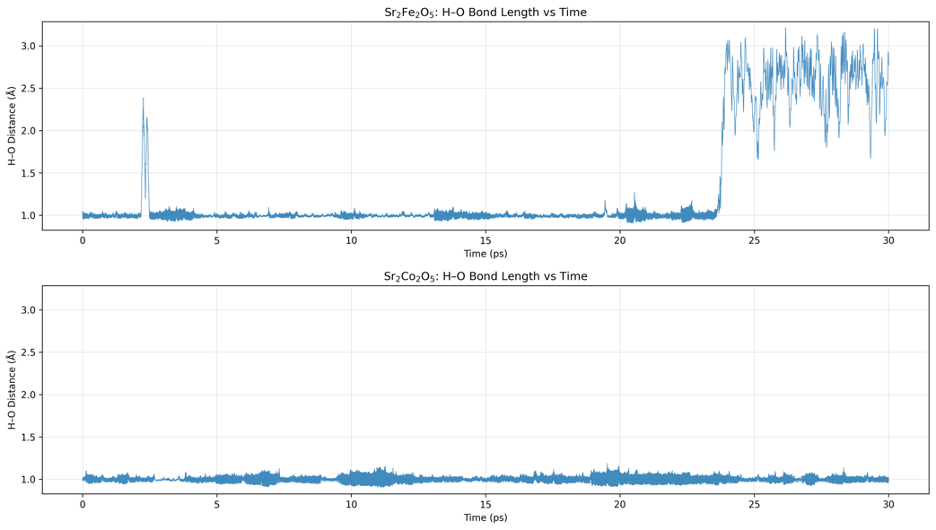}
    \caption{Hydrogen binding stability at 500 K in Sr\(_2\)Fe\(_2\)O\(_5\) and Sr\(_2\)Co\(_2\)O\(_5\). AIMD trajectories (30 ps) showing the H–O bond length as a function of time for Sr\(_2\)Fe\(_2\)O\(_5\) (top) and Sr\(_2\)Co\(_2\)O\(_5\) (bottom). In Sr\(_2\)Fe\(_2\)O\(_5\), hydrogen remains bonded near ~1.0 Å for most of the simulation, with intermittent excursions to larger H–O distances (e.g., near ~24 ps) consistent with local reorientation/hopping within a short-range O–H···O motif, rather than long-range diffusion. In contrast, Sr\(_2\)Co\(_2\)O\(_5\) exhibits tight fluctuations around ~1.0 Å throughout the trajectory with no pronounced excursions, indicating a more stable, persistent hydrogen bonding environment under identical thermal conditions.}
    \label{fig:HO_AIMD_500K}
\end{figure}

\begin{figure}[h!]
    \centering
    \includegraphics[width=\linewidth]{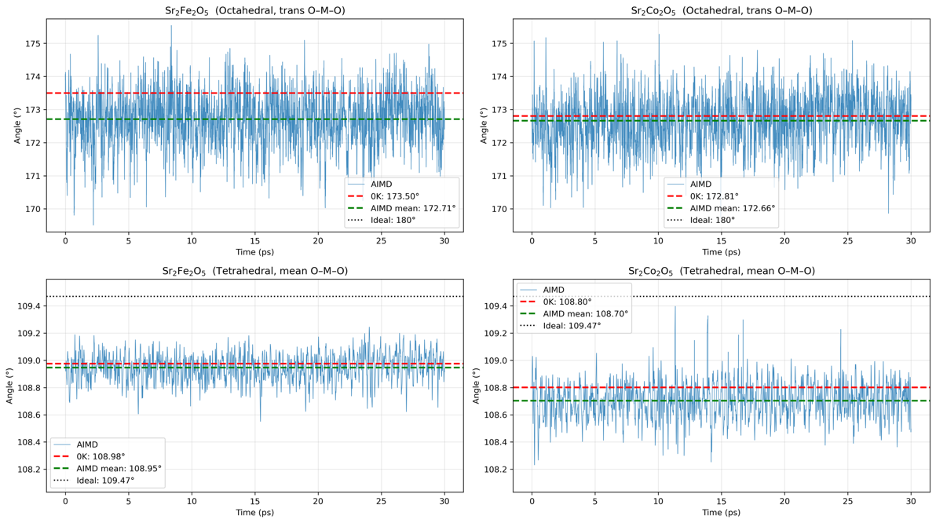}
    \caption{Thermal fluctuations of coordination polyhedra in Sr\(_2\)Fe\(_2\)O\(_5\) and Sr\(_2\)Co\(_2\)O\(_5\) at 500 K. Time evolution of the average O–M–O bond angles extracted from 30 ps AIMD trajectories for octahedral MO\(_6\) units (top row, trans O–M–O) and tetrahedral MO\(_4\) units (bottom row), shown for Sr\(_2\)Fe\(_2\)O\(_5\) (left column) and Sr\(_2\)Co\(_2\)O\(_5\) (right column). Blue curves show instantaneous AIMD values, while the 0 K DFT reference angles (red dashed) and AIMD time-averaged means (green dashed) are included for direct comparison; ideal reference angles are indicated by black dotted lines (180° for octahedra and 109.47° for tetrahedra). In both materials, the octahedral and tetrahedral coordination environments remain intact and exhibit only thermal fluctuations around their baseline values, with no evidence of coordination changes or framework instability over the simulated timescale.}
    \label{fig:polyhedra_fluctuations}
\end{figure}